\documentclass[%
 reprint,
 amsmath,amssymb,
 aps,
floatfix,
]{revtex4-1}
@article{TWPA2,
  title = {Resonant Phase Matching of Josephson Junction Traveling Wave Parametric Amplifiers},
  author = {O'Brien, Kevin and Macklin, Chris and Siddiqi, Irfan and Zhang, Xiang},
  journal = {Phys. Rev. Lett.},
  volume = {113},
  issue = {15},
  pages = {157001},
  numpages = {5},
  year = {2014},
  month = {Oct},
  publisher = {American Physical Society},
  doi = {10.1103/PhysRevLett.113.157001},
  url = {https://link.aps.org/doi/10.1103/PhysRevLett.113.157001}
}

@article{Planck,
      author         = "Ade, P. A. R. and others",
      title          = "{Planck 2013 results. XVI. Cosmological parameters}",
      collaboration  = "Planck",
      journal        = "Astron. Astrophys.",
      volume         = "571",
      year           = "2014",
      pages          = "A16",
      doi            = "10.1051/0004-6361/201321591",
      meprint         = "1303.5076",
      archivePrefix  = "arXiv",
      primaryClass   = "astro-ph.CO",
      reportNumber   = "CERN-PH-TH-2013-129",
      SLACcitation   = "
}

@article{Dine_1981,
      title = {A simple solution to the strong CP problem with a harmless axion},
  author = {Dine, Fischler, Srednicki},
  journal = {Physics Letters B},
  volume = {104},
  issue = {3},
  pages = {199--202},
  year = {1981},
  publisher = {ELSEVIER},
  doi = {https://doi.org/10.1016/0370-2693(81)90590-6}
}
@article{Kim_1979,
  title = {Weak-Interaction Singlet and Strong 
CP Invariance},
  author = {Kim, E. J.},
  journal = {Phys. Rev. Lett.},
  volume = {43},
  issue = {2},
  pages = {103},
  year = {1979},
  month = {July},
  publisher = {American Physical Society},
  doi = {https://doi.org/10.1103/PhysRevLett.43.103}
}

@article{Preskill_1983,
  title = {Cosmology of the invisible axion},
  author = {Preskill, J. and Wise, M. B. and Wilczek, F.},
  journal = {Phys. Rev. Lett. B},
  volume = {120},
  issue = {127},
  year = {1983},
  publisher = {American Physical Society},
  doi = {10.1016/0370-2693(83)90637-8}
 } 
 
 @article{Sikivie_1983,
  title = {A Cosmological bound on the invisible axion},
  author = {Abbott, L. F. and Sikivie, P.},
  journal = {Phys. Rev. Lett. B},
  volume = {120},
  issue = {133},
  year = {1983},
  publisher = {American Physical Society},
  doi = {10.1016/0370-2693(83)90638-X}
 }
 
 @article{Dine_1983,
  title = {The not-so-harmless axion},
  author = {Dine, M. and Fischler, W.},
  journal = {Phys. Rev. Lett. B},
  volume = {120},
  issue = {137},
  year = {1983},
  publisher = {American Physical Society},
  doi = {10.1016/0370-2693(83)90639-1}
}

@article{Vainshtein_1980,
  title = {Can confinement ensure natural CP invariance of strong interactions?},
  author = {Vainshtein, Shifman, Zakharov},
  journal = {Nuclear Physics B},
  volume = {166},
  issue = {3},
  pages = {493--506},
  year = {1980},
  month = {April},
  publisher = {ELSEVIER},
  doi = {https://doi.org/10.1016/0550-3213(80)90209-6}
}

@article{Sikivie1985,
  title = {Detection rates for ``invisible''-axion searches},
  author = {Sikivie, P.},
  journal = {Phys. Rev. D},
  volume = {32},
  issue = {11},
  pages = {2988--2991},
  numpages = {0},
  year = {1985},
  month = {Dec},
  publisher = {American Physical Society},
  doi = {10.1103/PhysRevD.32.2988},
  url = {https://link.aps.org/doi/10.1103/PhysRevD.32.2988}
}

@article{pig,
      author         = "Aune, S. and others",
      title          = "{CAST search for sub-eV mass solar axions with $^{3}He$ buffer
                        gas}",
      collaboration  = "CAST",
      journal        = "Phys. Rev. Lett.",
      volume         = "107",
      year           = "2011",
      pages          = "261302",
      doi            = "10.1103/PhysRevLett.107.261302",
      meprint         = "1106.3919",
      archivePrefix  = "arXiv",
      primaryClass   = "hep-ex",
      SLACcitation   = "
}

@article{CAST,
  title = {Search for Sub-eV Mass Solar Axions by the CERN Axion Solar Telescope with $^{3}\mathrm{He}$ Buffer Gas},
  author = {M. Arik and others},
  collaboration = {CAST Collaboration},
  journal = {Phys. Rev. Lett.},
  volume = {107},
  issue = {26},
  pages = {261302},
  numpages = {5},
  year = {2011},
  month = {Dec},
  publisher = {American Physical Society},
  doi = {10.1103/PhysRevLett.107.261302},
  url ={https://link.aps.org/doi/10.1103/PhysRevLett.107.261302}
}

@misc{attocube,
	author    = "Attocube",
	title     = "Premium Line Introduction",
	note     = "http://www.attocube.com/attomotion/",
	year      = "2018",
	journal   = "",
	volume   = "",
	number   = "",
	pages    = "",
	month    = "",
}

@misc{attocube2,
	author    = "attocube systems AG",
	title     = "",
	note     = "Munich, Germany",
	year      = "",
	journal   = "",
	volume   = "",
	number   = "",
	pages    = "",
	month    = "",
}

@misc{HFSS2,
	author  =  {{ANSYS$^{\tiny{\textregistered}}$ HFSS}},
	title     = "Release 19.0",
	url     = "https://www.ansys.com/products/electronics/ansys-hfss",
	year      = "2018"
}   

@software{HFSS,
  author = {{ANSYS HFSS, Release 19.0}},
  title = {},
  url = {https://www.ansys.com/products/electronics/ansys-hfss},
  version = {}
}

@techreport{AttocubeHeating,
     title = {{Power dissipation of a piezo}},
     author = {Tobias Lindenberg and Christoph B{\"o}defeld},
     group = {},
     year = {2013},
     institution = {attocube systems AG, Munich, Germany},
     month = {October},
     Date-Added = {},
     Date-Modified = {}
}

@article{DMdensity,
  author={J I Read},
  title={The local dark matter density},
  journal={Journal of Physics G: Nuclear and Particle Physics},
  volume={41},
  number={6},
  pages={063101},
  url={http://stacks.iop.org/0954-3899/41/i=6/a=063101},
  year={2014},
  abstract={I review current efforts to measure the mean density of dark matter near the Sun. This encodes valuable dynamical information about our Galaxy and is also of great importance for âdirect detectionâ dark matter experiments. I discuss theoretical expectations in our current cosmology; the theory behind mass modelling of the Galaxy; and I show how combining local and global measures probes the shape of the Milky Way dark matter halo and the possible presence of a âdark discâ. I stress the strengths and weaknesses of different methodologies and highlight the continuing need for detailed tests on mock dataâparticularly in the light of recently discovered evidence for disequilibria in the Milky Way disc. I collate the latest measurements of Ï dm and show that, once the baryonic surface density contribution Î£ b is normalized across different groups, there is remarkably good agreement. Compiling data from the literature, I estimate Î£ b = 54.2 Â± 4.9âM â pc â2 , where the dominant source of uncertainty is in theHÂi gas contribution. Assuming this contribution from the baryons, I highlight several recent measurements of Ï dm in order of increasing data complexity and prior, and, correspondingly, decreasing formal error bars. Comparing these measurements with spherical extrapolations from the Milky Wayâs rotation curve, I show that the Milky Way is consistent with having a spherical dark matter halo at R 0 âŒ 8 kpc. The very latest measures of Ï dm based on âŒ10Â000 stars from the Sloan Digital Sky Survey appear to favour little halo flattening at R 0 , suggesting that the Galaxy has a rather weak dark matter disc, with a correspondingly quiescent merger history. I caution, however, that this result hinges on there being no large systematics that remain to be uncovered in the SDSS data, and on the local baryonic surface density being Î£ b âŒ 55âM â pc â2 . I conclude by discussing how the new Gaia satellite will be transformative. We will obtain much tighter constraints on both Î£ b and Ï dm by having accurate 6D phase space data for millions of stars near the Sun. These data will drive us towards fully three dimensional models of our Galactic potential, moving us into the realm of precision measurements ofÂÏ dm .}
}

@article{dog,
      author         = "Arias, Paola and Cadamuro, Davide and Goodsell, Mark and
                        Jaeckel, Joerg and Redondo, Javier and Ringwald, Andreas",
      title          = "{WISPy Cold Dark Matter}",
      journal        = "JCAP",
      volume         = "1206",
      year           = "2012",
      pages          = "013",
      doi            = "10.1088/1475-7516/2012/06/013",
      meprint         = "1201.5902",
      archivePrefix  = "arXiv",
      primaryClass   = "hep-ph",
      reportNumber   = "DESY-11-226, MPP-2011-140, CERN-PH-TH-2011-323,
                        IPPP-11-80, DCPT-11-160, --DCPT-11-160",
      SLACcitation   = "
}

@article{ALPS,
  author={Paola Arias and Davide Cadamuro and Mark Goodsell and Joerg Jaeckel and Javier Redondo and Andreas Ringwald},
  title={WISPy cold dark matter},
  journal={Journal of Cosmology and Astroparticle Physics},
  volume={2012},
  number={06},
  pages={013},
  url={http://stacks.iop.org/1475-7516/2012/i=06/a=013},
  year={2012},
  abstract={Very weakly interacting slim particles (WISPs), such as axion-like particles (ALPs) or hidden photons (HPs), may be non-thermally produced via the misalignment mechanism in the early universe and survive as a cold dark matter population until today. We find that, both for ALPs and HPs whose dominant interactions with the standard model arise from couplings to photons, a huge region in the parameter spaces spanned by photon coupling and ALP or HP mass can give rise to the observed cold dark matter. Remarkably, a large region of this parameter space coincides with that predicted in well motivated models of fundamental physics. A wide range of experimental searches â exploiting haloscopes (direct dark matter searches exploiting microwave cavities), helioscopes (searches for solar ALPs or HPs), or light-shining-through-a-wall techniques â can probe large parts of this parameter space in the foreseeable future.}
}

@article{TWPA,
	Author = {CULLEN, A.  L. },
	Date = {1958/02/01/online},
	Date-Added = {2018-06-19 19:06:35 +0000},
	Date-Modified = {2018-06-19 19:06:35 +0000},
	Day = {01},
	Journal = {Nature},
	L3 = {10.1038/181332a0; },
	Month = {02},
	Pages = {332 EP  -},
	Publisher = {Nature Publishing Group SN  -},
	Title = {A Travelling-Wave Parametric Amplifier},
	Ty = {JOUR},
	Url = {http://dx.doi.org/10.1038/181332a0},
	Volume = {181},
	Year = {1958},
	Bdsk-Url-1 = {http://dx.doi.org/10.1038/181332a0}}

@phdthesis{thesisBoutan,
  author        = "Boutan, Christian",
  title         = "A Piezoelectrically Tuned RF-Cavity Search for Dark Matter Axions",
  school        = "University of Washington",
  address       = "",
  month         = "",
  year          = "2017",
  url           = "https://digital.lib.washington.edu/researchworks/handle/1773/38675"
}

@article{Sikivie:1983ip,
      author         = "Sikivie, P.",
      title          = "{Experimental Tests of the Invisible Axion}",
      journal        = "Phys. Rev. Lett.",
      volume         = "51",
      pages          = "1415-1417",
      doi            = "10.1103/PhysRevLett.51.1415",
      year           = "1983",
      reportNumber   = "PRINT-83-0597 (FLORIDA), UF-TP-83-13, C83-08-04",
      SLACcitation   = "
}

@article{Asztalos:2009yp,
  title = {SQUID-Based Microwave Cavity Search for Dark-Matter Axions},
  author = {Asztalos, S. J. and Carosi, G. and Hagmann, C. and Kinion, D. and van Bibber, K. and Hotz, M. and Rosenberg, L. J and Rybka, G. and Hoskins, J. and Hwang, J. and Sikivie, P. and Tanner, D. B. and Bradley, R. and Clarke, J.},
  journal = {Phys. Rev. Lett.},
  volume = {104},
  issue = {4},
  pages = {041301},
  numpages = {4},
  year = {2010},
  month = {Jan},
  publisher = {American Physical Society},
  doi = {10.1103/PhysRevLett.104.041301},
  url = {https://link.aps.org/doi/10.1103/PhysRevLett.104.041301}
}

@article{PhysRevD.42.3572,
  title = {Periodic signatures for the detection of cosmic axions},
  author = {Turner, Michael S.},
  journal = {Phys. Rev. D},
  volume = {42},
  issue = {10},
  pages = {3572--3575},
  numpages = {0},
  year = {1990},
  month = {Nov},
  publisher = {American Physical Society},
  doi = {10.1103/PhysRevD.42.3572},
  url = {https://link.aps.org/doi/10.1103/PhysRevD.42.3572}
}
@article{Kim:1979if,
      author         = "Kim, Jihn E.",
      title          = "{Weak Interaction Singlet and Strong CP Invariance}",
      journal        = "Phys. Rev. Lett.",
      volume         = "43",
      pages          = "103",
      doi            = "10.1103/PhysRevLett.43.103",
      year           = "1979",
      reportNumber   = "UPR-0120T",
      SLACcitation   = "
}
@article{Shifman:1979if,
      author         = "Shifman, Mikhail A. and Vainshtein, A.I. and Zakharov,
                        Valentin I.",
      title          = "{Can Confinement Ensure Natural CP Invariance of Strong
                        Interactions?}",
      journal        = "Nucl.Phys.",
      volume         = "B166",
      pages          = "493",
      doi            = "10.1016/0550-3213(80)90209-6",
      year           = "1980",
      reportNumber   = "ITEP-64-1979",
      SLACcitation   = "
}
@article{Dine:1981rt,
      author         = "Dine, Michael and Fischler, Willy and Srednicki, Mark",
      title          = "{A Simple Solution to the Strong CP Problem with a        Harmless Axion}",
      journal        = "Phys.Lett.",
      volume         = "B104",
      pages          = "199",
      doi            = "10.1016/0370-2693(81)90590-6",
      year           = "1981",
      reportNumber   = "Print-81-0320 (IAS,PRINCETON)",
      SLACcitation   = "
}
@article{DINE1983137,
title = "The not-so-harmless axion",
fulljournal = "Physics Letters B",
journal = "Phys. Lett. B",
volume = "120",
number = "1",
pages = "137 - 141",
year = "1983",
issn = "0370-2693",
doi = "https://doi.org/10.1016/0370-2693(83)90639-1",
url = "http://www.sciencedirect.com/science/article/pii/0370269383906391",
author = "Michael Dine and Willy Fischler"
}
@article{Zhitnitsky:1980tq,
      author         = "Zhitnitsky, A.R.",
      title          = "{On Possible Suppression of the Axion Hadron
                        Interactions. (In Russian)}",
      journal        = "Sov.J.Nucl.Phys.",
      volume         = "31",
      pages          = "260",
      year           = "1980",
      SLACcitation   = "
}

@article{Weinberg:1977ma,
      author         = "Weinberg, Steven",
      title          = "{A New Light Boson?}",
      journal        = "Phys. Rev. Lett.",
      volume         = "40",
      pages          = "223-226",
      doi            = "10.1103/PhysRevLett.40.223",
      year           = "1978",
      reportNumber   = "HUTP-77/A074",
      SLACcitation   = "
}
@article{Wilczek:1977pj,
      author         = "Wilczek, Frank",
      title          = "{Problem of Strong p and t Invariance in the Presence of
                        Instantons}",
      journal        = "Phys. Rev. Lett.",
      volume         = "40",
      pages          = "279-282",
      doi            = "10.1103/PhysRevLett.40.279",
      year           = "1978",
      reportNumber   = "Print-77-0939 (COLUMBIA)",
      SLACcitation   = "
}
@article{Abbott:1982af,
      author         = "Abbott, L.F. and Sikivie, P.",
      title          = "{A Cosmological Bound on the Invisible Axion}",
      journal        = "Phys.Lett.",
      volume         = "B120",
      pages          = "133-136",
      doi            = "10.1016/0370-2693(83)90638-X",
      year           = "1983",
      reportNumber   = "PRINT-82-0695 (BRANDEIS)",
      SLACcitation   = "
}
@article{Preskill:1982cy,
      author         = "Preskill, John and Wise, Mark B. and Wilczek, Frank",
      title          = "{Cosmology of the Invisible Axion}",
      journal        = "Phys.Lett.",
      volume         = "B120",
      pages          = "127-132",
      doi            = "10.1016/0370-2693(83)90637-8",
      year           = "1983",
      reportNumber   = "HUTP-82-A048, NSF-ITP-82-103",
      SLACcitation   = "
}
@article{doi:10.1063/1.366000,
author = {Daw,Edward  and Bradley,Richard F. },
title = {Effect of high magnetic fields on the noise temperature of a heterostructure field-effect transistor low-noise amplifier},
journal = {Journal of Applied Physics},
volume = {82},
number = {4},
pages = {1925-1929},
year = {1997},
doi = {10.1063/1.366000},
URL = { 
        https://doi.org/10.1063/1.366000
},
eprint = { 
        https://doi.org/10.1063/1.366000
    
}
}
@article{PhysRevLett.50.925,
  title = {Can Galactic Halos Be Made of Axions?},
  author = {Ipser, J. and Sikivie, P.},
  journal = {Phys. Rev. Lett.},
  volume = {50},
  issue = {12},
  pages = {925--927},
  numpages = {0},
  year = {1983},
  month = {Mar},
  publisher = {American Physical Society},
  doi = {10.1103/PhysRevLett.50.925},
  url = {https://link.aps.org/doi/10.1103/PhysRevLett.50.925}
}
@article{Turner:1989vc,
      author         = "Turner, Michael S.",
      title          = "{Windows on the Axion}",
      journal        = "Phys.Rept.",
      volume         = "197",
      pages          = "67-97",
      doi            = "10.1016/0370-1573(90)90172-X",
      year           = "1990",
      reportNumber   = "FERMILAB-CONF-89-104-A",
      SLACcitation   = "
}
@article{Sikivie:2006ni,
      author         = "Sikivie, Pierre",
      title          = "{Axion Cosmology}",
      journal        = "Lect. Notes Phys.",
      volume         = "741",
      pages          = "19-50",
      doi            = "10.1007/978-3-540-73518-2_2",
      year           = "2008",
      eprint         = "astro-ph/0610440",
      archivePrefix  = "arXiv",
      primaryClass   = "astro-ph",
      reportNumber   = "UFIFT-HEP-06-16",
      SLACcitation   = "
}
@article{PhysRevD.78.083507,
  title = {Axion cosmology and the energy scale of inflation},
  author = {Hertzberg, Mark P. and Tegmark, Max and Wilczek, Frank},
  journal = {Phys. Rev. D},
  volume = {78},
  issue = {8},
  pages = {083507},
  numpages = {9},
  year = {2008},
  month = {Oct},
  publisher = {American Physical Society},
  doi = {10.1103/PhysRevD.78.083507},
  url = {https://link.aps.org/doi/10.1103/PhysRevD.78.083507}
}
@Article{Bonati2016,
author="Bonati, Claudio
and D'Elia, Massimo
and Mariti, Marco
and Martinelli, Guido
and Mesiti, Michele
and Negro, Francesco
and Sanfilippo, Francesco
and Villadoro, Giovanni",
title="Axion phenomenology and $\theta$-dependence from N                                            f               = 2 + 1 lattice QCD",
journal="Journal of High Energy Physics",
year="2016",
volume="2016",
number="3",
pages="155",
abstract="We investigate the topological properties of N                                                  f                 = 2 + 1 QCD with physical quark masses, both at zero and finite temperature. We adopt stout improved staggered fermions and explore a range of lattice spacings a ∼ 0.05 − 0.12 fm. At zero temperature we estimate both finite size and finite cut-off effects, comparing our continuum extrapolated results for the topological susceptibility $\chi$ with predictions from chiral perturbation theory. At finite temperature, we explore a region going from T                                                  c                 up to around 4 T                                                  c                , where we provide continuum extrapolated results for the topological susceptibility and for the fourth moment of the topological charge distribution. While the latter converges to the dilute instanton gas prediction the former differs strongly both in the size and in the temperature dependence. This results in a shift of the axion dark matter window of almost one order of magnitude with respect to the instanton computation.",
issn="1029-8479",
doi="10.1007/JHEP03(2016)155",
url="http://dx.doi.org/10.1007/JHEP03(2016)155"
}
@article{PhysRevD.92.034507,
  title = {Lattice QCD input for axion cosmology},
  author = {Berkowitz, Evan and Buchoff, Michael I. and Rinaldi, Enrico},
  journal = {Phys. Rev. D},
  volume = {92},
  issue = {3},
  pages = {034507},
  numpages = {15},
  year = {2015},
  month = {Aug},
  publisher = {American Physical Society},
  doi = {10.1103/PhysRevD.92.034507},
  url = {https://link.aps.org/doi/10.1103/PhysRevD.92.034507}
}
@article{Borsanyi2016,
author = {Borsanyi, S and Fodor, Z and Guenther, J and Kampert, K.-H. and Katz, S D and Kawanai, T and Kovacs, T G and Mages, S W and Pasztor, A and Pittler, F and Redondo, J and Ringwald, A and Szabo, K K},
issn = {0028-0836},
journal = {Nature},
month = {Nov},
number = {7627},
pages = {69--71},
publisher = {Macmillan Publishers Limited, part of Springer Nature. All rights reserved.},
title = {{Calculation of the axion mass based on high-temperature lattice quantum chromodynamics}},
url = {http://dx.doi.org/10.1038/nature20115},
volume = {539},
year = {2016}
}
@article{PhysRevLett.118.071802,
  title = {Unifying Inflation with the Axion, Dark Matter, Baryogenesis, and the Seesaw Mechanism},
  author = {Ballesteros, Guillermo and Redondo, Javier and Ringwald, Andreas and Tamarit, Carlos},
  journal = {Phys. Rev. Lett.},
  volume = {118},
  issue = {7},
  pages = {071802},
  numpages = {7},
  year = {2017},
  month = {Feb},
  publisher = {American Physical Society},
  doi = {10.1103/PhysRevLett.118.071802},
  url = {https://link.aps.org/doi/10.1103/PhysRevLett.118.071802}
}

@article{DePanfilis:1987dk,
      author         = "De Panfilis, S. and Melissinos, A.C. and Moskowitz, B.E.
                        and Rogers, J.T. and Semertzidis, Y.K. and others",
      title          = "{Limits on the Abundance and Coupling of Cosmic Axions at
                        4.5-Microev < m(a) < 5.0-Microev}",
      journal        = "Phys. Rev. Lett.",
      volume         = "59",
      pages          = "839",
      doi            = "10.1103/PhysRevLett.59.839",
      year           = "1987",
      reportNumber   = "UR-994, FERMILAB-PUB-87-046, ER13065-482",
      SLACcitation   = "
}
@article{Wuensch:1989sa,
      author         = "Wuensch, Walter and De Panfilis-Wuensch, S. and
                        Semertzidis, Y.K. and Rogers, J.T. and Melissinos, A.C.
                        and others",
      title          = "{Results of a Laboratory Search for Cosmic Axions and
                        Other Weakly Coupled Light Particles}",
      journal        = "Phys. Rev.",
      volume         = "D40",
      pages          = "3153",
      doi            = "10.1103/PhysRevD.40.3153",
      year           = "1989",
      reportNumber   = "FERMILAB-PUB-89-185-E, BNL-43010",
      SLACcitation   = "
}
@article{Hagmann:1990tj,
      author         = "Hagmann, C. and Sikivie, P. and Sullivan, N. S. and
                        Tanner, D. B.",
      title          = "{Results from a search for cosmic axions}",
      journal        = "Phys. Rev.",
      volume         = "D42",
      pages          = "1297-1300",
      doi            = "10.1103/PhysRevD.42.1297",
      year           = "1990",
      reportNumber   = "PRINT-90-0420 (FLORIDA)",
      SLACcitation   = "
}
@article{PhysRevD.92.092003,
  title = {Revised experimental upper limit on the electric dipole moment of the neutron},
  author = {Pendlebury, J. M. et al.},
  journal = {Phys. Rev. D},
  volume = {92},
  issue = {9},
  pages = {092003},
  numpages = {0},
  year = {2015},
  month = {Nov},
  publisher = {American Physical Society},
  doi = {10.1103/PhysRevD.92.092003},
  url = {https://journals.aps.org/prd/abstract/10.1103/PhysRevD.92.092003}
}
@article{0004-637X-845-2-121,
  author={Erik W. Lentz and Thomas R. Quinn and Leslie J Rosenberg and Michael J. Tremmel},
  title={A New Signal Model for Axion Cavity Searches from N -body Simulations},
  fulljournal={The Astrophysical Journal},
  journal={Ap.J.},
  volume={845},
  number={2},
  pages={121},
  url={http://stacks.iop.org/0004-637X/845/i=2/a=121},
  year={2017},
  abstract={Signal estimates for direct axion dark matter (DM) searches have used the isothermal sphere halo model for the last several decades. While insightful, the isothermal model does not capture effects from a halo’s infall history nor the influence of baryonic matter, which has been shown to significantly influence a halo’s inner structure. The high resolution of cavity axion detectors can make use of modern cosmological structure-formation simulations, which begin from realistic initial conditions, incorporate a wide range of baryonic physics, and are capable of resolving detailed structure. This work uses a state-of-the-art cosmological N -body+Smoothed-Particle Hydrodynamics simulation to develop an improved signal model for axion cavity searches. Signal shapes from a class of galaxies encompassing the Milky Way are found to depart significantly from the isothermal sphere. A new signal model for axion detectors is proposed and projected sensitivity bounds on the Axion DM eXperiment (ADMX) data are presented.}
}
@article{ASZTALOS201139,
title = "Design and performance of the ADMX SQUID-based microwave receiver",
fulljournal = "Nuclear Instruments and Methods in Physics Research Section A: Accelerators, Spectrometers, Detectors and Associated Equipment",
journal = "Nucl. Instrum. Methods Phys. Res. A",
volume = "656",
number = "1",
pages = "39 - 44",
year = "2011",
issn = "0168-9002",
doi = "https://doi.org/10.1016/j.nima.2011.07.019",
url = "http://www.sciencedirect.com/science/article/pii/S0168900211014483",
author = "S.J. Asztalos and G. Carosi and C. Hagmann and D. Kinion and K. van Bibber and M. Hotz and L. J Rosenberg and G. Rybka and A. Wagner and J. Hoskins and C. Martin and N. S. Sullivan and D. B. Tanner and R. Bradley and John Clarke",
keywords = "Microwave cavity",
keywords = "SQUIDS",
keywords = "Axion",
keywords = "Dark matter"
}
@article{Brubaker:2017rna,
  title = {HAYSTAC axion search analysis procedure},
  author = {Brubaker, B. M. and Zhong, L. and Lamoreaux, S. K. and Lehnert, K. W. and van Bibber, K. A.},
  journal = {Phys. Rev. D},
  volume = {96},
  issue = {12},
  pages = {123008},
  numpages = {34},
  year = {2017},
  month = {Dec},
  publisher = {American Physical Society},
  doi = {10.1103/PhysRevD.96.123008},
  url = {https://link.aps.org/doi/10.1103/PhysRevD.96.123008}
}
@article{PhysRevD.64.092003,
  title = {Large-scale microwave cavity search for dark-matter axions},
  author = {Asztalos, S. and Daw, E. and Peng, H. and Rosenberg, L. J and Hagmann, C. and Kinion, D. and Stoeffl, W. and van Bibber, K. and Sikivie, P. and Sullivan, N. S. and Tanner, D. B. and Nezrick, F. and Turner, M. S. and Moltz, D. M. and Powell, J. and Andr\'e, M.-O. and Clarke, J. and M\"uck, M. and Bradley, Richard F.},
  journal = {Phys. Rev. D},
  volume = {64},
  issue = {9},
  pages = {092003},
  numpages = {28},
  year = {2001},
  month = {Oct},
  publisher = {American Physical Society},
  doi = {10.1103/PhysRevD.64.092003},
  url = {https://link.aps.org/doi/10.1103/PhysRevD.64.092003}
}
@article{PhysRevLett.118.061302,
  title = {First Results from a Microwave Cavity Axion Search at $24~\mathrm{\mu}\mathrm{eV}$},
  author = {Brubaker, B. M. and Zhong, L. and Gurevich, Y. V. and Cahn, S. B. and Lamoreaux, S. K. and Simanovskaia, M. and Root, J. R. and Lewis, S. M. and Al Kenany, S. and Backes, K. M. and Urdinaran, I. and Rapidis, N. M. and Shokair, T. M. and van Bibber, K. A. and Palken, D. A. and Malnou, M. and Kindel, W. F. and Anil, M. A. and Lehnert, K. W. and Carosi, G.},
  journal = {Phys. Rev. Lett.},
  volume = {118},
  issue = {6},
  pages = {061302},
  numpages = {5},
  year = {2017},
  month = {Feb},
  publisher = {American Physical Society},
  doi = {10.1103/PhysRevLett.118.061302},
  url = {https://link.aps.org/doi/10.1103/PhysRevLett.118.061302}
}
@article{SLOAN201695,
title = "Limits on axion–photon coupling or on local axion density: Dependence on models of the Milky Way’s dark halo ",
journal = "Physics of the Dark Universe ",
volume = "14",
number = "",
pages = "95 - 102",
year = "2016",
note = "",
issn = "2212-6864",
doi = {10.1016/j.dark.2016.09.003},
url = "https://www.sciencedirect.com/science/article/pii/S2212686416300504",
author = "J.V. Sloan and M. Hotz and C. Boutan and R. Bradley and G. Carosi and D. Carter and J. Clarke and N. Crisosto and E.J. Daw and J. Gleason and J. Hoskins and R. Khatiwada and D. Lyapustin and A. Malagon and S. O'Kelley and R. S. Ottens and Rosenberg, L. J and G. Rybka and I. Stern and N. S. Sullivan and D. B. Tanner and K. van Bibber and A. Wagner and D. Will",
keywords = "Axions",
keywords = "Halo models",
keywords = "Dark matter",
keywords = "Microwave cavity",
keywords = "Direct detections "
}

@article{PhysRevD.94.082001,
  title = {Modulation sensitive search for nonvirialized dark-matter axions},
  author = {Hoskins, J. and Crisosto, N. and Gleason, J. and Sikivie, P. and Stern, I. and Sullivan, N. S. and Tanner, D. B. and Boutan, C. and Hotz, M. and Khatiwada, R. and Lyapustin, D. and Malagon, A. and Ottens, R. and Rosenberg, L. J. and Rybka, G. and Sloan, J. and Wagner, A. and Will, D. and Carosi, G. and Carter, D. and Duffy, L. D. and Bradley, R. and Clarke, J. and O'Kelley, S. and van Bibber, K. and Daw, E. J.},
  journal = {Phys. Rev. D},
  volume = {94},
  issue = {8},
  pages = {082001},
  numpages = {13},
  year = {2016},
  month = {Oct},
  publisher = {American Physical Society},
  doi = {10.1103/PhysRevD.94.082001}, 
  url = {https://link.aps.org/doi/10.1103/PhysRevD.94.082001},
}

@article{doi:10.1063/1.121490,
author = {Michael M{\"u}ck and Marc-Olivier Andr{\'e} and John Clarke and Jost Gail and Christoph Heiden},
title = {Radio-frequency amplifier based on a niobium dc superconducting quantum interference device with microstrip input coupling},
fulljournal = {Applied Physics Letters},
journal = {Appl. Phys. Lett.},
volume = {72},
number = {22},
pages = {2885-2887},
year = {1998},
doi = {10.1063/1.121490},
URL = {https://doi.org/10.1063/1.121490}
}

@article{doi:10.1063/1.125383,
author = {Michael M{\"u}ck and Marc-Olivier Andr{\'e} and John Clarke and Jost Gail and Christoph Heiden},
title = {Microstrip superconducting quantum interference device radio-frequency amplifier: Tuning and cascading},
journal = {Applied Physics Letters},
volume = {75},
number = {22},
pages = {3545-3547},
year = {1999},
doi = {10.1063/1.125383},
eprint = {https://doi.org/10.1063/1.125383}
}

@article{doi:10.1063/1.98459,
author = {Tsutom Yotsuya and Masaaki Yoshitake and Junya Yamamoto},
title = {New type cryogenic thermometer using sputtered Zr‐N films},
fulljournal = {Applied Physics Letters},
journal = {App. Phys. Lett.},
volume = {51},
number = {4},
pages = {235-237},
year = {1987},
doi = {10.1063/1.98459},
URL = { 
        https://doi.org/10.1063/1.98459
    
}
}
@article{doi:10.1063/1.1770483,
author = {R. H. Dicke},
title = {The Measurement of Thermal Radiation at Microwave Frequencies},
fulljournal = {Review of Scientific Instruments},
journal = {Rev. Sci. Inst.},
volume = {17},
number = {7},
pages = {268-275},
year = {1946},
doi = {10.1063/1.1770483},
URL = { 
        https://doi.org/10.1063/1.1770483
},
meprint = { 
        https://doi.org/10.1063/1.1770483
}
}

@article{PhysRevD.69.011101,
  title = {Improved rf cavity search for halo axions},
  author = {Asztalos, S. J. and Bradley, R. F. and Duffy, L. and Hagmann, C. and Kinion, D. and Moltz, D. M. and Rosenberg, L. J and Sikivie, P. and Stoeffl, W. and Sullivan, N. S. and Tanner, D. B. and van Bibber, K. and Yu, D. B.},
  journal = {Phys. Rev. D},
  volume = {69},
  issue = {1},
  pages = {011101},
  numpages = {5},
  year = {2004},
  month = {Jan},
  publisher = {American Physical Society},
  doi = {10.1103/PhysRevD.69.011101},
  url = {https://link.aps.org/doi/10.1103/PhysRevD.69.011101}
}
@article{1538-4357-571-1-L27,
  author={S. J. Asztalos and E. Daw and H. Peng and L. J Rosenberg and D. B. Yu and C. Hagmann and D. Kinion and W. Stoeffl and K. van
Bibber and J. LaVeigne and P. Sikivie and N. S. Sullivan and D. B. Tanner and F. Nezrick and D. M. Moltz},
  title={Experimental Constraints on the Axion Dark Matter Halo Density},
  journal={The Astrophysical Journal Letters},
  volume={571},
  number={1},
  pages={L27},
  url={http://stacks.iop.org/1538-4357/571/i=1/a=L27},
  year={2002},

}
@article{PhysRevD.96.095001,
  title = {Axions, instantons, and the lattice},
  author = {Dine, Michael and Draper, Patrick and Stephenson-Haskins, Laurel and Xu, Di},
  journal = {Phys. Rev. D},
  volume = {96},
  issue = {9},
  pages = {095001},
  numpages = {8},
  year = {2017},
  month = {Nov},
  publisher = {American Physical Society},
  doi = {10.1103/PhysRevD.96.095001},
  url = {https://link.aps.org/doi/10.1103/PhysRevD.96.095001}
}

@article{PhysRevLett.80.2043,
  title = {Results from a High-Sensitivity Search for Cosmic Axions},
  author = {Hagmann, C. and Kinion, D. and Stoeffl, W. and van Bibber, K. and Daw, E. and Peng, H. and Rosenberg, Leslie J and LaVeigne, J. and Sikivie, P. and Sullivan, N. S. and Tanner, D. B. and Nezrick, F. and Turner, Michael S. and Moltz, D. M. and Powell, J. and Golubev, N. A.},
  journal = {Phys. Rev. Lett.},
  volume = {80},
  issue = {10},
  pages = {2043--2046},
  numpages = {0},
  year = {1998},
  month = {Mar},
  publisher = {American Physical Society},
  doi = {10.1103/PhysRevLett.80.2043},
  url = {https://link.aps.org/doi/10.1103/PhysRevLett.80.2043}
}
@Article{friis1944,
  Author         = "H.T. Friis",
  Title          = "{Noise Figures of Radio Receivers}",
  Journal        = "Proceedings of the IRE",
  Volume         = "32",
  Number         = "7",
  Pages          = "419--422",
  issn           = "0096-8390",
  doi            = "10.1109/JRPROC.1944.232049",
  year           = 1944
}
@article{PETRECZKY2016498,
title = {The topological susceptibility in finite temperature QCD and axion cosmology},
journal = {Physics Letters B},
volume = {762},
pages = {498 - 505},
year = {2016},
issn = {0370-2693},
url = {https://www.sciencedirect.com/science/article/pii/S0370269316305718?via
author = {Peter Petreczky and Hans-Peter Schadler and Sayantan Sharma}
}
@article{PhysRevLett.120.151301,
  title = {Search for Invisible Axion Dark Matter with the Axion Dark Matter Experiment},
  author = {Du, N. and Force, N. and Khatiwada, R. and Lentz, E. and Ottens, R. and Rosenberg, L. J and Rybka, G. and Carosi, G. and Woollett, N. and Bowring, D. and Chou, A. S. and Sonnenschein, A. and Wester, W. and Boutan, C. and Oblath, N. S. and Bradley, R. and Daw, E. J. and Dixit, A. V. and Clarke, J. and O'Kelley, S. R. and Crisosto, N. and Gleason, J. R. and Jois, S. and Sikivie, P. and Stern, I. and Sullivan, N. S. and Tanner, D. B and Hilton, G. C.},
  collaboration = {ADMX Collaboration},
  journal = {Phys. Rev. Lett.},
  volume = {120},
  issue = {15},
  pages = {151301},
  numpages = {5},
  year = {2018},
  month = {Apr},
  publisher = {American Physical Society},
  doi = {10.1103/PhysRevLett.120.151301},
  url = {https://link.aps.org/doi/10.1103/PhysRevLett.120.151301}
}
@article{cast2017,
  title={New CAST limit on the axion-photon interaction},
  collaboration={CAST Collaboration},
  author={Anastassopoulos, V. and others},
journal={Nature Physics},
year={2017},
month={May},
day={01},
volume={13},
pages={584},
url={http://dx.doi.org/10.1038/nphys4109}
}

@article{PhysRevD.97.092001,
  title = {Results from phase 1 of the HAYSTAC microwave cavity axion experiment},
  author = {Zhong, L. and Al Kenany, S. and Backes, K. M. and Brubaker, B. M. and Cahn, S. B. and Carosi, G. and Gurevich, Y. V. and Kindel, W. F. and Lamoreaux, S. K. and Lehnert, K. W. and Lewis, S. M. and Malnou, M. and Maruyama, R. H. and Palken, D. A. and Rapidis, N. M. and Root, J. R. and Simanovskaia, M. and Shokair, T. M. and Speller, D. H. and Urdinaran, I. and van Bibber, K. A.},
  journal = {Phys. Rev. D},
  volume = {97},
  issue = {9},
  pages = {092001},
  numpages = {7},
  year = {2018},
  month = {May},
  publisher = {American Physical Society},
  doi = {10.1103/PhysRevD.97.092001},
  url = {https://link.aps.org/doi/10.1103/PhysRevD.97.092001}
}
@article{PhysRevLett.121.261302,
  title = {Piezoelectrically Tuned Multimode Cavity Search for Axion Dark Matter},
  author = {Boutan, C. and Jones, M. and LaRoque, B. H. and Oblath, N. S. and Cervantes, R. and Du, N. and Force, N. and Kimes, S. and Ottens, R. and Rosenberg, L. J. and Rybka, G. and Yang, J. and Carosi, G. and Woollett, N. and Bowring, D. and Chou, A. S. and Khatiwada, R. and Sonnenschein, A. and Wester, W. and Bradley, R. and Daw, E. J. and Agrawal, A. and Dixit, A. V. and Clarke, J. and O'Kelley, S. R. and Crisosto, N. and Gleason, J. R. and Jois, S. and Sikivie, P. and Stern, I. and Sullivan, N. S. and Tanner, D. B. and Harrington, P. M. and Lentz, E.},
  collaboration = {ADMX Collaboration},
  journal = {Phys. Rev. Lett.},
  volume = {121},
  issue = {26},
  pages = {261302},
  numpages = {7},
  year = {2018},
  month = {Dec},
  publisher = {American Physical Society},
  doi = {10.1103/PhysRevLett.121.261302},
  url = {https://link.aps.org/doi/10.1103/PhysRevLett.121.261302}
}

@article{Peng2000569,
        title = "Cryogenic cavity detector for a large-scale cold dark-matter axion search ",
        journal = "Nuclear Instruments and Methods in Physics Research Section A: Accelerators, Spectrometers, Detectors and Associated Equipment ",
        volume = "444",
        number = "3",
        pages = "569 - 583",
        year = "2000",
        note = "",
        issn = "0168-9002",
        doi = "http://dx.doi.org/10.1016/S0168-9002(99)00971-7",
        url = "http://www.sciencedirect.com/science/article/pii/S0168900299009717",
        author = {H. Peng and S. Asztalos and E. Daw and N.A. Golubev and C. Hagmann and D. Kinion and J. LaVeigne and D.M. Moltz and F. Nezrick and J. Powell and others},
        keywords = "Axions",
        keywords = "Dark matter",
        keywords = "Microwave detector "
}
@article{SREDNICKI1985689,
title = "Axion couplings to matter: (I). CP-conserving parts",
journal = "Nuclear Physics B",
volume = "260",
number = "3",
pages = "689 - 700",
year = "1985",
issn = "0550-3213",
doi = "https://doi.org/10.1016/0550-3213(85)90054-9",
url = "http://www.sciencedirect.com/science/article/pii/0550321385900549",
author = "Mark Srednicki",
abstract = "The CP-conserving couplings of axions to photons, electrons, and nucleons are derived for an arbitrary axion model. The relevance of the results to proposed axion search experiments is briefly discussed."
}
@article{PhysRevB.83.134501,
  title = {Dispersive magnetometry with a quantum limited SQUID parametric amplifier},
  author = {Hatridge, M. and Vijay, R. and Slichter, D. H. and Clarke, John and Siddiqi, I.},
  journal = {Phys. Rev. B},
  volume = {83},
  issue = {13},
  pages = {134501},
  numpages = {8},
  year = {2011},
  month = {Apr},
  publisher = {American Physical Society},
  doi = {10.1103/PhysRevB.83.134501},
  url = {https://link.aps.org/doi/10.1103/PhysRevB.83.134501}
}
@Article{Eichler2014,
author={Eichler, Christopher
and Wallraff, Andreas},
title={Controlling the dynamic range of a Josephson parametric amplifier},
journal={EPJ Quantum Technology},
year={2014},
volume={1},
number={1},
pages={2},
abstract={One of the central challenges in the development of parametric amplifiers is the control of the dynamic range relative to its gain and bandwidth, which typically limits quantum limited amplification to signals which contain only a few photons per inverse bandwidth. Here, we discuss the control of the dynamic range of Josephson parametric amplifiers by using Josephson junction arrays. We discuss gain, bandwidth, noise, and dynamic range properties of both a transmission line and a lumped element based parametric amplifier. Based on these investigations we derive useful design criteria, which may find broad application in the development of practical parametric amplifiers.},
issn={2196-0763},
doi={10.1140/epjqt2},
url={https://doi.org/10.1140/epjqt2}
}
@phdthesis{beltran2010development,
  title={Development of a Josephson parametric amplifier for the preparation and detection of nonclassical states of microwave fields},
  author={Beltran, Manuel Angel Castellanos},
  year={2010},
  school={Citeseer}
}
@article{Du_2018,
   title={Search for Invisible Axion Dark Matter with the Axion Dark Matter Experiment},
   volume={120},
   ISSN={1079-7114},
   url={http://dx.doi.org/10.1103/PhysRevLett.120.151301},
   DOI={10.1103/physrevlett.120.151301},
   number={15},
   journal={Physical Review Letters},
   publisher={American Physical Society (APS)},
   author={Du, N. and Force, N. and Khatiwada, R. and Lentz, E. and Ottens, R. and Rosenberg, L. J and Rybka, G. and Carosi, G. and Woollett, N. and Bowring, D. and et al.},
   year={2018},
   month={Apr}
}

@misc{Janis,
  title = {Janis Standard Dilution Refrigerator Systems},
  howpublished = {\url{https://www.janis.com/Products/productsoverview/DilutionRefrigerators/StandardDilutionRefrigeratorSystems.aspx}},
  note = {Accessed: 2019-10-07}
}
@article{Peccei1977Sept,
  title = {Constraints imposed by $\mathrm{CP}$ conservation in the presence of pseudoparticles},
  author = {Peccei, R. D. and Quinn, Helen R.},
  journal = {Phys. Rev. D},
  volume = {16},
  issue = {6},
  pages = {1791--1797},
  numpages = {0},
  year = {1977},
  month = {Sep},
  publisher = {American Physical Society},
  doi = {10.1103/PhysRevD.16.1791},
  url = {https://link.aps.org/doi/10.1103/PhysRevD.16.1791}
}
@article{Peccei1977June,
  title = {$\mathrm{CP}$ Conservation in the Presence of Pseudoparticles},
  author = {Peccei, R. D. and Quinn, Helen R.},
  journal = {Phys. Rev. Lett.},
  volume = {38},
  issue = {25},
  pages = {1440--1443},
  numpages = {0},
  year = {1977},
  month = {Jun},
  publisher = {American Physical Society},
  doi = {10.1103/PhysRevLett.38.1440},
  url = {https://link.aps.org/doi/10.1103/PhysRevLett.38.1440}
}
@article{chambers1950,
  title={Anomalous skin effect in metals},
  author={Chambers, RG},
  journal={Nature},
  volume={165},
  number={4189},
  pages={239--240},
  year={1950},
  publisher={Springer}
}
@misc{braine2019extended,
    title={Extended Search for the Invisible Axion with the Axion Dark Matter Experiment},
    author={T. Braine and R. Cervantes and N. Crisosto and N. Du and S. Kimes and L. J Rosenberg and G. Rybka and J. Yang and D. Bowring and A. S. Chou and R. Khatiwada and A. Sonnenschein and W. Wester and G. Carosi and N. Woollett and L. D. Duffy and R. Bradley and C. Boutan and M. Jones and B. H. LaRoque and N. S. Oblath and M. S. Taubman and J. Clarke and A. Dove and A. Eddins and S. R. O'Kelley and S. Nawaz and I. Siddiqi and N. Stevenson and A. Agrawal and A. V. Dixit and J. R. Gleason and S. Jois and P. Sikivie and N. S. Sullivan and D. B. Tanner and E. Lentz and E. J. Daw and J. H. Buckley and P. M. Harrington and E. A. Henriksen and K. W. Murch},
    year={2019},
    eprint={1910.08638},
    archivePrefix={arXiv},
    primaryClass={hep-ex}
}
@article{siddiqi2004rf,
  title = {RF-Driven Josephson Bifurcation Amplifier for Quantum Measurement},
  author = {Siddiqi, I. and Vijay, R. and Pierre, F. and Wilson, C. M. and Metcalfe, M. and Rigetti, C. and Frunzio, L. and Devoret, M. H.},
  journal = {Phys. Rev. Lett.},
  volume = {93},
  issue = {20},
  pages = {207002},
  numpages = {4},
  year = {2004},
  month = {Nov},
  publisher = {American Physical Society},
  doi = {10.1103/PhysRevLett.93.207002},
  url = {https://link.aps.org/doi/10.1103/PhysRevLett.93.207002}
}
@phdthesis{slichter2011quantum,
    title    = {Quantum Jumps and Measurement Backaction in a Superconducting Qubit},
    school   = {University of California, Berkeley},
    author   = {Daniel Huber Slichter},
    year     = {2011},
    address  = {},
    month    = {},
    note     = {}
}
@article{josephson1962possible,
  added-at = {2014-04-11T19:39:03.000+0200},
  author = {Josephson, B.D.},
  biburl = {https://www.bibsonomy.org/bibtex/2626cab8b998fc5f40b2236183f35f997/tsunetomo},
  doi = {10.1016/0031-9163(62)91369-0},
  interhash = {d9b20df06f51ffbc018229d503f4b553},
  intrahash = {626cab8b998fc5f40b2236183f35f997},
  issn = {0031-9163},
  journal = {Physics Letters },
  keywords = {superconductivity},
  number = 7,
  pages = {251 - 253},
  timestamp = {2014-04-11T19:39:03.000+0200},
  title = {Possible new effects in superconductive tunnelling },
  url = {http://www.sciencedirect.com/science/article/pii/0031916362913690},
  volume = 1,
  year = 1962
}
@article{hatridge2011dispersive,
  title = {Dispersive magnetometry with a quantum limited SQUID parametric amplifier},
  author = {Hatridge, M. and Vijay, R. and Slichter, D. H. and Clarke, J. and Siddiqi, I.},
  journal = {Phys. Rev. B},
  volume = {83},
  issue = {13},
  pages = {134501},
  numpages = {8},
  year = {2011},
  month = {Apr},
  publisher = {American Physical Society},
  doi = {10.1103/PhysRevB.83.134501},
  url = {https://link.aps.org/doi/10.1103/PhysRevB.83.134501}
}
@article{yurke1989observation,
  title = {Observation of parametric amplification and deamplification in a Josephson parametric amplifier},
  author = {Yurke, B. and Corruccini, L. R. and Kaminsky, P. G. and Rupp, L. W. and Smith, A. D. and Silver, A. H. and Simon, R. W. and Whittaker, E. A.},
  journal = {Phys. Rev. A},
  volume = {39},
  issue = {5},
  pages = {2519--2533},
  numpages = {0},
  year = {1989},
  month = {Mar},
  publisher = {American Physical Society},
  doi = {10.1103/PhysRevA.39.2519},
  url = {https://link.aps.org/doi/10.1103/PhysRevA.39.2519}
}
@article{krantz2016single,
   title={Single-shot read-out of a superconducting qubit using a Josephson parametric oscillator},
   volume={7},
   ISSN={2041-1723},
   url={http://dx.doi.org/10.1038/ncomms11417},
   DOI={10.1038/ncomms11417},
   number={1},
   journal={Nature Communications},
   publisher={Springer Science and Business Media LLC},
   author={Krantz, Philip and Bengtsson, Andreas and Simoen, Michaël and Gustavsson, Simon and Shumeiko, Vitaly and Oliver, W. D. and Wilson, C. M. and Delsing, Per and Bylander, Jonas},
   year={2016},
   month={May}
}
@article{clerk2010introduction,
  abstract = {{The topic of quantum noise has become extremely timely due to the rise of
quantum information physics and the resulting interchange of ideas between the
condensed matter and AMO/quantum optics communities. This review gives a
pedagogical introduction to the physics of quantum noise and its connections to
quantum measurement and quantum amplification. After introducing quantum noise
spectra and methods for their detection, we describe the basics of weak
continuous measurements. Particular attention is given to treating the standard
quantum limit on linear amplifiers and position detectors using a general
linear-response framework. We show how this approach relates to the standard
Haus-Caves quantum limit for a bosonic amplifier known in quantum optics, and
illustrate its application for the case of electrical circuits, including
mesoscopic detectors and resonant cavity detectors.}},
  added-at = {2013-09-09T23:59:35.000+0200},
  archiveprefix = {arXiv},
  author = {Clerk, A. A. and Devoret, M. H. and Girvin, S. M. and Marquardt, F. and Schoelkopf, R. J.},
  biburl = {https://www.bibsonomy.org/bibtex/290cbaf4e5ddf303f478540abdb290569/jacksankey},
  citeulike-article-id = {3454550},
  citeulike-linkout-0 = {http://arxiv.org/abs/0810.4729},
  citeulike-linkout-1 = {http://arxiv.org/pdf/0810.4729},
  citeulike-linkout-2 = {http://dx.doi.org/10.1103/revmodphys.82.1155},
  day = 15,
  doi = {10.1103/revmodphys.82.1155},
  eprint = {0810.4729},
  interhash = {c376d4a588bc33979d2dad69cf79c99b},
  intrahash = {90cbaf4e5ddf303f478540abdb290569},
  issn = {0034-6861},
  journal = {Reviews of Modern Physics},
  keywords = {measurement, quantum-noise, theory amplification optomechanics review},
  month = apr,
  number = 2,
  pages = {1155--1208},
  posted-at = {2012-06-25 02:38:58},
  priority = {2},
  publisher = {American Physical Society},
  timestamp = {2013-09-10T00:19:56.000+0200},
  title = {{Introduction to Quantum Noise, Measurement and Amplification}},
  url = {http://dx.doi.org/10.1103/revmodphys.82.1155},
  volume = 82,
  year = 2010
}
@article{roy2018quantum,
  title = {Quantum-limited parametric amplification with Josephson circuits in the regime of pump depletion},
  author = {Roy, Ananda and Devoret, Michel},
  journal = {Phys. Rev. B},
  volume = {98},
  issue = {4},
  pages = {045405},
  numpages = {27},
  year = {2018},
  month = {Jul},
  publisher = {American Physical Society},
  doi = {10.1103/PhysRevB.98.045405},
  url = {https://link.aps.org/doi/10.1103/PhysRevB.98.045405}
}
@article{jaklevic1964quantum,
  added-at = {2014-04-16T10:01:42.000+0200},
  author = {Jaklevic, R. C. and Lambe, John and Silver, A. H. and Mercereau, J. E.},
  biburl = {https://www.bibsonomy.org/bibtex/246e950525a83823536e07e7d280bcd8b/tsunetomo},
  doi = {10.1103/PhysRevLett.12.159},
  interhash = {540fa1a8ddf71363adee99bdcf16f564},
  intrahash = {46e950525a83823536e07e7d280bcd8b},
  journal = {Phys. Rev. Lett.},
  keywords = {superconductivity},
  month = feb,
  number = 7,
  numpages = {0},
  pages = {159--160},
  publisher = {American Physical Society},
  timestamp = {2014-04-16T10:01:42.000+0200},
  title = {Quantum Interference Effects in {Josephson} Tunneling},
  url = {http://link.aps.org/doi/10.1103/PhysRevLett.12.159},
  volume = 12,
  year = 1964
}
@article{wahlsten1978arrays,
  title={Arrays of Josephson tunnel junctions as parametric amplifiers},
  author={Wahlsten, S and Rudner, S and Claeson, T},
  journal={Journal of Applied Physics},
  volume={49},
  number={7},
  pages={4248--4263},
  year={1978},
  publisher={AIP}
}

}
@article{caves1982quantum,
  title={Quantum limits on noise in linear amplifiers},
  author={Caves, Carlton M},
  journal={Physical Review D},
  volume={26},
  number={8},
  pages={1817},
  year={1982},
  publisher={APS}
}
@article{castellanos2007widely,
  title={Widely tunable parametric amplifier based on a superconducting quantum interference device array resonator},
  author={Castellanos-Beltran, MA and Lehnert, KW},
  journal={Applied Physics Letters},
  volume={91},
  number={8},
  pages={083509},
  year={2007},
  publisher={AIP}
}
@article{sandberg2008tuning,
  title={Tuning the field in a microwave resonator faster than the photon lifetime},
  author={Sandberg, Martin and Wilson, CM and Persson, Fredrik and Bauch, Thilo and Johansson, G{\"o}ran and Shumeiko, Vitaly and Duty, Tim and Delsing, Per},
  journal={Applied Physics Letters},
  volume={92},
  number={20},
  pages={203501},
  year={2008},
  publisher={AIP}
}
@article{yamamoto2008flux,
  title={Flux-driven Josephson parametric amplifier},
  author={Yamamoto, Tsuyoshi and Inomata, K and Watanabe, M and Matsuba, K and Miyazaki, T and Oliver, William D and Nakamura, Yasunobu and Tsai, Jaw Shen},
  journal={Applied Physics Letters},
  volume={93},
  number={4},
  pages={042510},
  year={2008},
  publisher={AIP}
}
@article{castellanos2008amplification,
  title={Amplification and squeezing of quantum noise with a tunable Josephson metamaterial},
  author={Castellanos-Beltran, MA and Irwin, KD and Hilton, GC and Vale, LR and Lehnert, KW},
  journal={Nature Physics},
  volume={4},
  number={12},
  pages={929},
  year={2008},
  publisher={Nature Publishing Group}
}
@article{olsson1988low,
  title={Ultra-low reflectivity 1.5 mu m semiconductor laser preamplifier},
  author={Olsson, NA and Oberg, MG and Tzeng, LD and Cella, T},
  journal={Electronics Letters},
  volume={24},
  number={9},
  pages={569--570},
  year={1988},
  publisher={IET}
}

@article{doi:10.1063/1.1321026,
author = {Mück,Michael  and Clarke,John },
title = {The superconducting quantum interference device microstrip amplifier: Computer models},
journal = {Journal of Applied Physics},
volume = {88},
number = {11},
pages = {6910-6918},
year = {2000},
doi = {10.1063/1.1321026},
URL = { 
        https://doi.org/10.1063/1.1321026},
eprint = { 
        https://doi.org/10.1063/1.1321026}
}

@Book{Clarke2004,
author={Clarke, J. and Braginski, A. I.},
title={The SQUID handbook Vol 1 Fundamentals and technology of SQUIDs and SQUID systems},
year={2004},
publisher={Wiley VCH},
address={Germany},
abstract={This two-volume handbook offers a comprehensive and coordinated presentation of SQUIDs (Superconducting QUantum Interference Devices), including device fundamentals, design, technology, system construction and multiple applications In Volume I the following topics were dealth with: INTRODUCTION (AI Braginski and J Clarke); SQUID THEORY (B Chesca, R Kleiner and D Koelle); SQUID FABRICATION TECHNOLOGY (R Cantor and F Ludwig); SQUID ELECTRONICS (D Drung and M Mueck); PRACTICAL DC SQUIDS: CONFIGURATION AND PERFORMANCE (R Cantor and D Koelle); PRACTICAL RF SQUIDS: CONFIGURATION AND PERFORMANCE (AI Braginski and Y Zhang); and SQUID SYSTEM ISSUES (C Foley, M Keene, H J M ter Brake, and J Vrba)},
isbn={3-527-40229-2},
url={http://inis.iaea.org/search/search.aspx?orig_q=RN:38047859}
}

@Book{Clarke2006,
author={Clarke, J. and Braginski, A. I.},
title={The SQUID handbook Vol 2 Applications of SQUIDs and SQUID systems},
year={2006},
publisher={Wiley-VCH},
address={Germany},
abstract={This two-volume handbook offers a comprehensive and coordinated presentation of SQUIDs (Superconducting QUantum Interference Devices), including device fundamentals, design, technology, system construction and multiple applications In Volume II the following topics were dealth with: SQUID Amplifiers (Clarke, Lee, Mueck, Richards); SQUIDs for Standards (Gallop, Piquemal); Generic Inverse Problem (Wikswo); Biomagnetism (Nenonen, Trahms, Vrba); Magnetic Measurements and Microscopy Using SQUIDs (Wellstood, Black); Nondestructive Evaluation of Materials and Structures Using SQUIDs (Donaldson, Krause); Geomagnetic and Ordnance Prospecting (Clem, Foley, Keene); and Gravity and Motion Detection (Paik)},
isbn={3-527-40408-2},
url={http://inis.iaea.org/search/search.aspx?orig_q=RN:38047858}
}

@Article{Tesche1977,
author="Tesche, Claudia D.
and Clarke, John",
title="dc SQUID: Noise and optimization",
journal="Journal of Low Temperature Physics",
year="1977",
month="Nov",
day="01",
volume="29",
number="3",
pages="301--331",
abstract="A computer model is described for the dc SQUID in which the two Josephson junctions are nonhysteretic, resistively shunted tunnel junctions. In the absence of noise, current-voltage(I--V) characteristics are obtained as functions of the applied flux, $\Phi$a, SQUID inductanceL, junction critical currentI0, and shunt resistanceR. The effects of asymmetry inL, I0, andR are discussed.I--V characteristics, flux-voltage transfer functions, and low-frequency spectral densities of the voltage noise are obtained at experimentally interesting values of the parameters in the presence of Johnson noise in the resistive shunts. The transfer functions and voltage spectral densities are used to calculate the flux and energy resolution of the SQUID operated as an open-loop, small-signal amplifier. The resolution of the SQUID with ac flux modulation is discussed. The flux resolution calculated for the SQUID of Clarke, Goubau, and Ketchen is1.6 {\texttimes} 10−5$\Phi$0Hz−1/2, approximately one-half the experimental value. Optimization of the SQUID resolution is discussed: It is shown that the optimum operating condition is $\beta$=2LI0/$\Phi$0≈1. Finally, some speculations are made on the ultimate performance of the tunnel junction dc SQUID. When the dominant noise source is Johnson noise in the resistive shunts, the energy resolution per Hz is4kBT($\pi$LC)1/2, whereC is the junction capacitance, and the constraintR=($\Phi$0/2$\pi$CI0)1/2has been imposed. This result implies that the energy resolution is proportional to (junction area)1/2. In the limiteI0R ≫kBT, the dominant noise source is shot noise in the junctions; for $\beta$=1, the energy resolution per Hz is then approximatelyh/2.",
issn="1573-7357",
doi="10.1007/BF00655097",
url="https://doi.org/10.1007/BF00655097"
}

@ARTICLE{403046,
author={J. E. {Sauvageau} and C. J. {Burroughs} and P. A. A. {Booi} and M. W. {Cromar} and R. P. {Benz} and J. A. {Koch}},
journal={IEEE Transactions on Applied Superconductivity},
title={Superconducting integrated circuit fabrication with low temperature ECR-based PECVD SiO/sub 2/ dielectric films},
year={1995},
volume={5},
number={2},
pages={2303-2309},
keywords={superconducting integrated circuits;integrated circuit technology;plasma CVD coatings;silicon compounds;dielectric thin films;sputter etching;superconducting integrated circuit fabrication;ECR PECVD SiO/sub 2/ films;Josephson voltage standards;VLSI scale array oscillators;SQUIDs;kinetic-inductance-based devices;Josephson junction;low temperature processing;interlayer dielectrics;statistical process control;experimental design;oxide;niobium;etching;endpoint detection;overetch;SiO/sub 2/;Nb;Superconducting integrated circuits;Fabrication;Plasma temperature;Etching;Voltage-controlled oscillators;Standards development;Very large scale integration;SQUIDs;Josephson junctions;Electrons},
doi={10.1109/77.403046},
ISSN={2378-7074},
month={June}
}

@article{doi:10.1063/1.93210,
author = {Ketchen,M. B.  and Jaycox,J. M. },
title = {Ultra‐low‐noise tunnel junction dc SQUID with a tightly coupled planar input coil},
journal = {Applied Physics Letters},
volume = {40},
number = {8},
pages = {736-738},
year = {1982},
doi = {10.1063/1.93210},
URL = {https://doi.org/10.1063/1.93210},
eprint = {https://doi.org/10.1063/1.93210}
}

@article{Du_2020,
Author = {Braine, T. and Cervantes, R. and Crisosto, N. and Du, N.and Kimes, S. and Rosenberg, L. J. and Rybka, G. and Yang, J. and Bowring, D. and Chou, A. S. and Khatiwada, R. and Sonnenschein, A. and Wester, W. and Carosi, G. and Woollett, N. and Duffy, L. D. and Bradley, R. and Boutan, C. and  Jones, M. and LaRoque, B. H. and Oblath, N. S. and Taubman, M. S. and Clarke, J. and Dove, A. and Eddins, A. and O’Kelley, S. R. and Nawaz, S. and Siddiqi, I. and  Stevenson, N. and Agrawal, A. and Dixit, A. V. and Gleason, J. R. and Jois, S. and Sikivie, P. and Sullivan, N. S. and  Tanner, D. B. and Lentz, E. and Daw, E. J.and Buckley, J. H. and Harrington, P. M. and Henriksen, E. A. and Murch, K. W.},
Title = {Extended Search for the Invisible Axion with the Axion Dark Matter Experiment},
Year = {2020},
journal = {Physical Review Letters},
volume = {124},
Issue = {10},
pages = {101303-101306},
URL = { 
        https://doi.org/10.1103/PhysRevLett.124.101303
    
},
Eprint = {arXiv:arXiv:1910.08638},
}

@misc{placeholder,
title = {Ref for inductance of slit. (To be completed.)},
year = {1492}
}

@book{MA-COM,
author = {MA-COM},
title = {MA46H120 GaAs varactor},
address={100 Chelmsford St, Lowell, MA, 01851}
}

@phdthesis{okelley_thesis,
author={Sean R. O'Kelley},
school={UC Berkeley},
year={2019},
title={The Microstrip SQUID Amplier in the Axion Dark Matter eXperiment (ADMX)},
journal={ProQuest Dissertations and Theses},
pages={284},
keywords={ADMX; Axion; Haloscope; MSA; SQUID; Physics; Astrophysics; 0605:Physics; 0596:Astrophysics},
isbn={9781392355336},
url={https://search.proquest.com/docview/2313355201?accountid=14496},
} 

@manual{lakeshore_sensor,
author = {Lakeshore},
title = {User's Manual Model 370 AC Resistance Bridge},
date = {May 14, 2009},
version = {Revision 1.14},
organization = {Lake Shore Cryotronics, Inc},
pagetotal = {184},
pubstate = {May 14, 2009},
}

@misc{EPICS,
  title = {Experimental Physics and Industrial Control System},
  howpublished = {\url{https://epics.anl.gov/base/index.php}},
  note = {Accessed: 2010-12-06}
}

@misc{LUA,
  title = {Lua Reference Manual},
  howpublished = {\url{https://www.lua.org/manual/}},
  note = {Accessed: 2010-12-06}
}
@misc{CST,
  title = {CST Studio Suite},
  howpublished = {\url{https://www.3ds.com/products-services/simulia/products/cst-studio-suite/}},
  note = {Accessed: 2019-12-06}
}
@misc{LNF,
  title = {Low Noise Factory},
  howpublished = {\url{http://www.lownoisefactory.com/}},
  note = {Accessed: 2019-12-06}
    
}
@article{Callen1951,
  title={Irreversibility and generalized noise},
  author={Callen, Welton},
  journal={Physics Review Letters},
  volume={83},
  number={1},
  pages={34-40},
  year={1951},
  publisher={APS}
}
@article{Devyatov1986,
  title={Quantum-statistical theory of microwave
detection using superconducting tunnel junctions},
  author={Devyatov, I. A. and Kuzmin, L.S. and Likharev, K. K. and Migulin, V. V. and Zorin, A. B.},
  journal={J. Appl. Phys.},
  volume={60},
  number={5},
  pages={1808-1828,},
  year={1986},
  publisher={American Institute of Physics}
}
}

@article{doi:10.1063/1.1651991,
author = {Stewart,W. C. },
title = {CURRENT‐VOLTAGE CHARACTERISTICS OF JOSEPHSON JUNCTIONS},
journal = {Applied Physics Letters},
volume = {12},
number = {8},
pages = {277-280},
year = {1968},
doi = {10.1063/1.1651991},
URL = {htps://doi.org/10.1063/1.1651991},
eprint = {https://doi.org/10.1063/1.1651991}
}

@article{doi:10.1063/1.1656743,
author = {McCumber,D. E. },
title = {Effect of ac Impedance on dc Voltage‐Current Characteristics of Superconductor Weak‐Link Junctions},
journal = {Journal of Applied Physics},
volume = {39},
number = {7},
pages = {3113-3118},
year = {1968},
doi = {10.1063/1.1656743},
URL = {https://doi.org/10.1063/1.1656743},
eprint = {https://doi.org/10.1063/1.1656743}
}

@article{doi:10.1063/1.3583380,
author = {Kinion,D.  and Clarke,John },
title = {Superconducting quantum interference device as a near-quantum-limited amplifier for the axion dark-matter experiment},
journal = {Applied Physics Letters},
volume = {98},
number = {20},
pages = {202503},
year = {2011},
doi = {10.1063/1.3583380},
URL = {https://doi.org/10.1063/1.3583380},
eprint = {https://doi.org/10.1063/1.3583380}
}

@article{doi:10.1063/1.4985384,
author = {Mück,Michael  and Schmidt,Bernd  and Clarke,John },
title = {Microstrip superconducting quantum interference device amplifier: Operation in higher-order modes},
journal = {Applied Physics Letters},
volume = {111},
number = {4},
pages = {042604},
year = {2017},
doi = {10.1063/1.4985384},
URL = {https://doi.org/10.1063/1.4985384},
eprint = {https://doi.org/10.1063/1.4985384}
}

@phdthesis{thesisKinion,
  author        = "Kinion, Darin",
  title         = "First Results from a Multiple-Microwave-Cavity Search for Dark-Matter Axions",
  school        = "University of California Davis",
  address       = "",
  month         = "",
  year          = "2001",
  url           = "https://search.proquest.com/docview/304685582"
}
\usepackage{dcolumn}
\usepackage{bm}
\usepackage[mathlines]{lineno}
\usepackage{color,soul}

\usepackage{subfigure}
\usepackage{amsmath}
\usepackage{feynmp}
\usepackage{tikz-feynman}
\usepackage{siunitx}
\usepackage{fancyhdr}
\tikzfeynmanset{compat=1.0.0}

\begin{document}

\title{Axion Dark Matter eXperiment: Detailed Design and Operations}

\author{R. Khatiwada}
 \email[Correspondence to:]{rkhatiwada@iit.edu}
\affiliation{Fermi National Accelerator Laboratory, Batavia IL 60510, USA}
 \affiliation{Illinois Institute of Technology, Chicago IL 60616, USA}
\author{D. Bowring}
  \affiliation{Fermi National Accelerator Laboratory, Batavia IL 60510, USA}
\author{A. S. Chou} 
  \affiliation{Fermi National Accelerator Laboratory, Batavia IL 60510, USA}
\author{A. Sonnenschein} 
  \affiliation{Fermi National Accelerator Laboratory, Batavia IL 60510, USA}
  \author{W. Wester} 
  \affiliation{Fermi National Accelerator Laboratory, Batavia IL 60510, USA}
  \author{D. V. Mitchell}
  \affiliation{Fermi National Accelerator Laboratory, Batavia IL 60510, USA}

 \author{T. Braine}
  \affiliation{University of Washington, Seattle, WA 98195, USA}
\author{C. Bartram}
  \affiliation{University of Washington, Seattle, WA 98195, USA}  
\author{R. Cervantes}
  \affiliation{University of Washington, Seattle, WA 98195, USA}
 \author{N. Crisosto}
   \affiliation{University of Washington, Seattle, WA 98195, USA}
\author{N. Du}%
  \affiliation{University of Washington, Seattle, WA 98195, USA}
\author{S. Kimes}
  \affiliation{University of Washington, Seattle, WA 98195, USA}
  \affiliation{currently Microsoft, Redmond, WA 98052, USA}
 \author{L. J Rosenberg}%
  \affiliation{University of Washington, Seattle, WA 98195, USA}
  \author{G. Rybka}%
  \affiliation{University of Washington, Seattle, WA 98195, USA}
    \author{J. Yang}%
  \affiliation{University of Washington, Seattle, WA 98195, USA}
    \author{D. Will}%
  \affiliation{University of Washington, Seattle, WA 98195, USA}  

\author{G. Carosi}
\affiliation{Lawrence Livermore National Laboratory, Livermore, CA 94550, USA}
\author{N. Woollett}
\affiliation{Lawrence Livermore National Laboratory, Livermore, CA 94550, USA}
\author{S. Durham}
\affiliation{Lawrence Livermore National Laboratory, Livermore, CA 94550, USA}
\author{L. D. Duffy}
  \affiliation{Los Alamos National Laboratory, Los Alamos, NM 87545, USA}

\author{R. Bradley}
  \affiliation{National Radio Astronomy Observatory, Charlottesville, Virginia 22903, USA}

\author{C. Boutan}
  \affiliation{Pacific Northwest National Laboratory, Richland, WA 99354, USA}
\author{M. Jones}
  \affiliation{Pacific Northwest National Laboratory, Richland, WA 99354, USA}
\author{B. H. LaRoque}
  \affiliation{Pacific Northwest National Laboratory, Richland, WA 99354, USA}
\author{N. S. Oblath}
  \affiliation{Pacific Northwest National Laboratory, Richland, WA 99354, USA}
\author{M. S. Taubman}
  \affiliation{Pacific Northwest National Laboratory, Richland, WA 99354, USA}
  \author{J. Tedeschi}
  \affiliation{Pacific Northwest National Laboratory, Richland, WA 99354, USA}

\author{John Clarke}
  \affiliation{University of California, Berkeley, CA 94720, USA}
\author{A. Dove}
   \affiliation{University of California, Berkeley, CA 94720, USA}
\author{A. Eddins}
\affiliation{University of California, Berkeley, CA 94720, USA}
 \affiliation{currently IBM Almaden Research Center, San Jose, CA 95120, USA}
  \author{A. Hashim}
  \affiliation{University of California, Berkeley, CA 94720, USA}
\author{S. R. O'Kelley}
  \affiliation{University of California, Berkeley, CA 94720, USA}
   \affiliation{currently Lawrence Livermore National Laboratory, Livermore, CA 94550, USA}
\author{S. Nawaz}
  \affiliation{University of California, Berkeley, CA 94720, USA}
  \affiliation{currently Global Communication Semiconductors Inc., Torrance, CA 90505, USA}
  
\author{I. Siddiqi}
  \affiliation{University of California, Berkeley, CA 94720, USA}
\author{N. Stevenson}
  \affiliation{University of California, Berkeley, CA 94720, USA}

\author{A. Agrawal}
\affiliation{University of Chicago, IL 60637, USA}
\author{A. V. Dixit}
\affiliation{University of Chicago, IL 60637, USA}
 
\author{J.~R. Gleason}
  \affiliation{University of Florida, Gainesville, FL 32611, USA}
\author{S. Jois}
  \affiliation{University of Florida, Gainesville, FL 32611, USA}
 \author{P. Sikivie}
  \affiliation{University of Florida, Gainesville, FL 32611, USA}
\author{N. S. Sullivan}
  \affiliation{University of Florida, Gainesville, FL 32611, USA}
\author{D. B. Tanner}
  \affiliation{University of Florida, Gainesville, FL 32611, USA}
  \author{J. A. Solomon}
  \affiliation{University of Florida, Gainesville, FL 32611, USA}
  \affiliation{currently University of North Carolina, Chapel Hill, NC 27599, USA}
  
\author{E. Lentz}
  \affiliation{University of G\"{o}ttingen, G\"{o}ttingen, Germany}  
  
\author{E. J. Daw}
  \affiliation{University of Sheffield, Sheffield, UK}
  \author{M. G. Perry}
  \affiliation{University of Sheffield, Sheffield, UK}
  
\author{J. H. Buckley}
  \affiliation{Washington University, St. Louis, MO 63130, USA}
\author{P. M. Harrington}
  \affiliation{Washington University, St. Louis, MO 63130, USA}
\author{E. A. Henriksen}
  \affiliation{Washington University, St. Louis, MO 63130, USA} 
\author{K. W. Murch}
  \affiliation{Washington University, St. Louis, MO 63130, USA}
\author{G. C. Hilton}
  \affiliation{National Institute of Standards and Technology, Gaithersburg, MD 20899, USA}   
  
\collaboration{ADMX Collaboration}\noaffiliation

\date{\today}

\vspace*{1cm}

\begin{abstract}
Axion Dark Matter eXperiment (ADMX) ultra low noise haloscope technology has enabled the successful completion of two science runs (1A and 1B) that looked for dark matter axions in the $2.66$ to $3.1$ $\mu$eV mass range with Dine-Fischler-Srednicki-Zhitnisky (DFSZ) sensitivity ~\cite{PhysRevLett.120.151301,Du_2020}. Therefore, it is the most sensitive axion search experiment to date in this mass range. We discuss the technological advances made in the last several years to achieve this sensitivity, which includes the implementation of components, such as state-of-the-art quantum limited amplifiers and a dilution refrigerator. Furthermore, we demonstrate the use of a frequency tunable Microstrip Superconducting Quantum Interference Device (SQUID) Amplifier (MSA), in Run 1A, and a Josephson Parametric Amplifier (JPA), in Run 1B, along with novel analysis tools that characterize the system noise temperature.

\end{abstract}

\keywords{dark matter, axion, quantum noise limited amplifiers}

\maketitle

\section{Introduction}
The goal of achieving sensitivity to Grand Unified Theory (GUT)-scale DFSZ dark matter axions has motivated the ADMX collaboration to implement new technologies in its recent axion searches with the primary focus on minimizing both amplifier and cavity blackbody noise. As such, we implemented two critical pieces of technology in the ADMX haloscope: state-of-the-art quantum amplifiers, and a dilution refrigerator. While these were broadly the most notable technological improvements, the details varied between the runs and will be described below.

In Run 1A, ADMX acquired data over an axion mass range from $2.66$ to $2.81$ $\mu$eV from January to June 2017, demonstrating the sustained use of a Microstrip Superconducting Quantum Interference Device (SQUID) Amplifier (MSA) in the frequency range $645$ to $680$ MHz. This was followed by the implementation of a Josephson Parametric Amplifier (JPA) in Run 1B (January to October 2018), covering $680$ to $790$ MHz, corresponding to an axion mass range of $2.81$ to $3.31$ $\mu$eV. Both sets of quantum amplifiers, combined with the order of magnitude reduction in physical temperature have dramatically improved ADMX performance over previous operations ~\cite{ASZTALOS201139}. We have refined techniques for measuring the reduced system noise temperature and have incorporated it into the analysis. 
The implementation of a Model JDR-$800$ $\mathrm{{}^{3}He-{}^{4}He}$ Dilution Refrigerator System to cool the cavity enabled us to minimize the thermal noise for both runs. In Run 1A, this led to an average cavity temperature of about 150 mK. In Run 1B, we achieved an average cavity temperature of about $130$ mK. In addition, we discuss a few other advances to improve our sensitivity. These include a complete update of the Data Acquisition System (DAQ) software, and the implementation of hardware for the blind injection of synthetic axion signals.
\par Meanwhile, ADMX has expanded the search to higher mass as a part of ongoing research and development efforts. The prototype ``Sidecar" cavity system attained new mass limits in three distinct frequency regions around $4.2, 5.4$ and $7.2$ GHz \cite{PhysRevLett.121.261302}. We briefly discuss instrumentation advancements for Sidecar, because it shares the detector volume with the ADMX experiment.
\\
\section{ADMX experiment overview}
Axions are hypothetical particles that exhibit behavior characteristic of dark matter; they are non-relativistic, highly stable, and weakly interacting ~\cite{Preskill_1983}~\cite{Sikivie_1983}~\cite{Dine_1983}. Axions were originally postulated to resolve the strong CP problem~\cite{Peccei1977June}~\cite{Peccei1977Sept}. Quantum chromodynamics (QCD) calculations suggest axions could have a mass range of $1$ to $100$ $\mu$$eV$~\cite{Bonati2016}~\cite{PhysRevD.92.034507}~\cite{PhysRevLett.118.071802}~\cite{PhysRevD.96.095001}. 

The existence of axions would modify Maxwell's equations as follows:
\begin{align}
\vec\nabla{\cdot}\vec{\mathrm{E}}&=\rho-g_{a\gamma\gamma}\vec{\mathrm{B}}{\cdot}\vec\nabla{a}\\
\vec\nabla{\cdot}\vec{\mathrm{B}}&=0\\
\vec\nabla\times\vec{\mathrm{E}}&=-\frac{\partial\vec{\mathrm{B}}}{\partial{t}}\\
\vec\nabla{\times}\vec{\mathrm{B}}&=\frac{\partial{\vec{\mathrm{E}}}}{\partial{t}}+\vec{\mathrm{J}}-g_{a\gamma\gamma}\left(\vec{\mathrm{E}}\times\vec\nabla{a}-\frac{\partial{a}}{\partial{t}}\vec{\mathrm{B}}
\right).
\end{align}

\begin{figure}
\begin{center}
\includegraphics[width=4 cm]{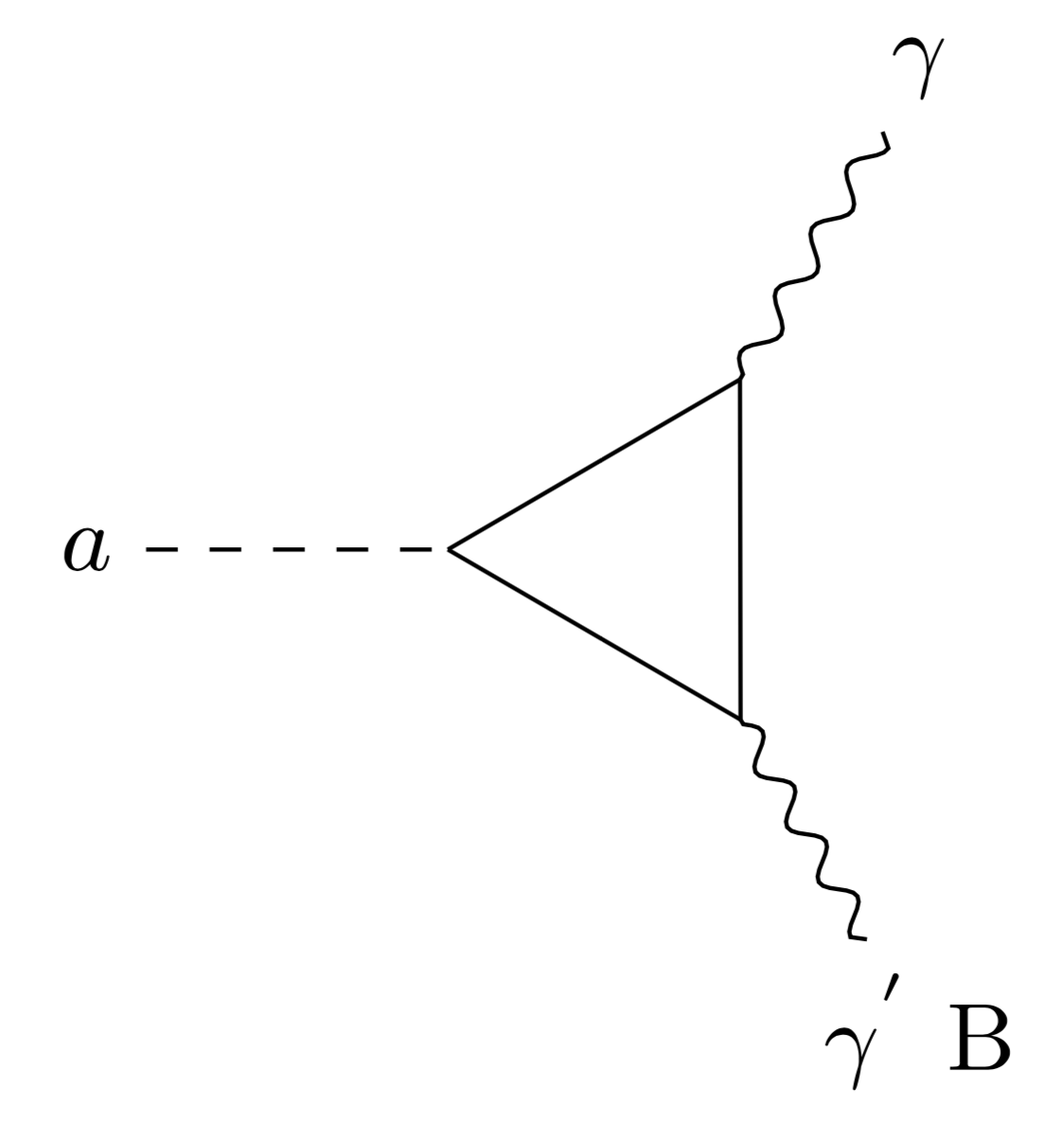}
\caption{Feynman diagram of the inverse Primakoff effect. An axion $a$ converts into a photon $\gamma$ by interacting with a virtual photon $\gamma^{'}$ in a static magnetic field B through fermionic loop. The coupling constant is denoted by $g_{a\gamma\gamma}$.}
\label{fig:axion_photon_Bfield}
\end{center}
\end{figure}

Here, $\vec{E}$ is the electric field, $\vec{B}$ is the magnetic field, $\rho$ is the charge density, $\vec{J}$ is the source current, $a$ is the scalar axion field and $g_{a\gamma\gamma}$ is the model-dependent axion-two photon coupling constant. The two primary models for axion-to-photon coupling are known as KSVZ (Kim-Shifman-Vainshtein-Zakaharov) ~\cite{Vainshtein_1980}~\cite{Kim_1979} and DFSZ (Dine-Fischler-Srednicki-Zhitnisky)~\cite{Dine_1981}. KSVZ couples only to hadrons, whereas DFSZ couples to both hadrons and leptons. These have values $-0.97$ and $0.36$ respectively. DFSZ couplings are about a factor of $3$ weaker than KSVZ couplings, so require greater experimental effort to detect. Therefore, reaching the DFSZ sensitivity has been a long sought after goal of axion experiments. The application of inhomogeneous magnetic field provides a new channel for axions to decay into a photon, whose frequency is given by, $f=E/h$ where $E$ corresponds to the total energy of the axion with contributions primarily from the rest mass energy and a small kinetic energy term and ``h" is the Plack's constant. This is known as the Inverse Primakoff Effect. The conversion is expressed by a Feynmann diagram in (Fig.~\ref{fig:axion_photon_Bfield}). 

\par In 1983, Pierre Sikivie introduced the axion haloscope, which uses a large density of virtual photons from a strong static magnetic field to allow the galactic axions to convert into real photons inside a microwave cavity. When the axion's frequency matches the resonance frequency of the microwave cavity, the conversion rate is enhanced to detectable levels. The power deposited in the cavity due to this conversion is given by, 

 \begin{widetext}
 \begin{equation}
 \label{eq:Eq.5}
P_{a\rightarrow\gamma}=(1.9{\times}10^{-22}\mathrm{W})\left(\frac{V}{136~ \mathrm{L}}\right)\left(\frac{B}{6.8~ \mathrm{T}}\right)^{2}\left(\frac{\mathrm{C}_{nlm}}{0.4}\right)\left(\frac{g_{\gamma}}{0.97}\right)^{2}\left(\frac{\rho_{a}}{\mathrm{0.45~  GeV/cm^3}}\right)\left(\frac{f_{a}}{\mathrm{650~MHz}}\right)\left(\frac{Q}{\mathrm{50000}}\right),
 \end{equation}
 \end{widetext}
 Here, $V$ is the volume of the cavity, $B$ is the magnetic field, $C_{nlm}$ is the form factor of the cavity, $\rho_{a}$ is the local dark matter density, $f_{a}$ is the frequency of the photon and $Q$ is the loaded quality factor of the cavity. The form factor is defined as the integral of the overlap between the electric field of the cavity transverse magnetic mode and the external magnetic field generated by the magnet~\cite{Sikivie1985}. For any given mode in an empty cylindrical cavity, the TM$_{010}$ mode has the highest form factor and the cavity radial dimension corresponds to approximately one-half of the photon wavelength. In practice, the geometry of the cavity is more complicated because of the presence of tuning rods, so simulation is necessary to understand the form factor.

\par From Eq.~\ref{eq:Eq.5}, it is clear that experimentalists have several handles which can be used to optimize the power extracted by the receiver. Cavity volume, magnetic field and quality factor can all be maximized, whereas the remaining parameters ($g_{\gamma}$, $\rho_{a}$) are fixed by nature. The signal-to-noise ratio (SNR) is defined by the Dicke radiometer equation~\cite{doi:10.1063/1.1770483}:
 \begin{equation}
 \frac{S}{N}=\frac{P_{axion}}{k_{B}T_{sys}}\sqrt{\frac{t}{b}}.
 \end{equation}
Here $S$ is the signal, $N$ is the noise, $P_{axion}$ is the power that would be deposited in the cavity in the event of an axion signal, $k_{B}$ is the Boltzmann constant, $T_{sys}$ is the system noise temperature, $t$ is the integration time, and $b$ is the measurement frequency bandwidth. The total system noise temperature $T_{sys}$ is composed of cavity blackbody noise and amplifier noise, which should be minimized to achieve the highest possible SNR.

\section{The detector}
\begin{figure}
\begin{center}
\includegraphics[width=8 cm]{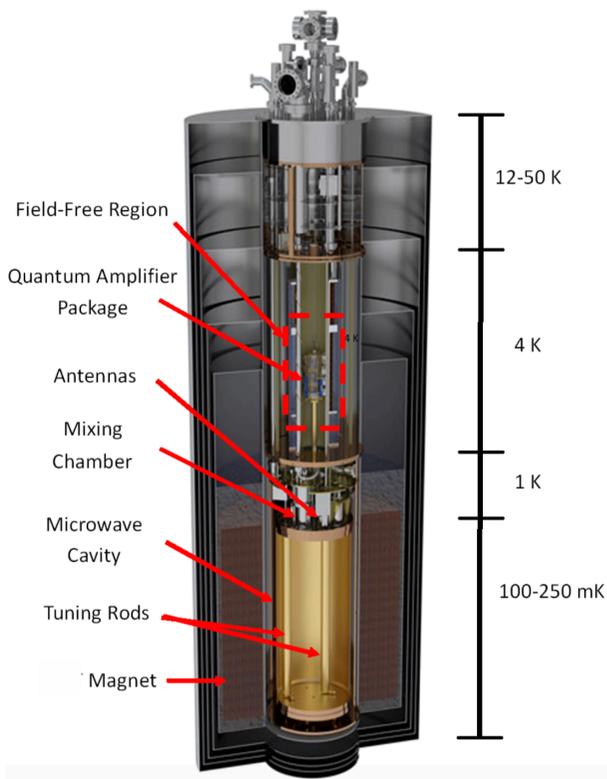}
\caption{Schematic of ADMX detector components. The microwave cavity can be seen at the center, with tuning rods. The central cylindrical structure containing cavity and electronics that is inserted into the magnet bore is called ``the insert''. Various temperature stages are indicated on the right hand side. The quantum amplifier package is thermalized to the microwave cavity}
\label{fig:insert}
\end{center}
\end{figure}

ADMX is located at the Center for Experimental Nuclear Physics and Astrophysics (CENPA) at the University of Washington, Seattle. The ADMX detector consists of several components collectively referred to as ``the insert'' shown in Fig.~\ref{fig:insert}. The insert is lowered into the bore of a superconducting solenoid magnet, which is operated typically at just under $8$ T, for data-taking operations. The cylindrical insert ($0.59$ m diameter, $3$ m height) contains the microwave  cavity, motion control system for the antenna and cavity tuning rods, cryogenic and quantum electronics, a dilution refrigerator, a liquid $\mathrm{{}^{4}He}$ reservoir, a bucking magnet and the Sidecar cavity and electronics. The insert is designed such that the field sensitive quantum amplifiers, switches and circulators are housed in a field free region, with a volume $0.22$ m height by $0.15$ m diameter, provided by a bucking coil. The  cavity is inserted concentrically in the magnet bore to maximize the form factor. The insert also involves a Cryomech PT60 pulse tube cooler that cools the top of the insert to $50$ K. Below that, a liquid $\mathrm{{}^{4}He}$ reservoir maintains the bucking coil and second stage High Electron Mobility Transistor (HEMT) amplifiers near 4 K. Two pumped $\mathrm{{}^{4}He}$ refrigerators are used, one is thermally tied to the motion control system and the thermal shields surrounding the cavity and counters the thermal load of moving tuning rods. The other pre-cools the $\mathrm{{}^{3}He}$/$\mathrm{{}^{4}He}$ mixture used in the dilution refrigerator before it enters the still. The dilution refrigerator mixing chamber is thermally anchored to both the first stage cryogenic electronics and the microwave cavity.

\subsection{Magnets}
ADMX operated the superconducting magnet at $6.8$ and $7.6$ T respectively for Runs 1A and 1B. The magnet requires approximately $2,000$ L of liquid helium per month for continuous cooling during data taking operations (supplied by a closed loop Linde liquifier system). The applied magnetic field is along the axis of the cavity. The bucking magnet reduces the magnetic field at the site of the electronics package to below $0.1$ mT. Two Hall probes are located on each end of the bucking coil to monitor the field at the electronics site during data acquisition to ensure it is within tolerable limits. The Hall probes are both model HGCT-$3020$ InAs Hall Generators from Lakeshore. 
\subsection{Cavity}
\begin{figure}
\begin{center}
\includegraphics[width=8 cm]{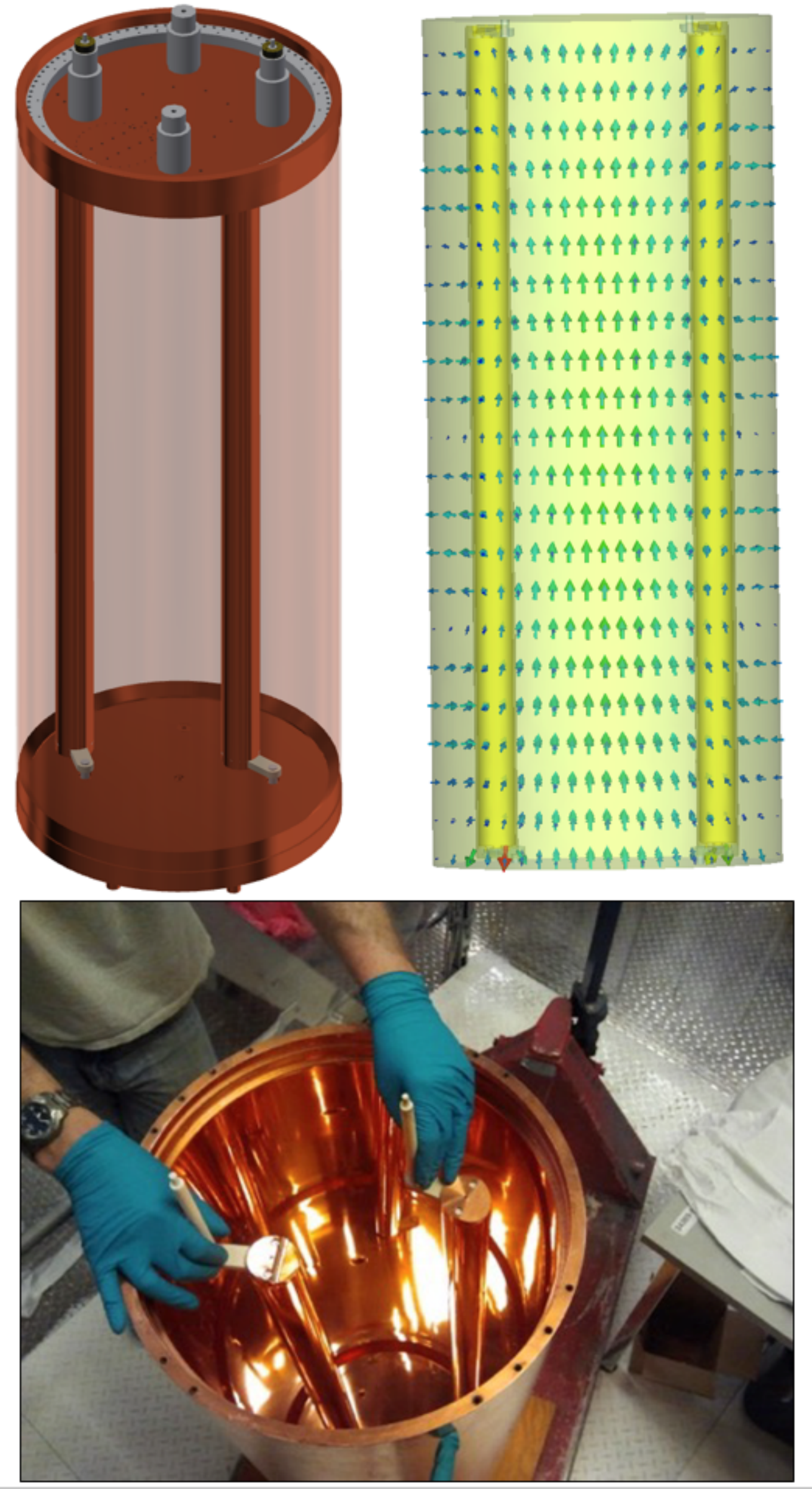}
\caption{ADMX cavity system. Top left shows a cutaway view of the CAD model. Top right is a Computer Simulation Technology (CST) Microwave Studio simulation~\cite{CST} of the TM$_{010}$ mode with each of the rods at $116^{\circ}$ from the center. Bottom  is a picture of system with top endcap removed showing two $0.05$ m diameter tuning rods and their aluminum-oxide ceramic axles. }
\label{fig:cavity_system}
\end{center}
\end{figure}

\begin{figure}
\begin{center}
\includegraphics[width=8 cm]{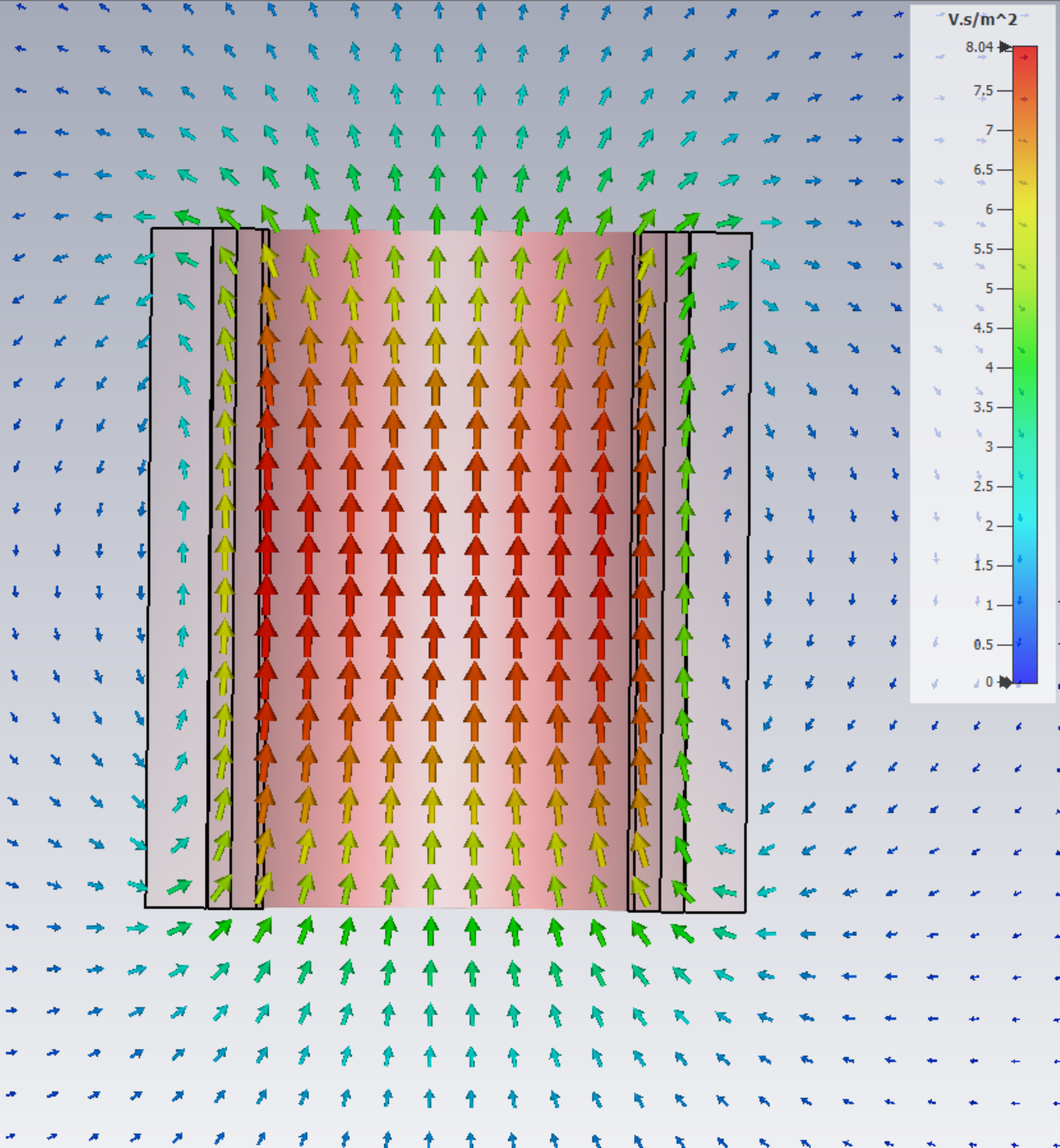}
\caption{ADMX magnet system simulated using CST Magnetic Field Solver~\cite{CST}. This shows the magnet field profile with the maximum ($8$ T) at the center. This field profile is used to convolve the form-factor for the resonant mode in Fig.~\ref{fig:cavity_system}.}
\label{fig:admx_main_magnet_field}
\end{center}
\end{figure}

The ADMX cavity is a copper-plated stainless steel ($136$ L) right-circular cylindrical cavity approximately 1 m in length and $0.4$ m in diameter. Two $0.05$ m diameter copper tuning rods run through the length of the cavity and are translated from the walls to near the center using aluminum oxide rotary armatures. This allows the fundamental TM$_{010}$-like mode that couples to the detector to be varied from $580$ MHz to $890$ MHz. Both the stainless steel cavity and the copper tuning rods are plated with OFHC copper to a minimum thickness of $0.08$ mm and then annealed for $8$ hours at $400^{\circ}\mathrm{C}$ in vacuum. The annealing process increases the grain-size of the copper crystals leading to longer electron scattering path lengths as the copper is cooled into the anomalous skin depth regime and thus producing high Q-factors for the detector~\cite{chambers1950}. The cavity system and the magnetic field profile of the main magnet can be seen in Figs. \ref{fig:cavity_system} and \ref{fig:admx_main_magnet_field}. 

The quality factor of the cavity modes are frequency dependent and are measured periodically via a transmission measurement made through the weakly coupled port to the output port. The presence of the tuning rod lowers the quality factor of an empty cavity. A quality factor between $40,000$ and $80,000$ was typically achieved in Runs 1A and 1B. The form-factor for the cavity is calculated from simulation. The mode structure of the simulation is compared to that measured from the cavity to ensure accuracy. The resulting E-field is convolved with a model of the B-field produced by the magnet. The form-factor of the $TM_{010}$ varies with frequency and rod position with an average value of $0.4$. 

\subsection{Mechanical/motion control system}

\begin{figure}
\begin{center}
\includegraphics[width=8 cm]{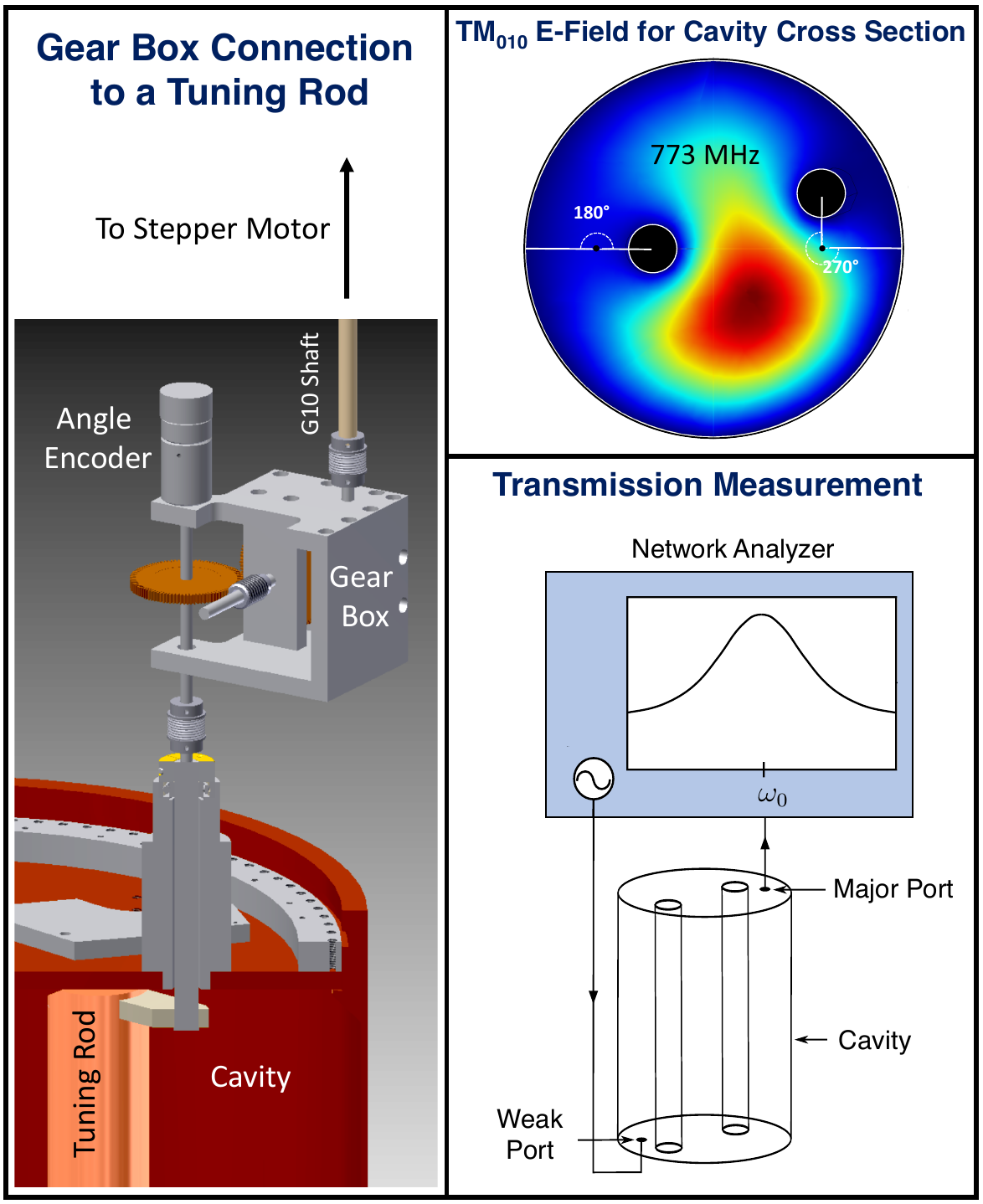}
\caption{Tuning mechanism of the main cavity \cite{thesisBoutan}. Left: a room temperature stepper motor rotates a G$10$ fiberglass shaft connected to a gear box. This motion undergoes a $1:19600$ gear reduction and rotates the tuning rod in the cavity. Top right: rotation of the tuning rods changes the boundary conditions in the cavity and alters the resonant frequency. Bottom right: the resonant frequency of the cavity is determined by injecting swept power from a network analyzer into a weakly coupled antenna. The swept power is picked up by the critically coupled antenna to measure the transfer function of the cavity.}
\label{fig:rod}
\end{center}
\end{figure}

Two copper-plated tuning rods swept out the $0.09$ m radius circular trajectories shown in Fig~\ref{fig:rod}. They are rotated from the walls of the cavity ($\theta = 0$), where they minimally impact the electromagnetic boundary conditions of the resonator, to the center where the TM$_{010}$ frequency is at its highest ($\theta = 180$). The armatures that protrude through the end caps and offset the rods are made of alumina. This prevents the rods from acting as antennae and thus radiating power out of the system.
Mechanical motion is translated to the rods via room temperature stepper motors mounted on the top plate of the insert. Acting through vacuum feed-throughs, these stepper motors communicate motion to long G10 fiberglass shafts that connect to gear boxes (upper right panel of Fig.~\ref{fig:rod}). The gear boxes have minimal backlash (a $19,600:1$ gear reduction), allowing for micro-radian positioning of the tuning rods. 
\begin{figure}
\begin{center}
\includegraphics[width=8 cm]{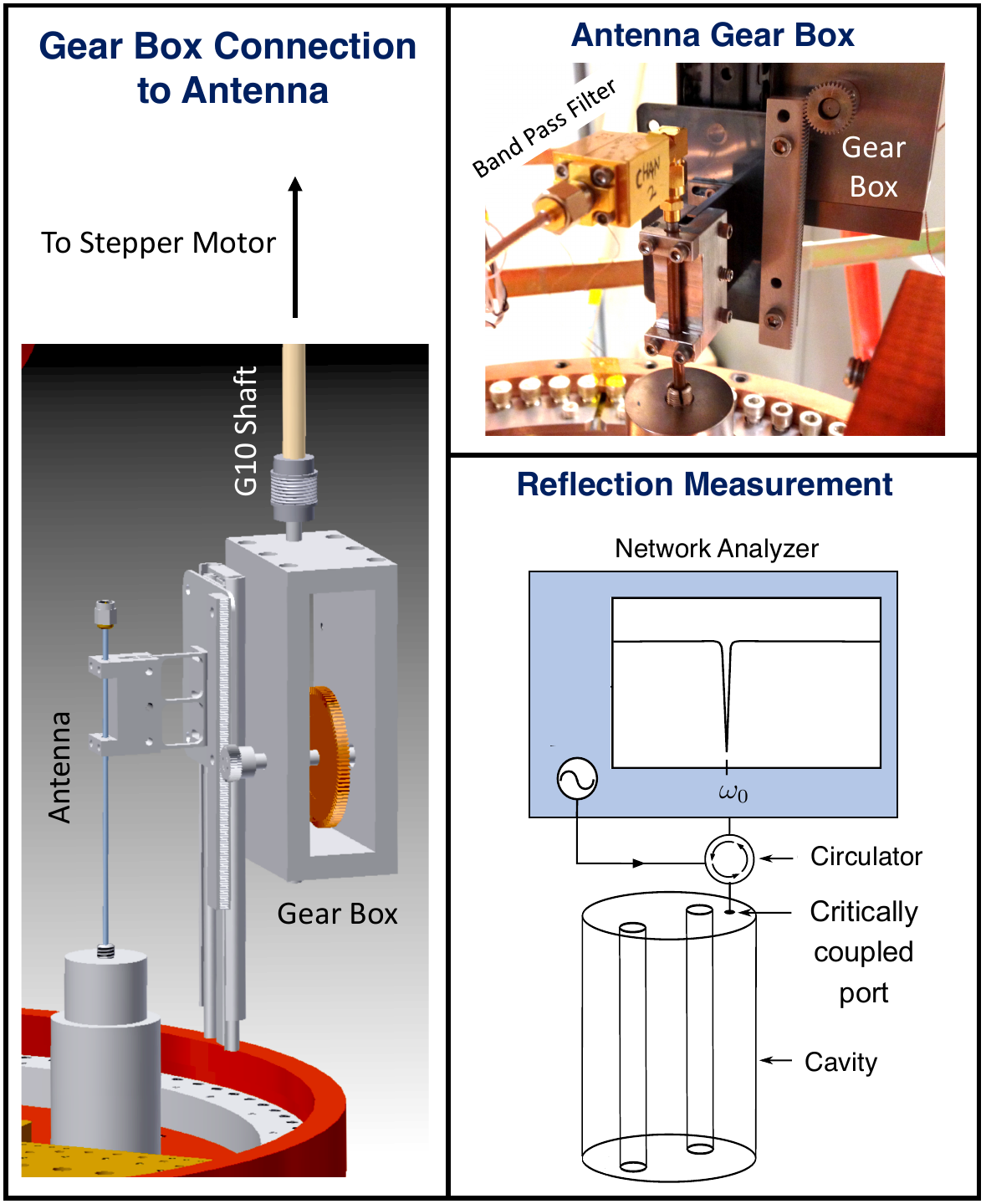}
\caption{Antenna coupling mechanism for the main experiment \cite{thesisBoutan}. Left: a room temperature stepper motor rotates a G$10$ fiberglass shaft connected to a gear box. Rotary motion is converted to linear motion which inserts or removes the antenna from the cavity. Top right: an image of the linear gear box holding the coaxial antenna. Bottom right: the antenna coupling is determined by reflecting swept power from a network analyzer off of the antenna. If the antenna is critically coupled, all of the incident power on resonance will be absorbed into the cavity, as indicated by a deep trough in the reflection measurement.}
\label{fig:ant}
\end{center}
\end{figure}
A variable depth antenna located on the top of the cavity picks up axion signal from the cavity and transmits it to the amplifiers. This semi-rigid, coaxial antenna attaches to a different gearbox which turns rotary motion from the room temperature stepper motor into linear depth control. The depth of the antenna is adjusted to maintain a critical coupling to the TM$\mathrm{_{010}}$ mode. When it is critically coupled (or impedance matched) to the cavity on resonance, the power escaping through the antenna equals the power lost in the walls of the cavity, whereas off-resonance most of the power is reflected. The coupling is determined by reflecting swept power from the cavity and measuring its magnitude. This was achieved with the network analyzer and a circulator shown in Fig.~\ref{fig:ant}. The swept output of the network analyzer is directed by the circulator towards the cavity. Power reflected from the cavity then travels back to the circulator, up the receiver chain and back to the input of the network analyzer. A good impedance match is marked by a deep trough in the reflected baseline on resonance. The depth of the antenna is adjusted to minimize the depth of this trough. Conventionally, when the difference between the minima of the trough and the off-resonance baseline reaches $-30$ dB, the antenna is considered critically coupled. This means that only $0.001$ $\%$ of the on-resonance incident power is reflected.

\subsection{Cryogenic electronics package}
\begin{figure}
\begin{center}
\includegraphics[width=8.5 cm]{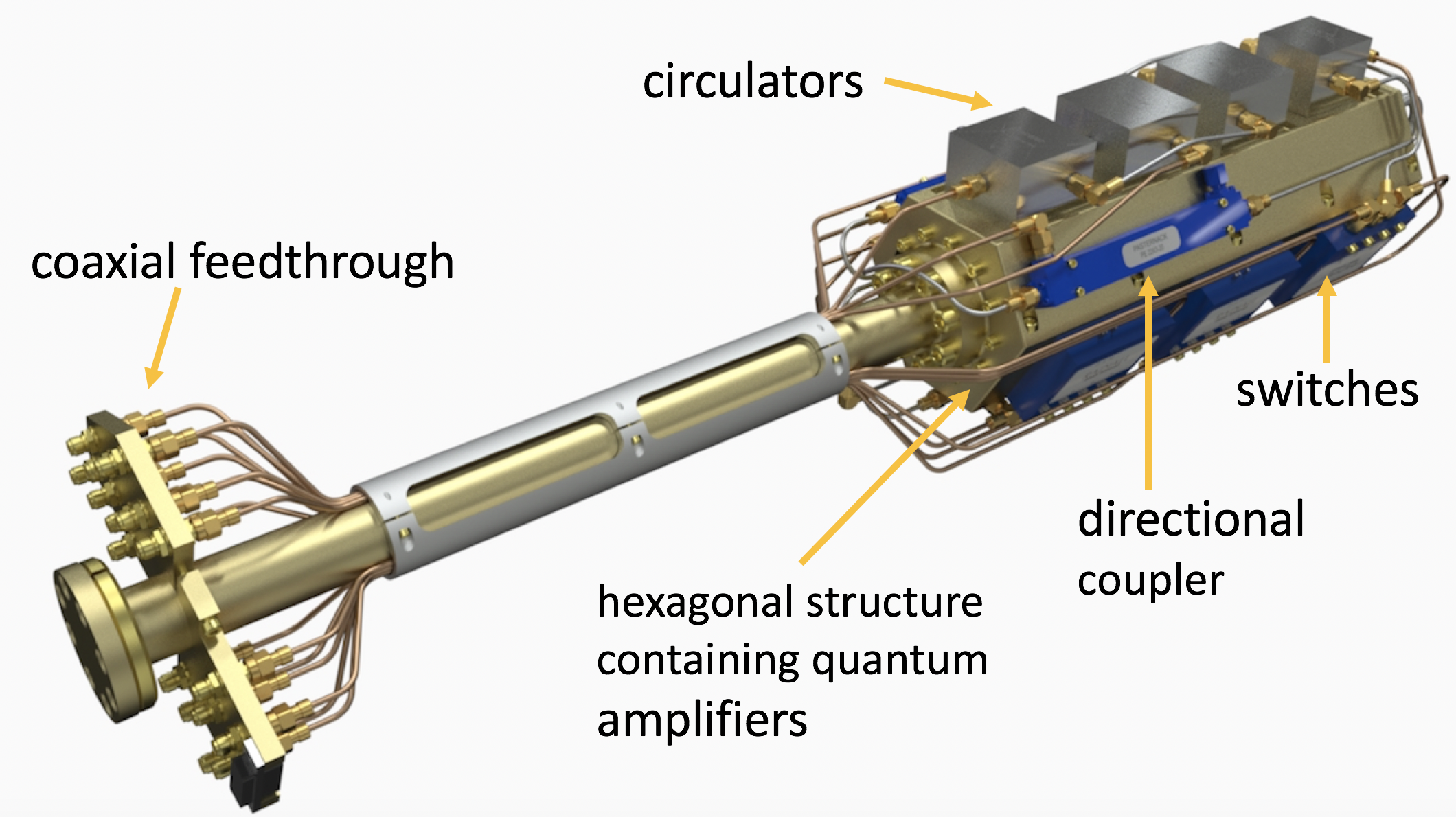}
\caption{Latest cryogenic electronics package design with the electronics and accessories being used in Run 1C (2020). The quantum electronics are housed inside the hexagonal chamber in a mu-metal shield. The hexagonal part is located in a zero magnetic field region of the ADMX detector to avoid damage to the delicate quantum electronics. The antenna go to the cavity and higher temperature electronics such as HEMT through the feed-through flange shown at the bottom of the cryogenic electronics package shaft.}
\label{fig:squidadel2019}
\end{center}
\end{figure}

The main body of the cryogenic electronic package system sits inside the field-free region (Fig.~\ref{fig:insert}) and contains the most delicate part of the experiment, the cryogenic radiofrequency (RF) electronics and quantum amplifier package. This includes quantum noise limited amplifiers (University of California Berkeley), circulators (Run 1A: Quinstar UTB$1262$KCS and Run 1B: QCY-$007020$UM00), directional couplers (Pasternack), switches (Radiall R$585433210$), and a dc block for Run 1A. Oxygen-free high thermal conductivity (OFHC) copper frame houses these electronics in the cryogenic electronic package. Fig.~\ref{fig:squidadel2019} shows an example of the cryogenic electronic package. Since most of the RF electronics are frequency dependent, they are swapped for different runs according to the target frequency range corresponding to different axion masses. 
\par Physical and amplifier noise temperatures of the cryogenic electronics housed in the package determine the total noise temperature of the system. Thus, keeping the cryogenic electronic package thermalized to the dilution refrigerator and characterizing the electronics is extremely important in determining the sensitivity of the experiment. For Run 1A (1B), the cryogenic electronic package was at a physical temperature of $300$ mK ($230$ mK) despite being heat sunk to the cavity which was at $150$ mK ($130$ mK). A newly designed cryogenic electronics package for Run 1C incorporated better thermalization of the various electronics to its frame and also to the  dilution refrigerator as well as removed the thermal short caused by misalignment to the liquid $^4$He reservoir. As a result, this design reduced the physical temperature of the electronics and components to $120$ mK (with a mixing chamber temperature of $100$ mK) in 2020 during Run 1C. In the following sub-sections, differences between the Runs 1A and Run 1B cryogenic electronics package circuitry will be discussed. 

The cryogenic electronics package initially housed two separate antennae extracting axion power from the cavity: the first coupled to the main cavity $\mathrm{TM_{010}}$ mode and the second coupled to the $\mathrm{TM_{020}}$ mode. In addition, a third antenna coupled to the Sidecar $\mathrm{TM_{010}}$ or $\mathrm{TM_{020}}$ mode. The $\mathrm{TM_{020}}$ main cavity antenna was not used since its HEMT amplifier failed to operate early into Run 1A.

\begin{figure}
\begin{center}
\includegraphics[width=7 cm]{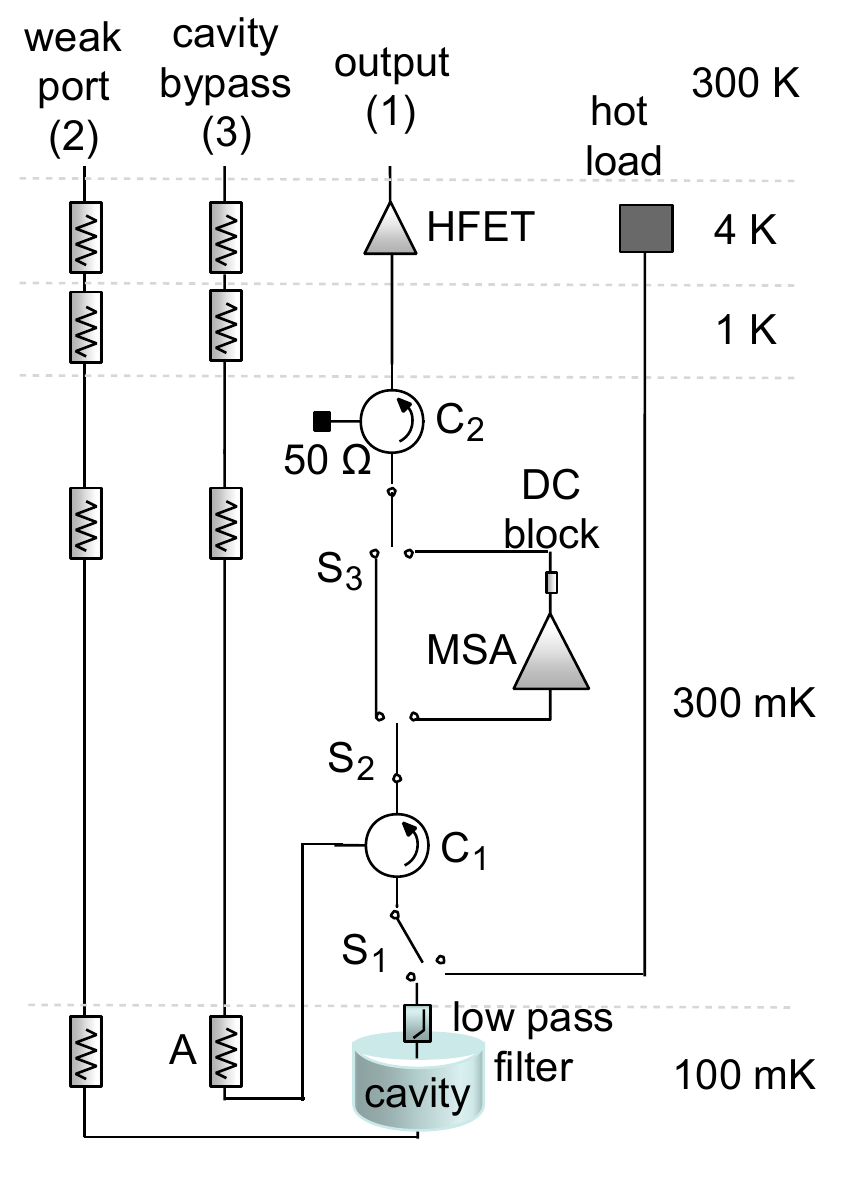}
\caption{Insert schematic of Run 1A antenna. It consisted of attenuators in the input lines, circulators (C$_{1}$ and C$_{2}$, a low pass filter, switch (S$_{1}$), MSA, dc block, variable resistor (hot load) and a HEMT amplifier connected to the cavity through coaxial cable. The switch allows the characterization of the system noise by switching to the heated ``hot load" from the cavity.}
\label{fig:rflayout_run1a}
\end{center}
\end{figure}

\begin{figure}
\begin{center}
\includegraphics[width=8.4 cm]{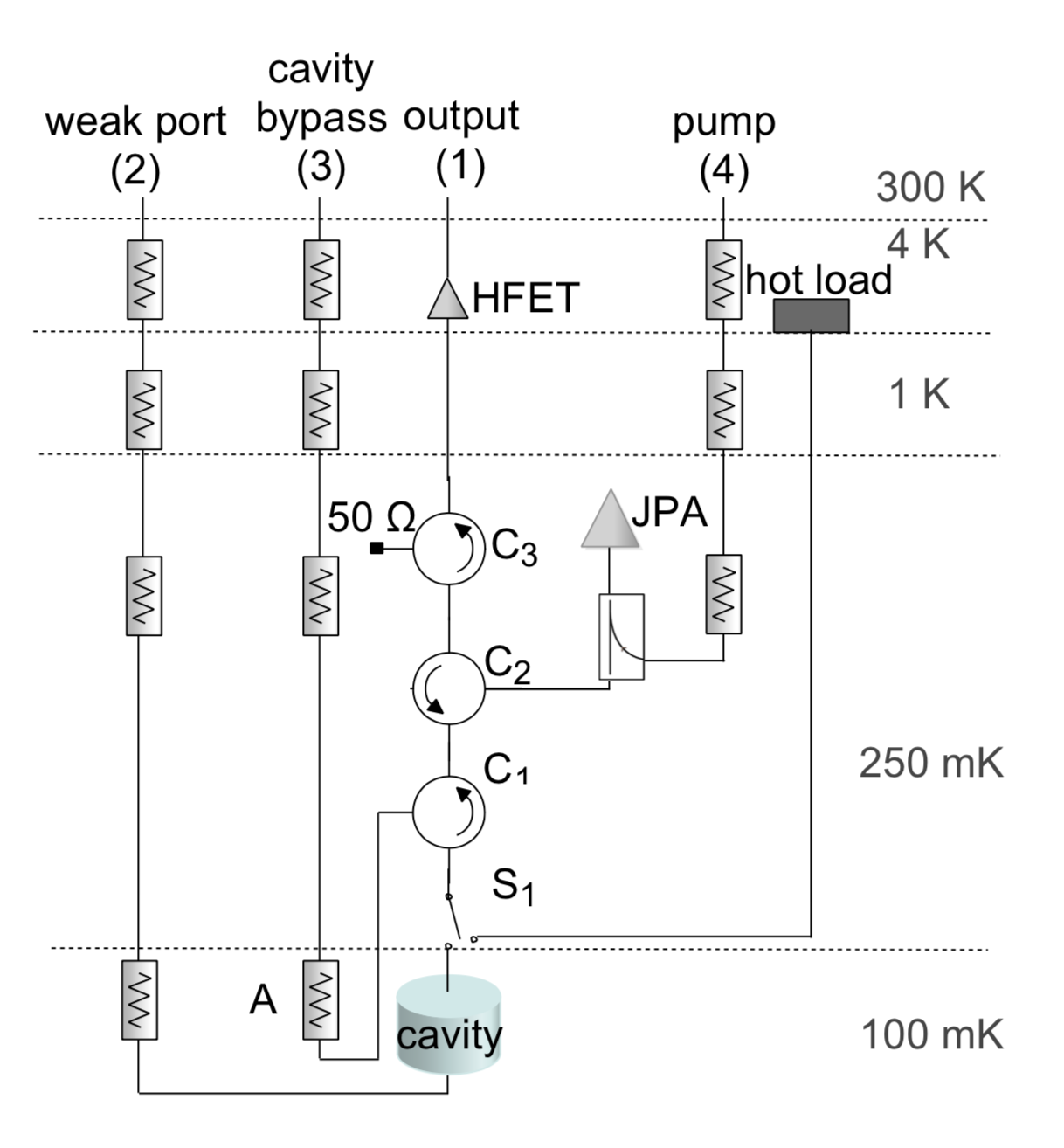}
\caption{Insert schematic of Run 1B antenna. It consisted of attenuators in the input lines, circulators (C$_{1}$, C$_{2}$ and C$_{3}$), switch S$_{1}$, directional coupler, JPA, variable resistor (hot load) and a HEMT amplifier connected to the cavity through coaxial cable. The switch enables characterization of the system noise by switching from the cavity to the ``hot load".}
\label{fig:rflayout_run1b}
\end{center}
\end{figure}

Fig.~\ref{fig:rflayout_run1a} shows the schematic of the antenna used for data taking in Run 1A (discussed in Ref\cite{Du_2018}). Similarly, Fig.~\ref{fig:rflayout_run1b} shows the main cavity $\mathrm{TM_{010}}$ antenna used to take data in Run 1B (published in Ref\cite{Du_2020}). Any photon signal generated by axions deposited in the cavity passes through the output chain electronics containing a series of switches and circulators  to the first stage quantum amplifiers and further to the HEMT (Low Noise Factory, (Run 1A: LNC$03{\_}14$A, Run 1B: LNF-LNC$0.6{\_}2$A) with an additional series of room temperature amplifiers (Minicircuits) before being digitized (Signatech). The cryogenic electronics package was wired with copper coaxial cables (Pasternack), whereas the cavity antenna was made from semi-rigid $0.0022$ m diameter NbTi superconducting coax (Keycom NbTiNbTi$085$) with approximately $0.025$ m of the center conductor exposed. The NbTi provided a thermal disconnect from the $100$ mK cavity to the $1$ K linear drive that the antennas were attached to. Flexible coaxial lines were used to connect to the cryogenic electronics package inputs which was coupled to the first stage quantum amplifiers via NbTi cables in the RF output chain. Coaxial cables in the input chain were stainless steel (Keycom ULT-$03$).

Any photon signals emerging from the cooled cavity are amplified by quantum amplifiers: the MSA for Run 1A, and the JPA for Run 1B. The building block of modern quantum amplifiers, the DC SQUID will be discussed in the subsequent sub-sections. Furthermore, both the MSA and JPA were fabricated by University of California Berkeley specifically for ADMX frequency range and their fabrication and properties will be discussed in detail below.
\subsubsection{DC SQUID}
\begin{figure*}
\begin{center}
\includegraphics{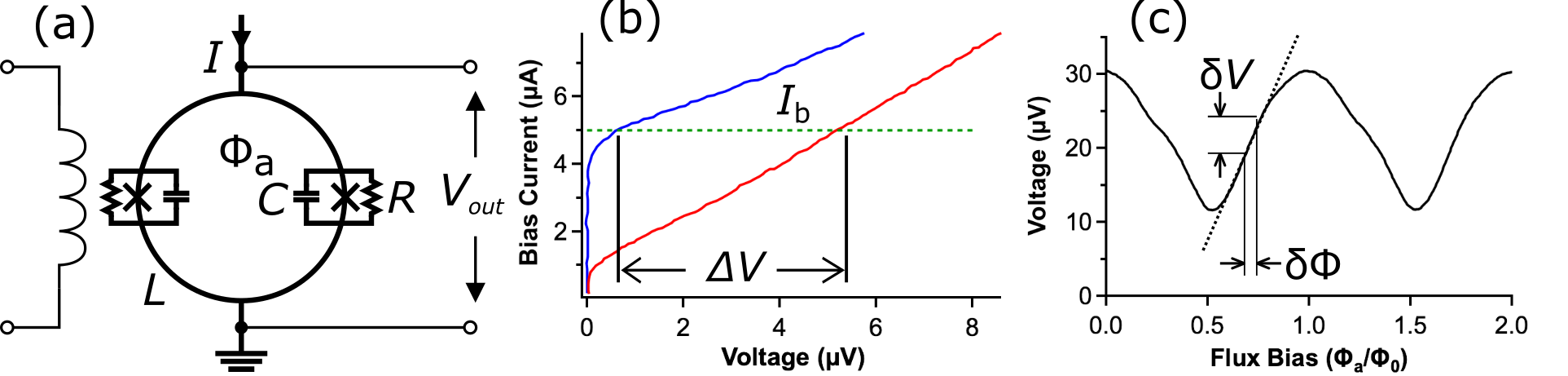}
\caption{DC SQUID. (a) Schematic, with Josephson junctions indicated with ``X". (b) Voltage $V$ vs. current $I$ for applied magnetic flux $\Phi_a=n\Phi_0$ and $(n+1/2)\Phi_0$. Voltage change $\Delta V$ at a constant bias current $I_b$. (c) Voltage $V$ vs. normalized magnetic flux $\Phi_a/\Phi_0$ at constant $I_b$.}
\label{fig:MSA_principles}
\end{center}
\end{figure*}

The MSA consists of a conventional dc Superconducting QUantum Interference Device (SQUID) \cite{jaklevic1964quantum,Clarke2004,Tesche1977}, shown schematically in Fig.~\ref{fig:MSA_principles}(a), with an integrated, tuned RF input coil. 
Since the SQUID is operated in the voltage state, the Stewart-McCumber parameter \cite{doi:10.1063/1.1651991,doi:10.1063/1.1656743} $\beta_c\equiv2\pi R^2I_0C/\Phi_0$ must be less than unity to ensure a non-hysteretic current-voltage ($I$-$V$) characteristic.
Here, $R$ is the Josephson junction \cite{josephson1962possible} shunt resistance, $I_0$ is the junction critical current, $C$ is the intrinsic junction capacitance and $\Phi_0\equiv h/2e\approx 2.07\times 10^-15$ Tm$^2$ is the flux quantum; $h$ is the Planck's constant and $e$ the electronic charge.
The lowest noise energy of the SQUID, $S_{\Phi}(f)/2L$ , is obtained when $\beta_L\equiv 2LI_0/\Phi_0=1$ and $\beta_c$ is at a value just below the onset of hysteresis \cite{Tesche1977}; here, $L$ is the geometric loop inductance and $S_{\Phi}(f)$ the spectral density of the flux noise. For our MSAs we designed $R$ to ensure $\beta_c$ is strictly less than unity and $L$ as a compromise between rf coupling and low-noise performance.
Figure~\ref{fig:MSA_principles}(b) shows a typical $I$-$V$ characteristic for the SQUID, biased with a constant current, illustrating the change in critical current and observed change in output voltage for applied flux, $\Phi_a=n\Phi_0$ and $(n+1/2)\Phi_0$.
In typical amplifier operation, the flux bias is set close to $(n\pm 1/4)\Phi_0$ to maximize the flux-to-voltage transfer coefficient $V_\Phi\equiv\partial V/\partial \Phi_a$, as illustrated in Fig.~\ref{fig:MSA_principles}(c).

\begin{figure*}
\begin{center}
\includegraphics{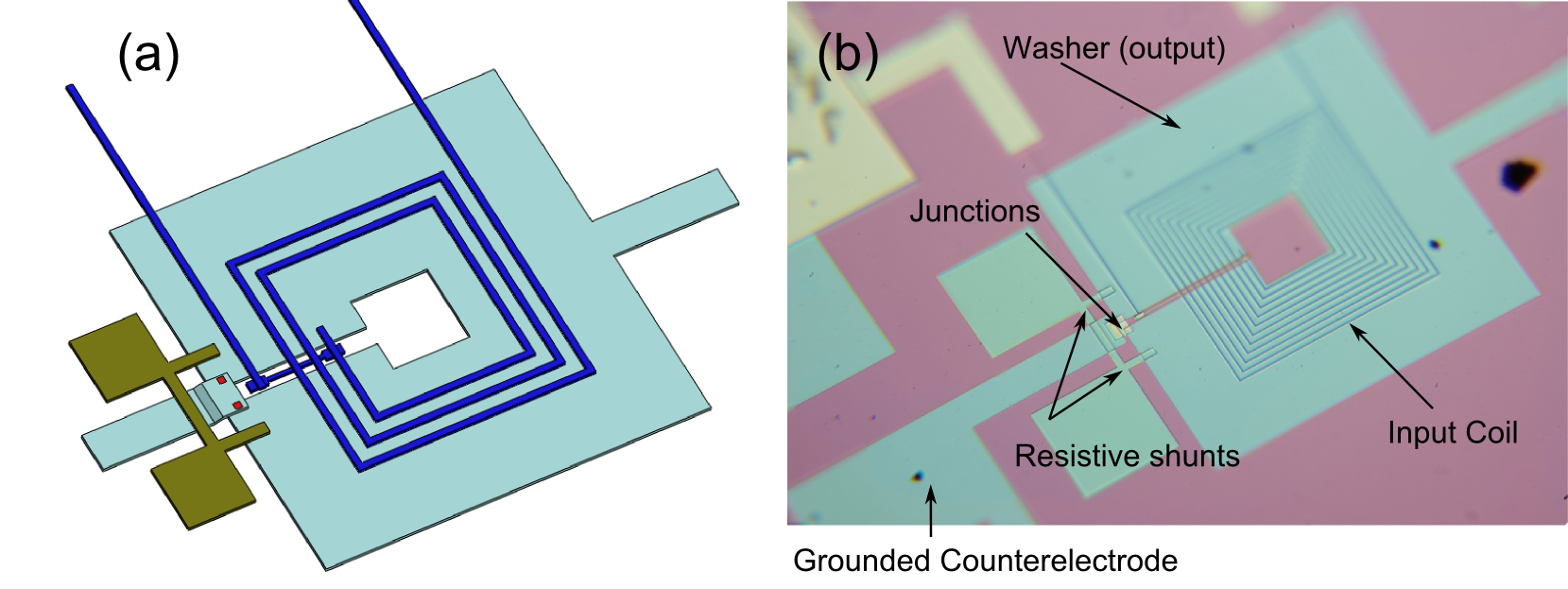}
\caption{Microstrip SQUID Amplifier (MSA). (a) Simplified schematic diagram (not to scale). Small red squares are Josephson junctions. (b)  Optical photograph of 11-turn MSA. Microstrip leads enter from the top. Junctions and resistors at lower left edge. }
\label{fig:MSA_geometry}
\end{center}
\end{figure*}

The layout of a typical SQUID, fabricated from photo-lithography patterned thin superconducting films, is shown schematically in Fig.~\ref{fig:MSA_geometry}(a) \cite{doi:10.1063/1.93210}.
The superconducting loop is deposited as a square washer interrupted by a narrow gap. The gap is bridged by a second superconducting layer, the ``counterelectrode'' connecting to each side of the gap via a Josephson junction.
The input coil is deposited on top of an electrically insulating film overlaying the SQUID washer, so that current in the input coil efficiently couples flux to the SQUID loop.
Figure~\ref{fig:MSA_geometry}(b) shows a SQUID with an 11-turn input coil.
By coupling the input loop to an appropriate input circuit one can realize a highly sensitive amplifier.
With conventional flux coupling between the input coil and the SQUID, however, the highest practical operating frequency of such amplifiers is limited to a few 100 MHz. This is because, with increasing frequency the parasitic capacitance between the input coil and the SQUID washer conducts a larger fraction of the input signal as a displacement current, reducing the magnetic flux coupled into the SQUID.
This limitation is eliminated with the Microstrip SQUID Amplifier (MSA) \cite{doi:10.1063/1.121490,doi:10.1063/1.1321026,Clarke2006}, which makes the washer-coil capacitance an integral component of a resonant microstrip input. The MSA was invented specifically to meet the needs of ADMX.

\subsubsection{Microstrip SQUID Amplifier (MSA): Principles and Fabrication}
\begin{figure}
\begin{center}
\includegraphics[width=8 cm]{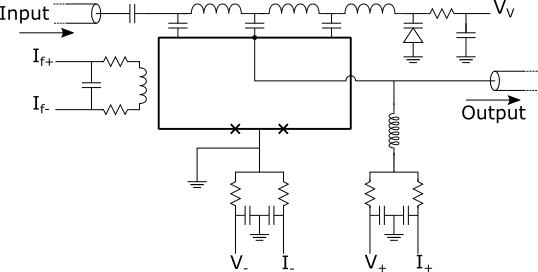}
\caption{Schematic diagram of MSA with grounded counter electrode.
Signal enters from left through coupling capacitor to the input coil, represented as a distributed inductance and capacitance to the SQUID loop.
End of the input coil is terminated by varactor diode to ground at top right.
DC varactor bias voltage $V_\text{V}$ determines varactor capacitance and input coil resonant frequency. Bias current $I_\text{b}$ greater than $I_\text{c}$ applied from $\mathrm{I_+}$ to $\mathrm{I_-}$ maintains SQUID in voltage state.
The terminals $\mathrm{V_+}$ and $\mathrm{V_-}$ are for diagnostic purposes, not essential for operation. Inductor at left couples static flux to the SQUID to set $\Phi_a\approx(n\pm1/4)\Phi_0$ to maximize the transfer coefficient $\partial V/\partial \Phi_a$. SQUID output voltage is connected via a $50$-$\Omega$ line to a transistor-based amplifier.
}
\label{fig:MSA_schematic}
\end{center}
\end{figure}

The circuit configuration of the MSA is shown schematically in Fig.~\ref{fig:MSA_schematic}.
The microstrip is represented as a distributed inductance and capacitance between the input coil and SQUID washer \cite{doi:10.1063/1.121490, doi:10.1063/1.1321026}.

The spiral input microstrip behaves as a $\lambda/2$ resonator of length $\ell$ when its termination is open, provided the coupling to the input line is weak, for instance due to an impedance mismatch and large coupling capacitor reactance.
The capacitance per unit length is well approximated by $C_\ell=(\epsilon/\epsilon_0)(w/d)$, where $\epsilon$ is the dielectric constant of the oxide between the washer and microstrip, $\epsilon_0$ the vacuum permittivity, $w$ the line-width of the microstrip, and $d$ the oxide thickness.
To a good approximation the inductance per unit length is $L_\ell=N^2L/\ell$, where $N$ is the number of turns on the input coil.
Because of the very strong flux coupling between the input coil and SQUID washer other inductances, such as the line inductance and kinetic inductance, are negligible.
The group velocity is then $c'=(1/L_\ell C_\ell)^{1/2}$, the characteristic impedance is $Z_0=(L_\ell/C_\ell)^{1/2}$, and the $\lambda/2$ resonance frequency is $c'/2\ell$.
One tunes the MSA by terminating the microstrip with a voltage-controlled capacitor (varactor diode), enabling one to change the electrical length of the microstrip without changing $Z_0$ or $c'$.

Although the dc SQUID is conventionally operated with the washer grounded, the MSA may be operated with the washer at the output potential \cite{doi:10.1063/1.1321026}, resulting in feedback through the washer-coil capacitance.
The feedback may be either positive or negative since $V_\Phi$, visualized as the slope of the curve in Fig.~\ref{fig:MSA_principles}(c), may be positive or negative depending on the choice of dc flux bias.
Qualitatively, positive feedback results in greater gain, greater noise, and higher resonant frequency, with opposite effects for negative feedback. A detailed account appears in Ref\cite{doi:10.1063/1.1321026}.
At low temperatures, the MSA can achieve quantum limited amplification.\cite{doi:10.1063/1.3583380}
\par Fabrication of the MSA largely follows the standard process \cite{403046} at the National Institute of Science and Technology, Boulder.  This process utilizes an \textit{in situ} sputtered trilayer of Nb, Al and Nb for the Josephson junction, e-beam evaporated PdAu shunt resistors, plasma-enhanced chemical vapor deposited $\mathrm{SiO_2}$ dielectric and an additional sputtered Nb wiring layer.
\par In the MSA shown in Fig.~\ref{fig:MSA_geometry}(b), the washer has inner and outer dimensions $d_1$ = 0.2 mm and $d_2$ = 1.0 mm, respectively, corresponding to an estimated self-inductance $L$ = 1.25 $\mu_{0}d_1$ + 0.3(\si{\pico\henry}/\si{\um})$(d_2-d_1)\approx$ 430 \si{\pico\henry}.
Here, the first term is the inductance of the square hole in the washer and the second the slit inductance \cite{doi:10.1063/1.93210}.
The 11 turns on the microstrip are 2-\si{\um} wide with a 6-\si{\um} pitch.
For each $2.0 \times 2.0$ \si{\um\squared} Josephson junction the shunt resistance is approximately $R\approx$ \SI{11}{\ohm}, the estimated self-capacitance $C\approx$ \SI{310}{\femto\farad}, and the critical current is typically $I_0\approx$ \SI{3.5}{\micro\ampere}.
These values lead to $\beta_{L}\approx1.5$ and $\beta_{c}\approx0.4$ \cite{okelley_thesis}.

\subsubsection{The Microstrip SQUID Amplifier (MSA): Operation and tuning }
\begin{figure}
\begin{center}
\includegraphics[width=8 cm]{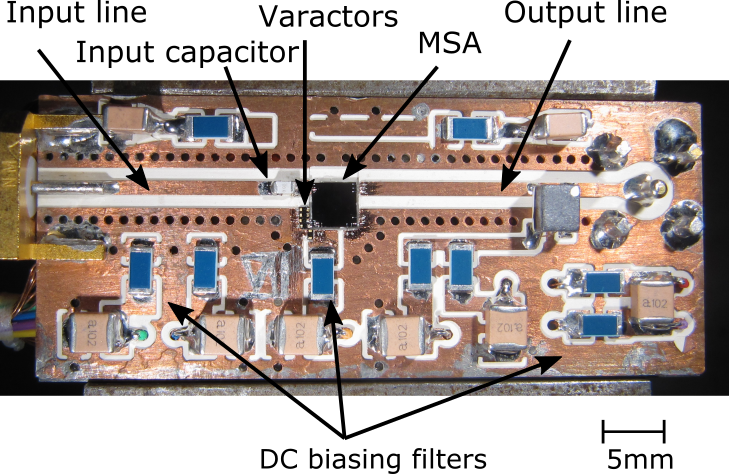}
\caption{Photograph of MSA.}
\label{fig:MSA_circuit_photo}
\end{center}
\end{figure}

Fig. \ref{fig:MSA_circuit_photo} shows the MSA mounted on its RF carrier board.
The signal from the cavity is transmitted to the MSA via a $50$-$\Omega$ coaxial transmission line to the RF board, followed by a $50$-$\Omega$ coplanar waveguide and a fixed capacitor sized for optimal coupling to the resonant microstrip.
When the ADMX cavity frequency is changed by positioning tuning rods, the MSA resonant frequency is tuned by applying a dc bias to the terminating varactor.
A $\lambda/2$ resonance assumes an open circuit ($\pi$ reflection) at both ends, but if either end of the MSA input coil is not an open circuit, the $\lambda/2$ resonant frequency $\omega_0$ is altered by $\omega=\omega_0(\phi_{in}+\phi_{end})/2\pi$, where the reflected phases $\phi_x$ are given by $\tan (\phi_x/2)=iZ_{x}/Z_0$.
Here, $Z_x$ is the loading impedance at either end of the microstrip and $Z_0$ the input coil ac impedance \cite{okelley_thesis}.
This formula is general--purely reactive loads generate a real reflected phase, but a real (resistive) load component generates an imaginary  (lossy) reflected phase.
The varactors \cite{MA-COM} appear at the top-right corner of the schematic in Fig.~\ref{fig:MSA_schematic} and at the lower-left corner of the MSA in Fig.~\ref{fig:MSA_circuit_photo}.
All dc bias signals (current, flux, and varactor tuning) pass through discrete RC low-pass filters mounted on the RF carrier.

It is challenging to achieve substantial gain with a conventional MSA at frequencies above $1$ GHz. Subsequent development of the MSA, however, in which it is operated in higher order modes than the $\lambda/2$ mode, enables high gain at gigahertz frequencies, for example, $24$ dB at $2.6$ GHz \cite{doi:10.1063/1.4985384}.

\begin{figure}
\begin{center}
\includegraphics[width=8.5 cm]{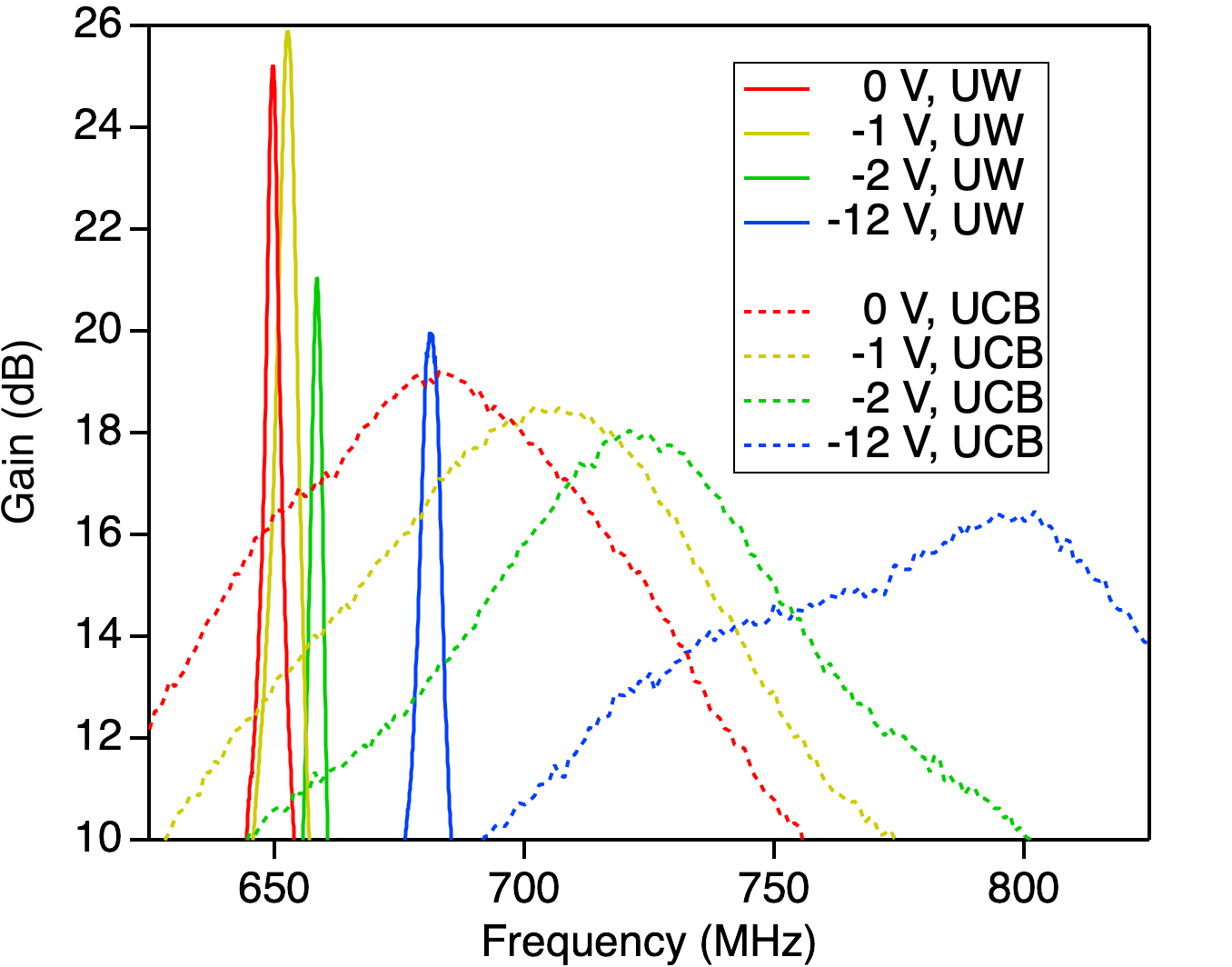}
\caption{MSA tuning vs. varactor voltage bias from $0$ to $-12$ V. (dashed) At Berkeley. (solid) In haloscope at Seattle.
}
\label{fig:MSA_tune_UW_UCB}
\end{center}
\end{figure}

Fig. ~\ref{fig:MSA_tune_UW_UCB} shows the gain vs. frequency of the MSA as measured at Berkeley (UCB) and as measured at Seattle (UW).
 The difference in performance is stark.
We note that at UCB the MSA was connected to ideal $50$-$\si{\ohm}$ loads, required magnetic shielding only from the ambient $50$-$\si{\micro\tesla}$ geomagnetic field, and operated at a bath temperature of $\SI{60}{\milli\kelvin}$. At UW the MSA was connected to switches and circulators that may have non-negligible S$_{11}$ parameters. Furthermore, it required both active and passive cancellation of the $6.8$ $\si{\tesla}$ haloscope magnetic field, and operated at a bath temperature of $\SI{300}{\milli\kelvin}$. Despite the unexpected performance, the noise temperature was sufficiently low to achieve sensitivity to the DFSZ threshold (see Sec. \ref{sec:1A_Noise_temp} ``Run 1A Noise Temperature").

In the transition from Run 1A to Run 1B, ADMX switched from an MSA to a JPA.

\subsubsection{Josephson Parametric Amplifier (JPA): Device and Operation}
The JPA is a low-noise microwave amplifier based on the Josephson junction \cite{josephson1962possible}. JPAs have been developed to achieve quantum limited amplification \cite{roy2018quantum}, adding only a minimum amount of noise required by quantum mechanics \cite{clerk2010introduction}. JPAs used by ADMX were fabricated at University of California, Berkeley. A typical JPA can achieve $20+$ dB of power gain over an instantaneous bandwidth of $10-20$ MHz.
\begin{figure*}
    \centering
    \includegraphics[width=15 cm]{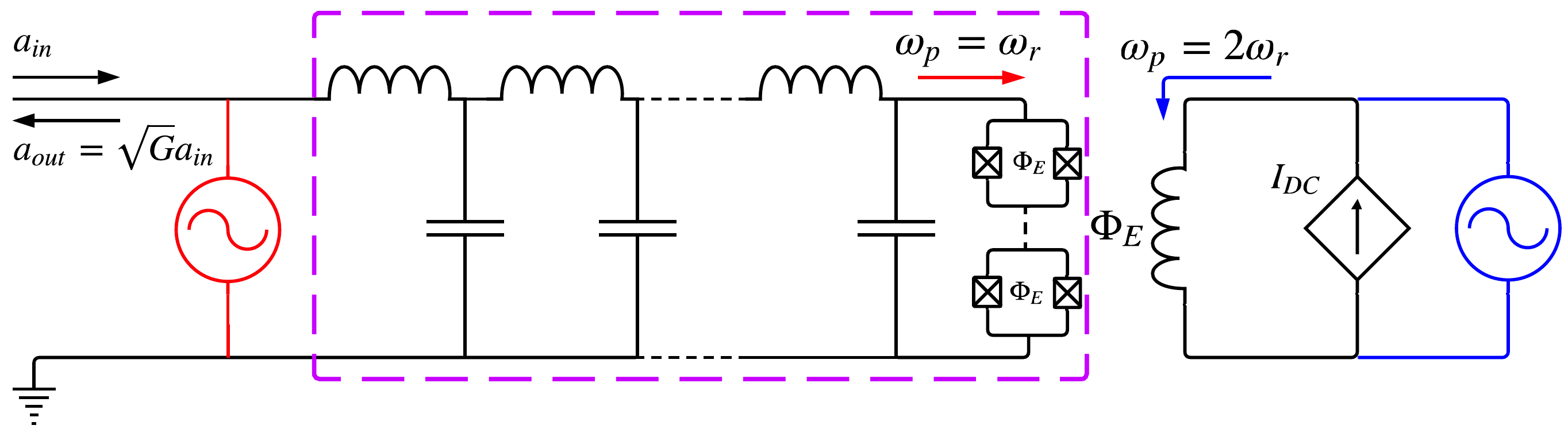}
    \caption{The circuit diagram of a JPA consists of an array of SQUIDs in series with a geometric inductance and shunted by a capacitor (outlined in purple). When the capacitance and inductance are lumped elements in the transmission line leading to the SQUIDs, such as in coplanar waveguides, then it is typically called a lumped-JPA, or LJPA. An external DC magnetic field is inductively coupled to the SQUIDs for flux tuning the resonant frequency. When flux-pumping a JPA at frequency $\omega_p = 2\omega_r$ (blue), where $\omega_r$ is the resonant frequency of the JPA, the pump tone is inductively coupled to the SQUIDs through an external coil or transmission line. When current-pumping a JPA at frequency $\omega_p = \omega_r$ (red), the pump tone propagates down the main transmission line. In both cases, the input signal $a_{in}$ enters through the main transmission line, mixes with the pump tone in the JPA, and then is reflected back down the transmission line. For phase-preserving amplification, such as is done in the ADMX experiment, the outgoing signal has been amplified by a factor $\sqrt{G}$ with respect to the input signal, where $G$ is the power gain of the JPA.
    }
    \label{fig:JPA_schematic}
\end{figure*}

\begin{figure}
    \centering
    \includegraphics[width=8 cm]{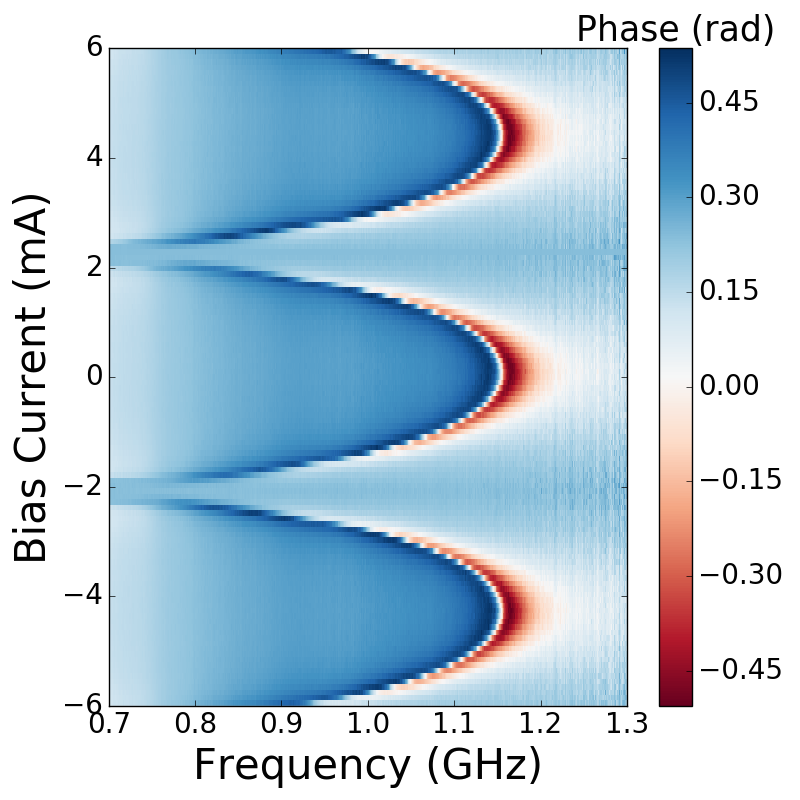}
    \caption{A typical tuning curve produced for characterizing a JPA, demonstrating that the resonant frequency can be tuned downward from its zero-bias state. Each horizontal line is a single trace on a VNA for a given DC bias current. The resonant frequency is measured as a sharp phase shift in the phase of the reflected signal.}
    \label{fig:JPA}
\end{figure}

The JPA is a non-linear oscillator which consists of two Josephson junctions placed in a SQUID loop \cite{jaklevic1964quantum, Clarke2004} (or an array of SQUID loops) shunted by a geometric capacitance (Fig.~\ref{fig:JPA_schematic}). The resonant frequency of a JPA is given by $\omega_0 = 1/\sqrt{(L_J + L_G)C}$, where $L_J$ is the total Josephson inductance of the SQUID loop, $C$ is the shunting capacitance, and $L_G$ is the geometric or stray inductance in the circuit. The inductance of a single Josephson junction can be expressed as
\begin{equation}
    L_J = \frac{L_{J0}}{\sqrt{1 - (I/I_0)^2}} = L_{J0} \bigg( 1 + \frac{1}{2}(I/I_0)^2 + ... \bigg),
\end{equation}
where $I_0$ is the critical current of the junction, $L_{J0} = \Phi_0/2\pi I_0$ is the Josephson inductance in the absence of any supercurrent flow $I$ through the junction, and $\Phi_0$ is the flux quantum. The non-linearity of the junction inductance can be understood through the series expansion of $L_J$, which, for $I \ll I_0$, can be truncated at the quadratic term. The behavior of two Josephson junctions in a SQUID loop can be modeled as single Josephson junction but with a flux-tunable critical current. The critical current $I_c$ of a SQUID can be expressed as a function of an externally applied magnetic flux $\Phi_E$:
\begin{equation}
    I_c(\Phi_E) = 2I_0\lvert\cos{\left( \frac{\pi\Phi_E}{\Phi_0} \right)}\rvert,
\end{equation}
where $I_0$ is taken to be identical for both Josephson junctions. Since $L_J$ is a function of the critical current $I_c$, and since $I_c$ increases non-linearly with an external flux bias through the loop (until the field enclosed is equal to half a flux quantum), the resonant frequency of the device can be tuned downward from its zero-bias state as shown in Fig.~\ref{fig:JPA}.

The operation of a JPA can be understood using the classical picture of parametric amplification: a strong pump tone at $\omega_p$ mixes with the weak signal at $\omega_0$ generating more photons at $\omega_0$. 

A JPA can be operated as a phase-preserving or phase-sensitive amplifier. In phase-preserving amplification, the phase difference between the pump tone and the signal tone is random, leading to an amplification of both quadratures of the readout signal by a factor of $\sqrt{G}$, where $G$ is the power gain of the JPA. Phase-preserving amplification adds at least a half a photon of noise to the readout signal \cite{caves1982quantum}, as required by quantum mechanics. In phase-sensitive amplification, the pump tone is in phase with one of the signal quadratures, leading to an amplification of $2\sqrt{G}$ for the in-phase quadrature and a de-amplification by the same factor for the out-of-phase quadrature. The JPA used in Run 1B was operated in phase-preserving mode for the duration of the data-taking.

Depending on how the JPA is designed and operated, two different types of wave-mixing processes can occur. 

\begin{figure}
  \centering
   \begin{subfigure}
  \centering
  \includegraphics[width=8 cm]{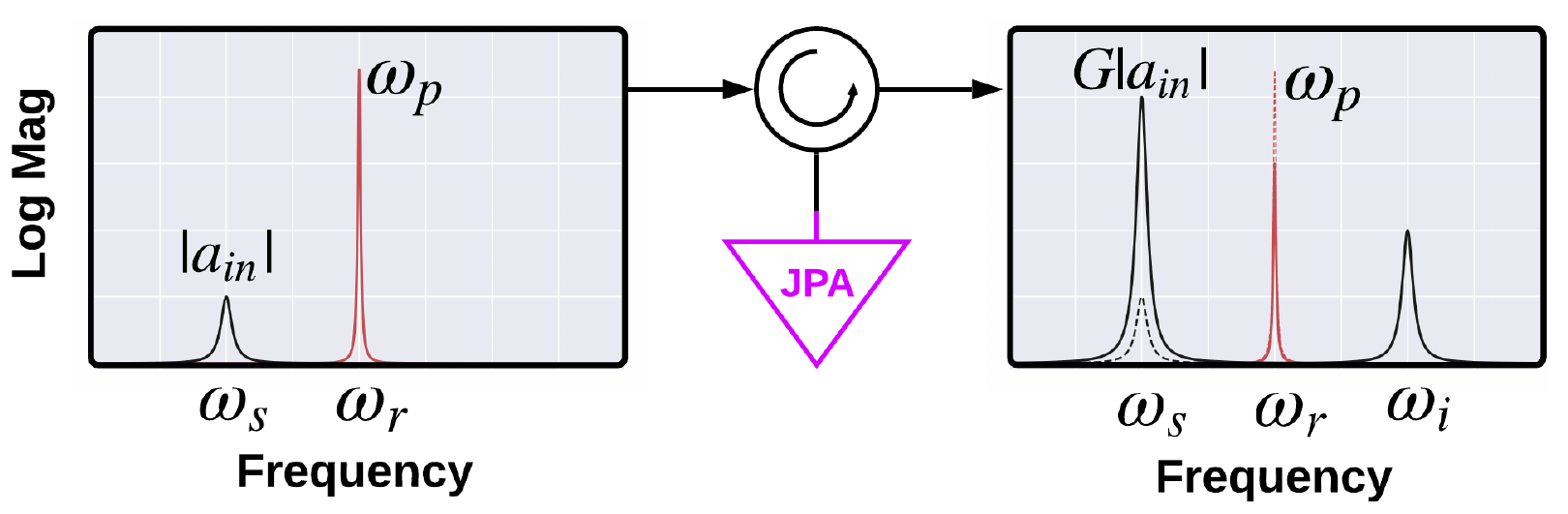}
 \end{subfigure}
  
  \begin{subfigure}
   \centering
   \includegraphics[width=8 cm]{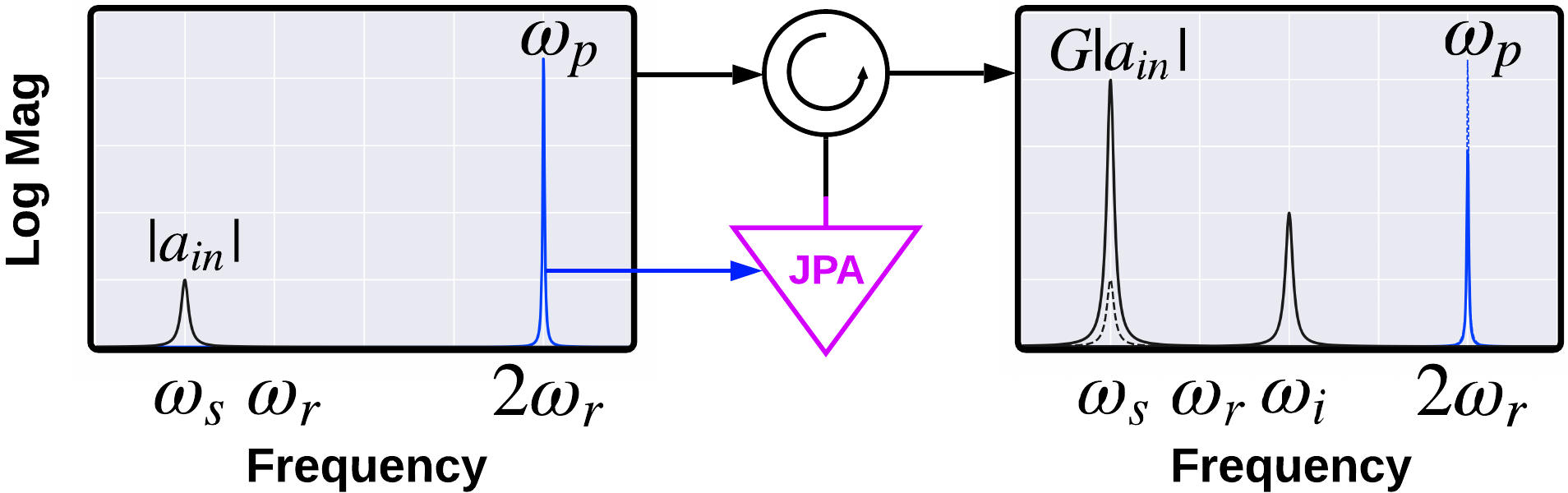}
   \caption{Top: Current-pumping a JPA. In current-pumped JPAs, the small signal tone and strong pump tone both enter the JPA through the main transmission line.
The output spectra contains the amplified signal, the depleted pump tone, and an idler signal. Bottom: Flux-pumping a JPA. In flux-pumped JPAs, the small signal tone enters the JPA through the main transmission line, but the strong pump tone is introduced through an external line inductively coupled to the SQUIDs. The output spectra contains the amplified signal, the depleted pump tone, and an idler signal.}
 \label{subfig:JPAs}
  \end{subfigure}
  \hfill
\end{figure}

In the ``current-pump" design \cite{wahlsten1978arrays, yurke1989observation, olsson1988low, siddiqi2004rf}, as shown in Fig.~\ref{subfig:JPAs}, both the signal tone and the pump tone enter the JPA through the main RF transmission line. In this design, parametric amplification is achieved by modulating the current through the Josephson junctions in the SQUID at the resonant frequency of the device. Since the non-linearity of the device in this scheme is due to a fourth-order Kerr non-linearity, the pump tone is not equal to twice the signal tone, but rather the two are approximately equal to each other, with only a slight detuning on the order of tens of MHz. This process results in four-wave mixing, in which two pump photons are converted into one signal photon and one idler photon. Energy conservation gives $2\omega_p = \omega_s + \omega_i$. The JPA used in the ADMX experiment for Run 1B was of the current-pump design.

In the ``flux-pump" design \cite{castellanos2007widely, sandberg2008tuning, yamamoto2008flux, castellanos2008amplification} as shown in Fig.~\ref{subfig:JPAs}, the signal tone enters the JPA through the main RF transmission line, but the pump tone is inductively coupled to the SQUID. Thus, parametric amplification occurs by modulating the frequency of the resonator by means of an additional external AC magnetic flux. This design results in a three-wave mixing process, in which a single pump photon is converted into signal and idler photons. Energy conservation gives $\omega_p = \omega_s + \omega_i$, where $\omega_p$ is the pump frequency, $\omega_s$ is the signal frequency, and $\omega_i$ is the idler frequency. For flux-pump designs, the pump frequency is approximately twice the signal frequency, $\omega_p \approx 2\omega_s$, thus $\omega_i \approx \omega_s \approx \frac{1}{2}\omega_p$. The main advantage of this design is that the pump tone is largely detuned from the signal tone, so it is easy to filter downstream in the readout line such that the readout is not contaminated by the strong pump tone. Flux-pump design is being investigated for future research and development. 

JPAs are characterized in reflection via a transmission line, which carries the signal tone, the amplified signal, and the pump tone if it is current-pumped. A circulator is needed to route the input signal into the JPA and the amplified signal down the output line and to isolate the pump from reflecting off the cavity and interfering (Fig.~\ref{fig:rflayout_run1b}). The performance of a JPA is controlled by tuning three parameters: flux bias, pump power, and the detuning of the pump tone from the signal frequency. At optimal performance, JPAs typically provide at least $20$ dB of gain over tens of MHz of bandwidth. This was found to be true of the ADMX JPA after installation during Run 1B. If a JPA adds the minimum amount of noise possible, then the SNR improvement of the amplifier is slightly lower than its power gain. Fig.~\ref{fig:jpa_performance} shows UCB fabricated JPA tuning for optimized gain, increase in noise floor when the JPA is on and noise temperature optimization as a function of different relevant parameters used at the UW for Run 1B.

\subsubsection{Josephson Parametric Amplifier (JPA): Fabrication}

\begin{figure*}
\begin{center}
\includegraphics[width=16cm]{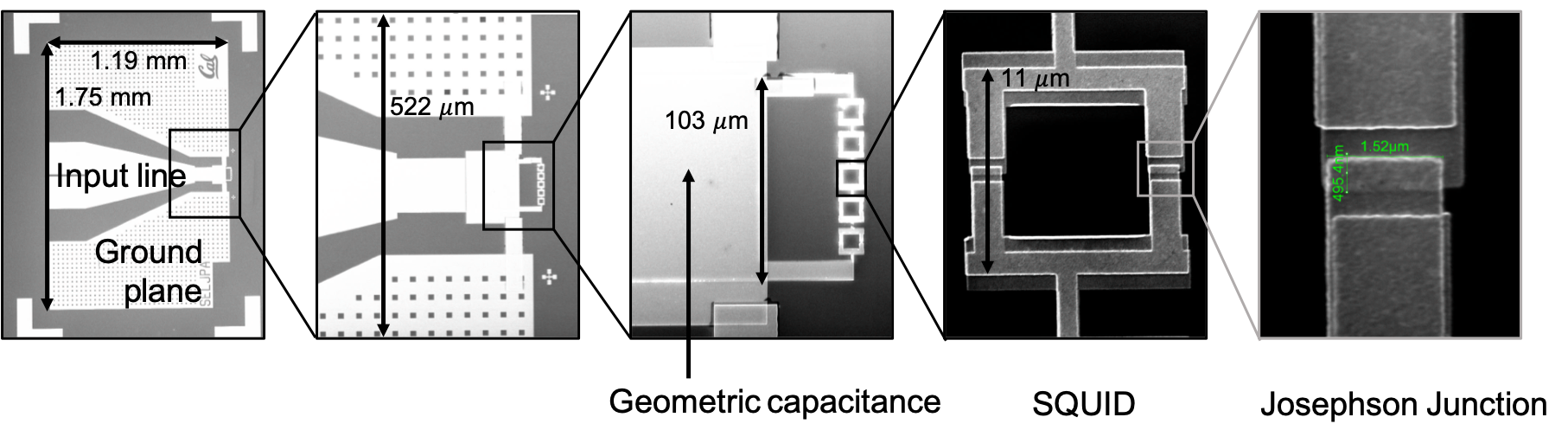}
\caption{JPA chip. The input line and ground plane form a coplanar waveguide leading to an array of five SQUIDs, which is shunted by a large parallel plate capacitor. The dimensions of a typical Josephson junction can be precisely fabricated on the order of tens to hundreds of nanometers. From left to right, the images are increasingly enlarged to show the array of SQUIDS and Josephson junction.}
\label{fig:JPA_chip}
\end{center}
\end{figure*}
JPAs are typically fabricated using electron-beam lithography to expose resist spun on top of a substrate. First, large structures, such as parallel plate capacitors, geometric inductors and RF launches are defined, typically in a lift-off process. Then Josephson junctions, most commonly of the superconductor-insulator-superconductor (S-I-S) variety, are added. The insulating barrier is typically made from a thermally grown oxide (e.g. Al-AlOx-Al, as is the case for the ADMX JPAs). After exposure and development, a double-angle evaporation is used to deposit a thin film of aluminum and the first layer is allowed to oxidize before depositing the second layer. This method is sometimes referred to as the Niemeyer-Dolan technique for fabricating very small overlapping structures. (For a full explanation of this technique, the reader is referred to Ref\cite{slichter2011quantum}.) For an image of a full JPA chip and enlarged images of the smaller structures, such as the SQUIDs and Josephson junctions, see Fig.~\ref{fig:JPA_chip}.

\subsection{Dilution refrigerator}
We use a dilution refrigerator as the final cooling stage of the ADMX detector. Cooling is provided by the circulation of $\mathrm{{}^{3}He}$ for the $\mathrm{{}^{3}He}$ -rich phase to a  dilute phase in the mixing chamber of the refrigerator. The rate is controlled by heater in the still which evaporates $\mathrm{{}^{3}He}$ from the dilute phase  to the gas phase which is almost pure $\mathrm{{}^{3}He}$. 

\begin{figure}[ht!]
\begin{center}
\includegraphics[width=8 cm]{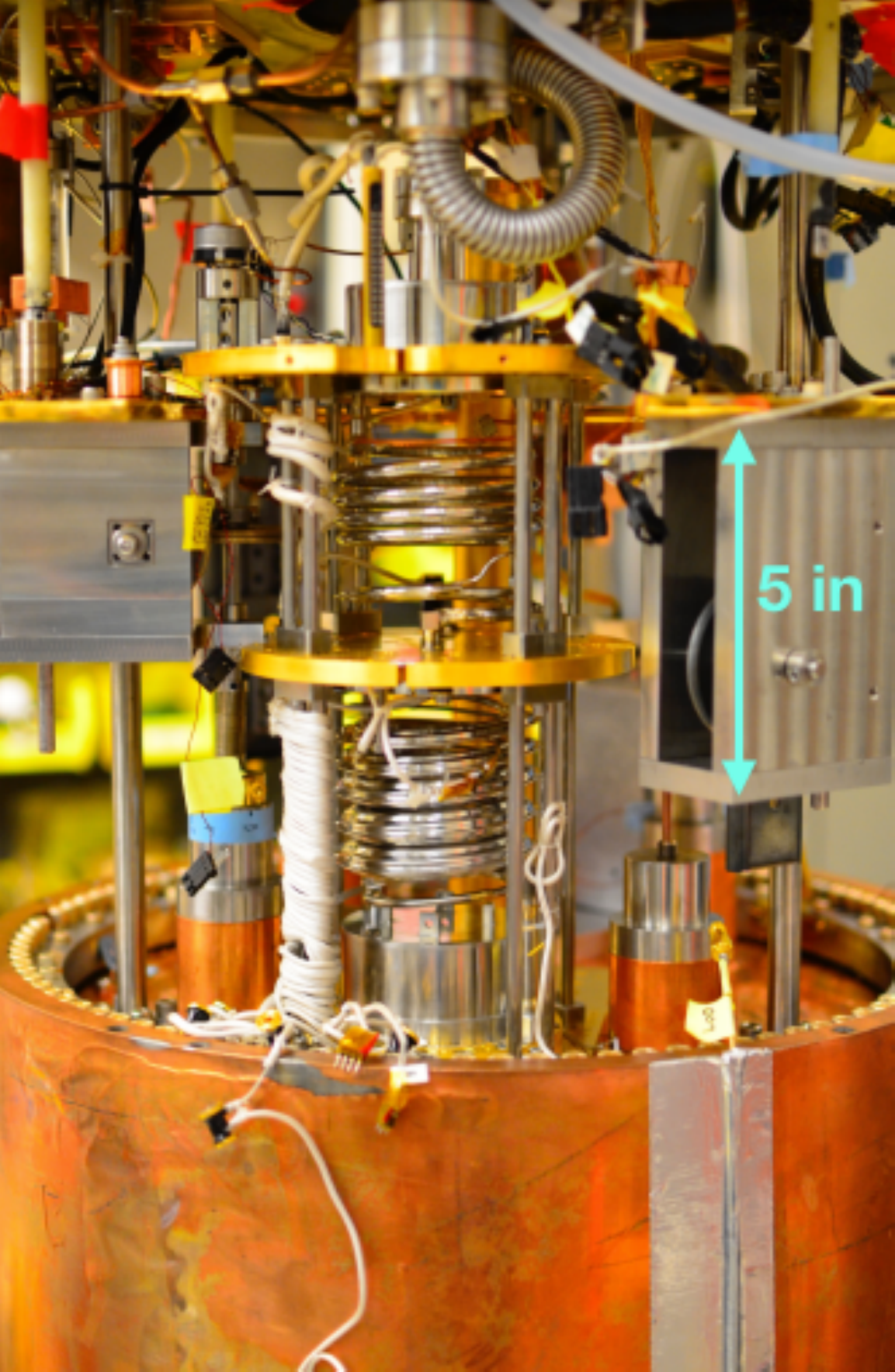}
\caption{Dilution Refrigerator bolted to the microwave cavity.}
\label{fig:DR_Cavity}
\end{center}
\end{figure}

The dilution refrigerator (Model JDR-$800$) was custom built by Janis Research Company~\cite{Janis}. Based on the anticipated heat loads of operating the haloscope, we designed it to have $750$ $\mu$W of cooling power at $100$ mK. An actively driven still heater is used to control the $\mathrm{{}^{3}He}$ flow rate.  A pumped $\mathrm{{}^{4}He}$ refrigerator pre-cools the $\mathrm{{}^{3}He}$ returned to the dilution refrigerator before it enters the still. 
The still is pumped by a Pfeiffer CombiLineTM, Model WS$1250$WA (OKTA-$2000/$A$100$L) oil-free pumping station. Roots pumps use Fomblin\textregistered ~oil to reduce hydrogen sources in the dilution refrigerator system.  At room temperature, LN2 traps and a hydrogen getter (MC$1500902$F from saesgroup Pure Gas) are used to clean the mixture.  A gold-plated intermediate plate is used to bolt the cavity and cryogenic electronics tower  to the mixing chamber of the refrigerator to ensure good thermalization. The attachment of the cavity to the dilution refrigerator mixing chamber is shown in Fig.~\ref{fig:DR_Cavity}.

\subsection{Sidecar, a high frequency prototype}
\begin{figure}
\begin{center}
\includegraphics[width=8 cm]{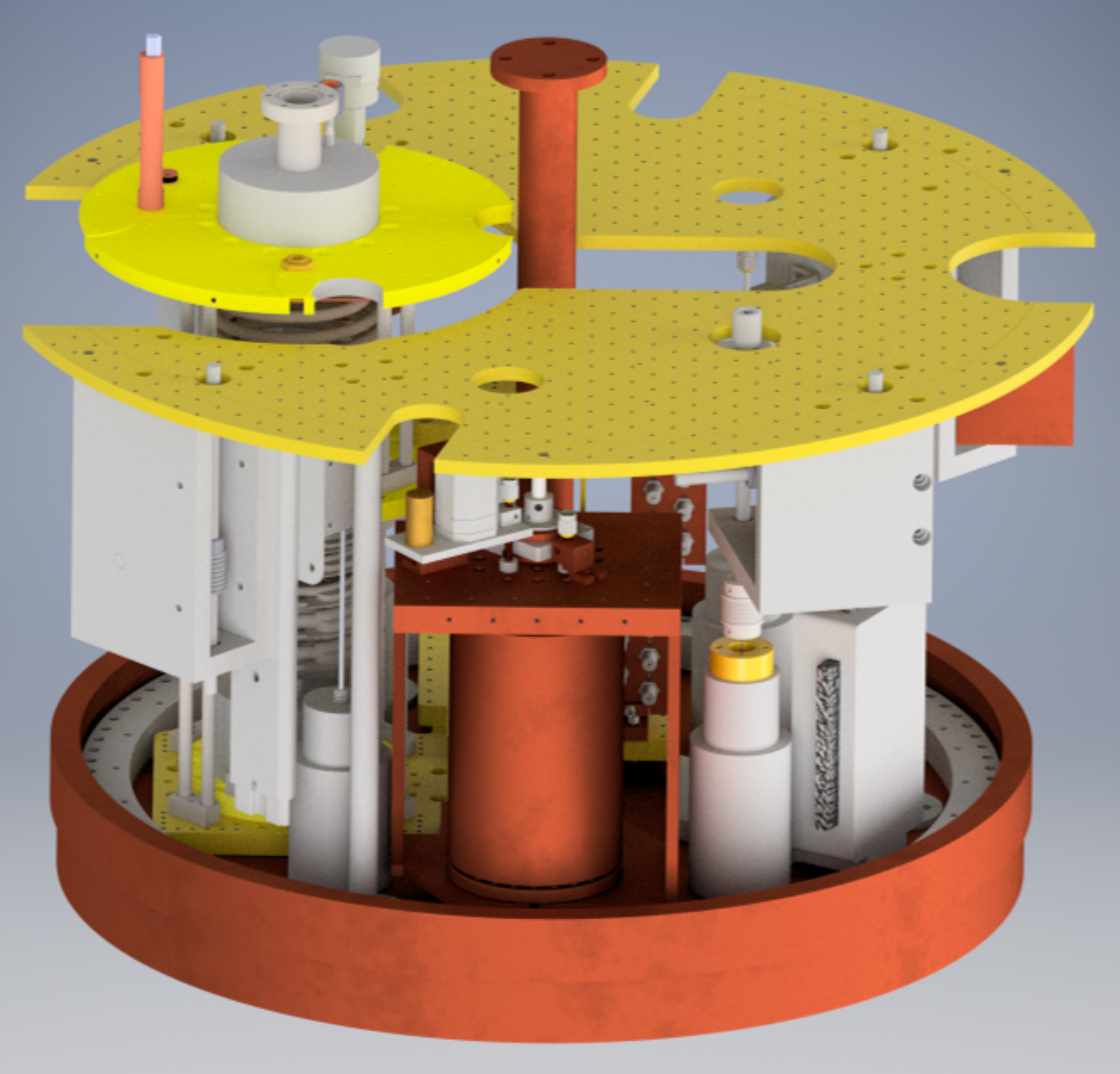}
\caption{CAD rendering of the Sidecar cavity in situ between the larger main cavity top plate (bottom) and $1$ K plate (top). The cavity and piezo-electric actuators are thermally sunk to the top of the Sidecar cavity and the large copper support frame minimizes thermal spikes associated with actuators stepping.}
\label{fig:sidecarLocation}
\end{center}
\end{figure}

Ultimately, we want to operate ADMX at higher frequencies of several GHz to probe higher mass axions. As a preliminary test, we allocated a small amount of space inside the insert to operate the Sidecar cavity (Fig.~\ref{fig:sidecarLocation}). Sidecar operated a $4-6$ GHz cavity in the TM$_{010}$ mode and has been used to demonstrate that data can be taken on the TM$_{020}$ mode which extends the cavity frequency range to $7.2$ GHz. One of the main differences between Sidecar and the ADMX main cavity is that Sidecar is tuned using attocube piezo-electric actuators~\cite{attocube}. Consequently, Sidecar is a prototype test-bed for future motion control system for the main cavity. The attocube actuators are less bulky and dissipate less heat than the currently used stepper motors, and will be implemented in future ADMX runs. In its location on top of the ADMX cavity, Sidecar experiences a mean field that is one half that of the main cavity. Thus, Sidecar also acts as a fully operational haloscope operating at higher frequency. For further information on the Sidecar cavity see Ref\cite{PhysRevLett.121.261302}.

\section{Helium Liquefaction System and Infrastructure}

To meet the high liquid $\mathrm{{}^{4}He}$ demands of the two solenoid magnet systems and the two $1$ K pots for the dilution refrigerator mixture condensing line and cavity thermal shields, a Linde L$1410$ $\mathrm{{}^{4}He}$ liquefaction system was installed at CENPA, which includes a closed loop system of the liquefier, Rotary Screw Compressor, screw compressor, and two Bauer compressors. Any helium vapor produced by the system is collected into a gas bag with a volume sufficient to hold $25$ liquid liters of $\mathrm{{}^{3}He}$ gas (equivalent to $19,000$ L gas at STP). From the gas bag, a screw compressor is used to compress the gaseous $\mathrm{{}^{4}He}$ into a medium pressure tank up to $10^{6}$ N/m$^2$. The medium pressure helium is then regulated into a pair of Bauers high pressure compressors to a system of $12, 1.4$ m tall, standardized high pressure rated cylinders (T-bottles) up to $10^{7}$ N/m$^2$. These T-bottles supply the L$1410$ liquefier with the required high pressure helium via two T-bottles acting as a surge tank to quell any instabilities in the supply pressure. 
\par The liquifier produces $15$ to $45$ liquid liters of $\mathrm{{}^{4}He}$ per hour, depending on whether liquid nitrogen pre-cooling is used and the purity of helium flowing into the purifier of the liquefier. From the liquefier, the liquid $\mathrm{{}^{4}He}$ is routed via a remote delivery tube (RDT) to a Mother Dewar of 2500 liquid liters volume. Stored liquid $\mathrm{{}^{4}He}$ is then transferred into either the main magnet or a reservoir.

In addition, University of Washington CENPA's cryogen infrastructure includes a large liquid nitrogen (LN2) tank that is used to provide LN2 for dilution refrigerator traps as well as to provide a $77$ K thermal shield for the main ADMX magnet. 

\section{DAQ Infrastructure and warm electronics}
The data acquired for ADMX can be divided into periodically sampled experimental state information and RF measurements taken during the axion search. The experimental state information consists of readings from temperature, pressure, magnetic field, and current sensors.
\subsection{Sensors}
For temperatures above 1 K, we used an assortment of platinum resistance (Lakeshore PT-$102$), Cernox (Lakeshore CX-$1010$ and CX-$1050$), and Ruthenium Oxide (Lakeshore RX-$102$ and RX$202$) sensors for read out. Sensor resistances were measured by performing a four-wire resistance measurements with an Agilent Multifunction Measure Unit. For temperatures below $1$ K, resistance measurement were made with a Lakeshore $370$ Alternating Current (AC) resistance bridge \cite{lakeshore_sensor}. To ensure heating of the sensors from the resistance bridge was minimized, the excitation voltage from the resistance bridge was reduced until the excitation voltage had no noticeable effect on the resulting temperature measurement (while still maintaining a high enough voltage to minimize noise). For temperature sensors at the 100 mK stage of the experiment, the excitation voltage that minimized heating was $20$ $\mu$V. 
In Run 1A, the temperatures of the cavity and quantum electronics package were measured using Cernox temperature sensors (Lakeshore CX-$1010$), while the temperature of the mixing chamber mounted to the cavity was measured using a Ruthenium Oxide temperator sensor (Scientific Instruments RO$600$). During operations, we observed that the Cernox sensors had a large magneto-resistance at temperatures below $1$ K. When the magnet was ramped to $6.8$ T, it was observed that temperature readings on the Cernox temperature sensors on the cavity increased by $70\%$ compared with the Ruthenium Oxide temperature sensor on the mixing chamber, which increased by $2\%$. Thus, in Run 1A, the temperature of the cavity was determined by the Ruthenium Oxide temperature sensor mounted to the mixing chamber. Because the quantum electronics package was kept in a field-free region, the Cernox temperature sensors located on the package did not suffer from magnetic field effects, and were used to measure the physical temperature of the quantum amplifier. 

In Run 1B, the Cernox sensors on the cavity and quantum amplifier package were replaced with Ruthenium Oxide sensors (Scientific Instruments RO-$600$). 

The vacuum insulation space between the insert and the magnet bore was monitored with an ionization gauge which is kept below $10^{-7}$ torr during operation. When the main magnetic field is changed, the magnetic field cancellation near the quantum electronics is verified by Hall probe measurements, but these probes are not energized during data taking due to the excess heat they generate.

\subsection{RF chain and Run Cadence}

\begin{figure}
\begin{center}
\includegraphics[width=8 cm]{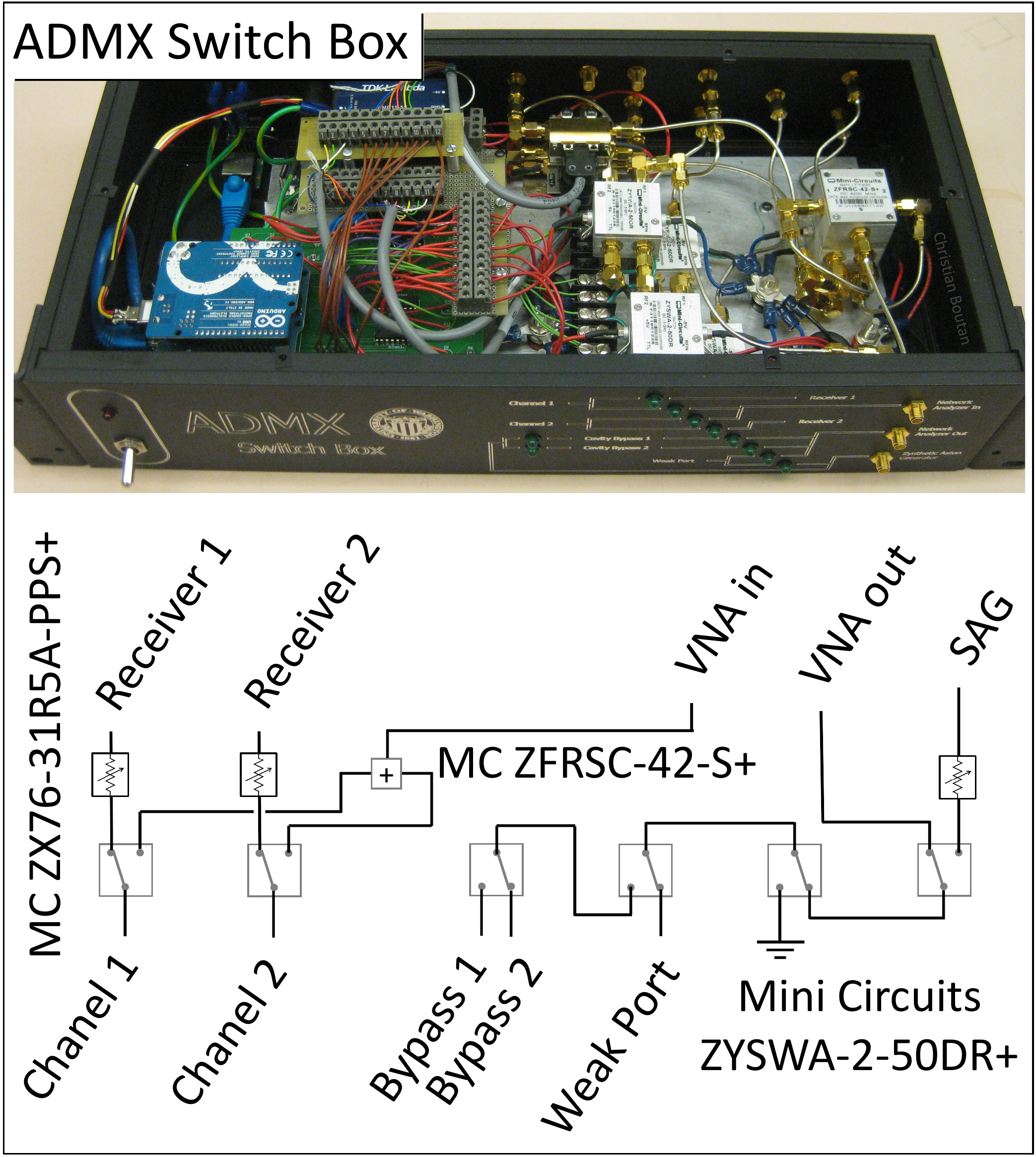}
\caption{Custom ADMX switch box that has been directing RF traffic since 2013. Signals from the vector network analyzer or synthetic axion generator are sent to either the weakly coupled antenna or to bypass lines used for antenna coupling measurements. Signals from the cavity are directed to either the ambient temperature receiver or back to the network analyzer. Programmable attenuators allow the option to easily correct for receiver compression, digitizer clipping or cavity input power.}
\label{fig:swbx}
\end{center}
\end{figure}

\begin{figure}
\begin{center}
\includegraphics[width=8 cm]{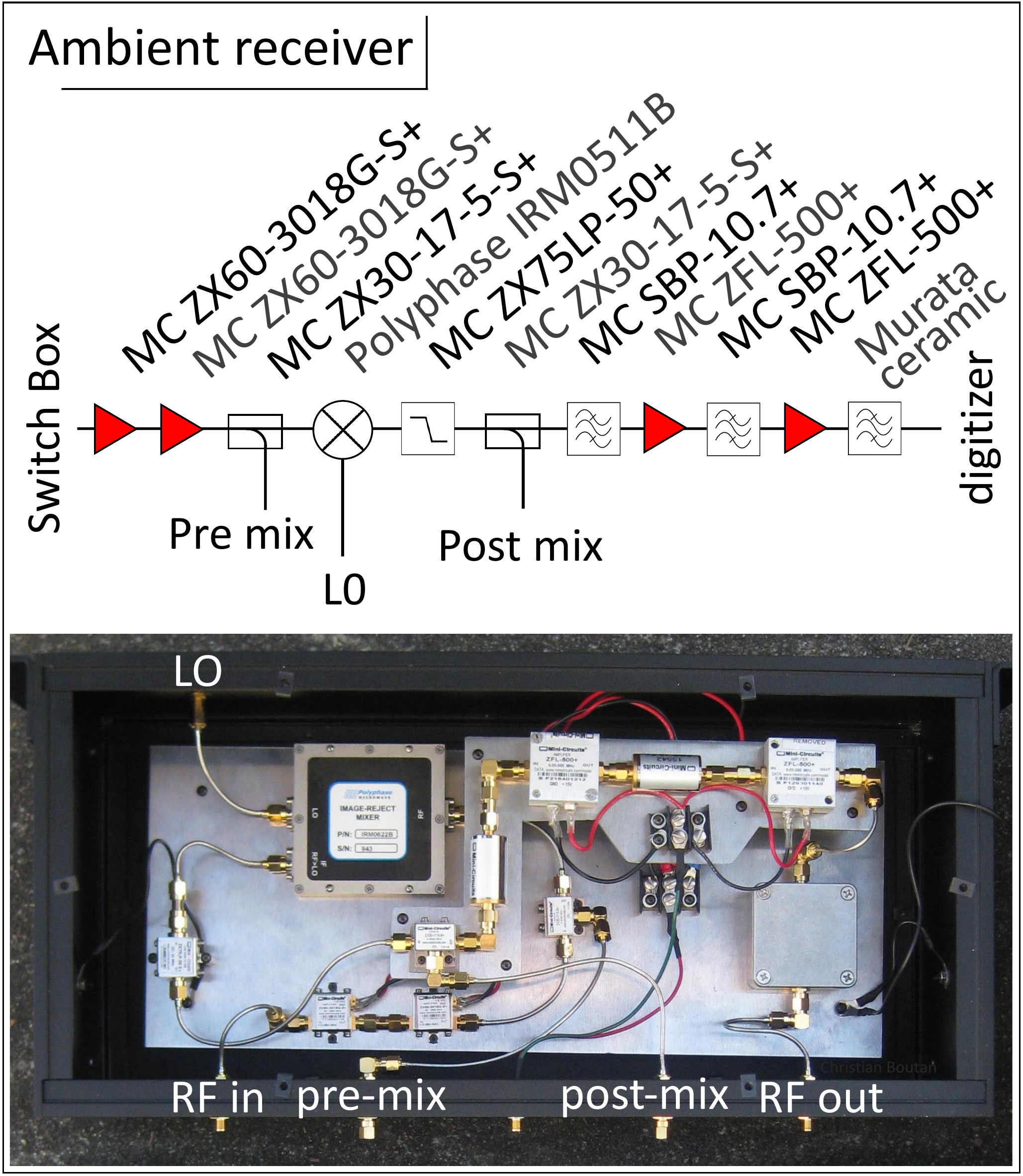}
\caption{Ambient receiver built for data from the TM$_{010}$ mode. Signal is amplified, mixed to 10.7 MHz, using an image reject mixer, bandpass filtered.}
\label{fig:rec}
\end{center}
\end{figure}

Thermal power from the cold space was amplified by a chain of amplifiers with the following approximate characteristic gains: quantum amplifiers (Figs.~\ref{fig:rflayout_run1a} MSA and \ref{fig:rflayout_run1b} JPA), $20-30$ dB, HEMT amplifiers $30$ dB and a series of room temperature amplifiers (Minicircuit), $40$ dB. This power was directed with a custom switch box (Fig.~\ref{fig:swbx}) to a variable-frequency superheterodyne receiver (Fig.~\ref{fig:rec}).

This ambient receiver mixed the signal down to $10.7$ MHz using a Polyphase image reject mixer to remove the higher frequency sideband. Additional amplification compensated for loss along the chain and anti aliasing, narrow bandpass filters centered on $10.7$ MHz ensured the removal of harmonics. The signal was then time series digitized at $200$ MegaSamples/s with an $8$ bit digitizer (Signatech). In software, this signal was digitally mixed down again and filtered, retaining the power spectrum with a bandwidth of $25$ and $50$ kHz in Runs 1A and 1B respectively. If detected, an axion signal would appear as an excess in this power spectrum or as a nearly coherent oscillation in the time series. 

A Keysight E5071C vector network analyzer (VNA) was used to make active measurements of the RF system. It sent frequency-swept signals through the switch box (Fig.~\ref{fig:swbx}) to the weakly coupled cavity port and measured the complex response of signals transmitted through the cavity ($S21$) and off the antenna ($S11$). Transmission measurements (Figs.~\ref{fig:rflayout_run1a} and ~\ref{fig:rflayout_run1b}) provided information about the cavity mode structure as well as the frequency and Q of the TM$_{010}$ mode. Knowledge about the antenna coupling was obtained with reflection measurements, performed by directing swept power through a bypass line and circulator to the antenna. When critically coupled, on resonance power was absorbed by the cavity. Off-resonance, the signal was reflected off the antenna and up through the receiver chain (see Figs. \ref{fig:ant}, \ref{fig:rflayout_run1a} and \ref{fig:rflayout_run1b}). This off-resonance baseline was also used for wide-band measurements of system gain for noise calibrations.
\par The data-taking process is fully automated via custom DAQ software tools that provide a number of useful features such as remote monitoring of state information. The lowest layer of the DAQ software is based on Experimental Physics and Industrial Control System (EPICS) \cite{EPICS}, which provides a uniform software interface for interaction with the instruments.
As data are acquired through EPICS, they are periodically logged in a SQL database.  The on-site database is synchronized with an off-site database mirror that allows for backup and analysis access.
The experiment was automated through a series of scripts written in Lua \cite{LUA}.  These scripts controlled the serial measurements made during the course of normal operations. Individual scripts customized for a specific task could also be developed for the purpose of engineering studies throughout the run.
During the course of a run, experiment operators interacted with the DAQ software through a web interface that enabled remote monitoring and plotting of experimental state information.

\subsection{Synthetic Axion Generator}
RF signals that were indistinguishable from an isothermally modeled axion signal were injected into the experiment through the weakly coupled port in the cavity to ensure the robustness of the experiment in detecting axions. Known as synthetic axions, these signals were generated using an arbitrary waveform generator (Keysight AG$33220$A) which produced a Maxwell-Boltzmann-like line-shape approximately 500 Hz in width. This signal was then mixed up to hundreds of MHz frequencies corresponding to the search range of ADMX. The injected power was varied by changing the output power of the arbitrary function generator and calibrated to the range of power predicted for QCD axions within the experiment. The synthetic axion injection system was implemented in Runs 1A and 1B with a blind injection scheme in Run 1B to introduce artificial axion candidates into the data. 

\subsection{System Noise temperature}
The detectability of an axion signal depends on the magnitude of the noise background. Because the background is almost entirely thermal noise, it is of paramount importance to understand the system noise temperature which includes thermal fluctuations from the photon occupation of the cavity, power fluctuations due to amplification electronics, as well as attenuation that decreases the SNR. It should be noted that the contribution to the system noise from the fluctuation of photon occupation of the cavity is given by the fluctuation-dissipation theorem or generalized Nyquist theorem ~\cite{Callen1951} as
\begin{equation}
    P_{n} = k_{B}Tb\left(\frac{hf/k_{B}T}{exp(hf/k_{B}T)-1}\right) + \frac{hfb}{2}.
    \label{eq:Eq.8}
\end{equation}
Here, $k_{B}$ is the Boltzman constant, $T$ is the physical temperature of the thermal source, $b$ is the bandwidth over which the noise is measured, $h$ is Planck's constant and $f$ is the frequency. The first term in the equation corresponds to the thermal noise power radiated into a single waveguide mode by a blackbody. The second term denotes the zero point fluctuation noise \cite{Devyatov1986}. In the thermal limit $hf<<k_{B}T$, $P_{n}$ converges to $k_{B}Tb$, the background noise contribution arises solely from the physical temperature of the cavity. From Eq.\ref{eq:Eq.5}, the system noise temperature $T_{sys}$ can be defined such that the ratio of the signal power of an axion signal coming from the cavity with a bandwidth $b$ to nearby background noise Gaussian power is
\begin{equation}
    SNR=\frac{P_{axion}}{k_{B}T_{sys}b}.
\end{equation}
It is important to recall the following two equations:  The noise from a thermal source of temperature $T$, followed by an amplifier with gain $G$ and noise $T_a$ is equivalent to a thermal source with equivalent noise temperature \cite{friis1944}
\begin{equation}
    GT_{\mathrm{equiv}}=G\left(T+T_a\right),
\end{equation}
while a thermal source of temperature $T$ followed by an attenuation $\alpha$ held at temperature $T_\alpha$ is equivalent to a thermal source with equivalent noise temperature
\begin{equation}
    T_{\mathrm{equiv}}=T\alpha+T\left(1-\alpha\right).
\end{equation}
These two equations can be combined for an arbitrary cascade of components; in general the earlier stage components like the first stage amplifiers have a more significant effect on the equivalent noise than that of later stage components.
\begin{figure}
    \centering
    \includegraphics[width=\linewidth]{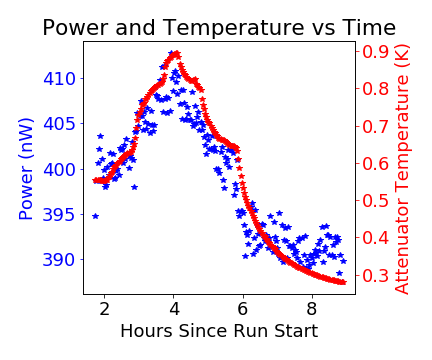}
      \includegraphics[width=\linewidth]{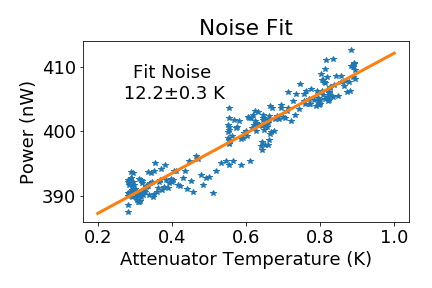}
    \caption{An example of the noise measurement of the HEMT at $705$ MHz performed by heating the millikelvin space. Top plot shows the heating and cooling of the millikelvin space whereas bottom shows the noise fit to the heating and cooling process. These measurements were performed four times during data taking, every 5 MHz throughout the frequency band covered.  They were consistent with uniform noise performance across the band.}
    \label{fig:Hot_Load_Test}
\end{figure}

\begin{figure*}[htb!]
    \centering
    \includegraphics[width=\textwidth]{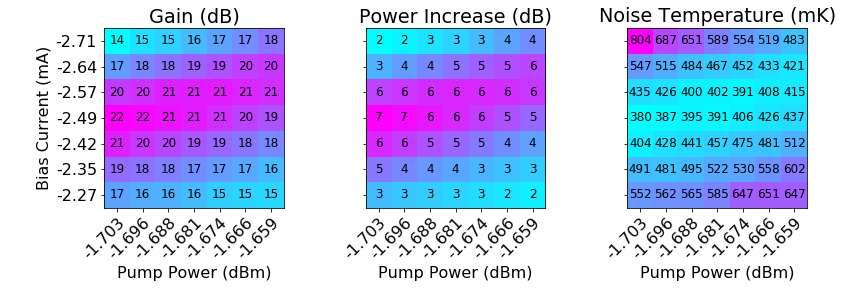}
    \caption{A typical SNRI scan used during operations to select the optimal JPA parameters.  \textit{Left:} The increase in gain between the JPA pump on and off. \textit{Center:} The increase in power between JPA pump on and off. \textit{Right:}  The resultant noise temperature from the combination of the two measurements and the known noise temperature of the downstream electronics.  In this case, a bias of $-2.49$ mA and a pump power of $-1.703$ dBm ( \textless $-100$ dBm on chip) with a system noise temperature of $380$ mK were selected as best for the data taking cycles in that measurement bandwidth.}
    \label{fig:jpa_performance}
\end{figure*}
In the ADMX setup described by Figs. \ref{fig:rflayout_run1a} and \ref{fig:rflayout_run1b} far from resonance, where the cavity is reflective, the primary source of thermal noise is attenuator A.  Near resonance, the thermal noise power is a combination of the power radiated at A and the power radiated from the cavity. In the case of critical coupling, the entire initial thermal noise comes from the cavity.  Beyond the cavity, there are contributions from the attenuation between the cavity and first-stage amplifier, the first-stage amplifier (MSA or JPA), the attenuation between the first and second stage amplifiers and finally the second stage amplifier (HEMT). Beyond the second-stage amplifier, the noise contributions of the receiver components are minor. 
\par As will be discussed in detail later, Runs 1A and 1B were designed such that several in situ measurements could be made of the noise from these various components in the RF chain which will be discussed in further detail.

\subsubsection{Heated Load Measurements}
A heated load measurement refers to physically changing the temperature of one part of the system while monitoring the power over a certain bandwidth.  The system noise can be divided into two components: the part that varies with the physical temperature and the part that does not.  

In ADMX, the performance of the JPA and MSA amplifiers was extremely sensitive to the physical temperature, so that a heated load measurement including the first-stage amplifier proved unreliable.  Two alternative configurations were found, however, both reliable and useful to measuring the system noise temperature.  In the first configuration, the first-stage amplifier was disabled or bypassed and the temperature of the millikelvin electronics was varied while the power far from resonance was measured.  In this case, measured power is
\begin{equation}
P \propto T_{\mathrm{m}}+T_{\mathrm{h}},
\end{equation}
where $T_m$ is the temperature of the millikelvin electronics and $T_h$ is the noise contribution of the HEMT and downstream electronics.  An example of this measurement is shown in Fig. \ref{fig:Hot_Load_Test}.

In the second configuration, the RF system switch, shown in Figs.~\ref{fig:rflayout_run1a} and \ref{fig:rflayout_run1b}, was switched to the hot load.  The temperature of the load could be varied independently from the temperature $T_m$ of the millikelvin electronics stage.  In this case the output power can be expressed as 
\begin{equation}
P \propto \alpha T_{L}+(1-\alpha) T_{\mathrm{m}}+T_{\mathrm{h}},
\label{eq:Eq.11}
\end{equation}
where $T_{L}$ is the temperature of the load, $\alpha$ is the attenuation of all of the components in the millikelvin electronics stage, and $T_h$ is the noise contribution of the HEMT and downstream electronics. Both of these configurations were used in Runs 1A and 1B for noise temperature studies of the HEMTs. 

\subsubsection{Signal-to-Noise Improvement Measurements}

The Signal-to-Noise Ratio Improvement (SNRI) measurement is commonly used in characterizing the performance of ultra-low noise amplifiers. To perform a SNRI measurement, one measures the total system gain and power output in the desired frequency band with the amplifier included and excluded from the RF chain (Figs.~\ref{fig:rflayout_run1a} and \ref{fig:rflayout_run1b}). The ratio of the system noise of the RF system with the amplifier included ($T_{\text{included}}$) to the system noise with the amplifier excluded ($T_{\text{excluded}}$) is
\begin{equation}
    \frac{T_{\text{included}}}{T_{\text{excluded}}}= \frac{G_{\text{excluded}} P_{\text{included}}}{G_{\text{included}} P_{\text{excluded}}}.
    \label{Eq:12}
\end{equation}
Here $G_{\text{excluded}}$ ($G_{\text{included}}$) is the gain when the amplifier is excluded (included), similarly, $P_{\text{included}}$ ($P_{\text{excluded}}$) is the measured power when the amplifier is included (excluded).

An SNRI measurement of the first stage amplifier combined with the heated load measurement of the HEMT and downstream noise yields the total system noise. The values of variables involved in obtaining the SNRI like gain and power increase are shown in Fig. \ref{fig:jpa_performance} as a function of various amplifier bias parameters such as current and pump power for the JPA. In addition, the system noise temperature is also shown as a function of the pump power bias values. 

\begin{figure}
\begin{center}
\includegraphics[width=8 cm]{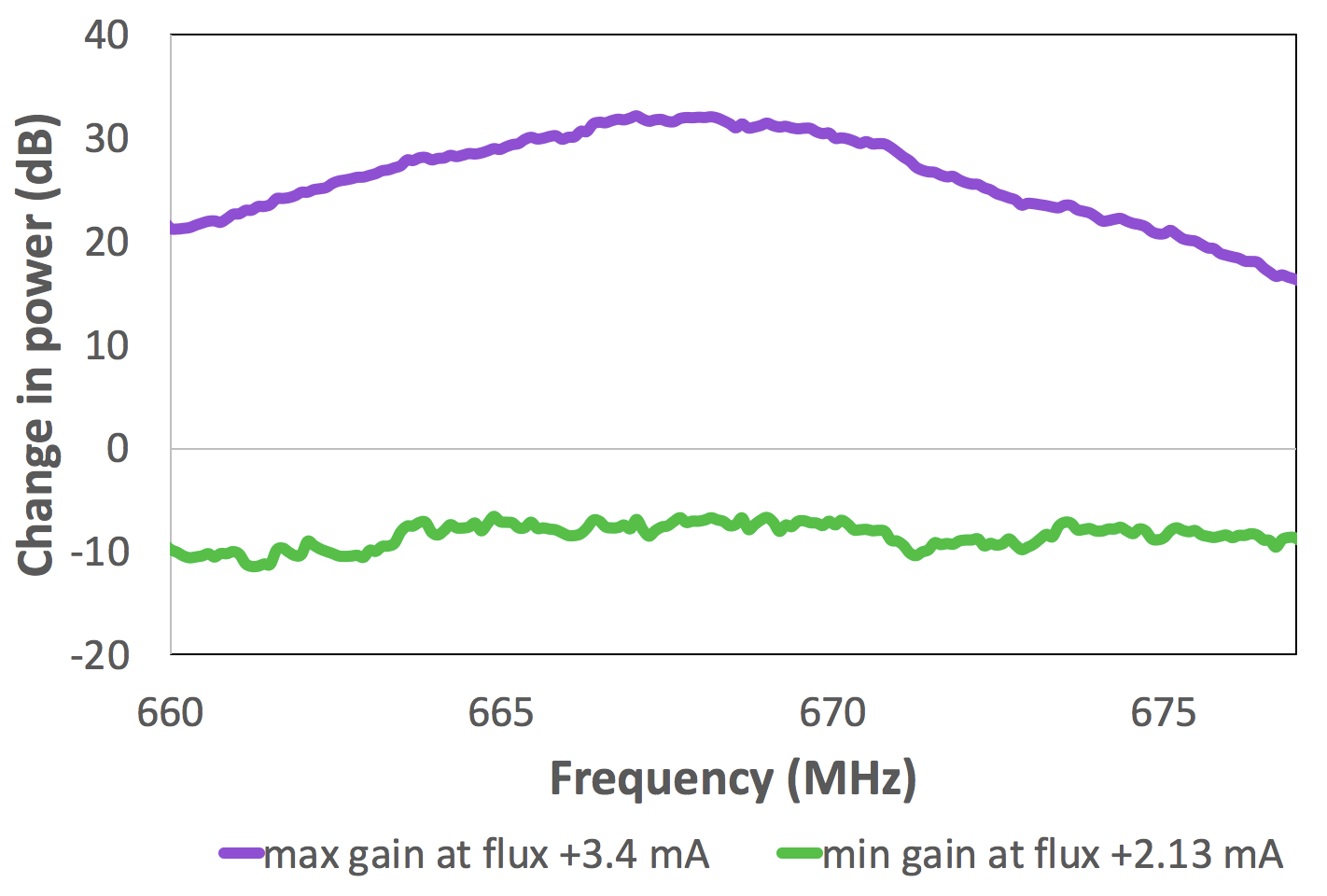}
\caption{Change in power as seen with a network analyzer when the MSA flux was optimized to yield maximum gain (violet) and minimum gain (green) for the constant bias values of current and varactor voltage. This difference between the violet and green curve was subtracted from the gain of the MSA to obtain the SNR increase used in the noise temperature analysis.}
\label{fig:MSA_SNRI}
\end{center}
\end{figure}

In case of the MSA in Run 1A, switching it out of the signal path required actuating two switches shown in Fig.~\ref{fig:rflayout_run1a}, which changed the MSA temperature significantly and impeded proper performance.  Therefore, the power and gain with the MSA switched out were measured infrequently, introducing uncertainty into the measurement of the overall system gain or output power varied over time. However, SNRI measurements could also be made by comparing minimum and maximum power transmitted through the MSA by changing the flux bias while keeping the current and varactor biases constant. An example of this is shown in Fig.~\ref{fig:MSA_SNRI}. 

\begin{figure}
\begin{center}
\includegraphics[width=8 cm]{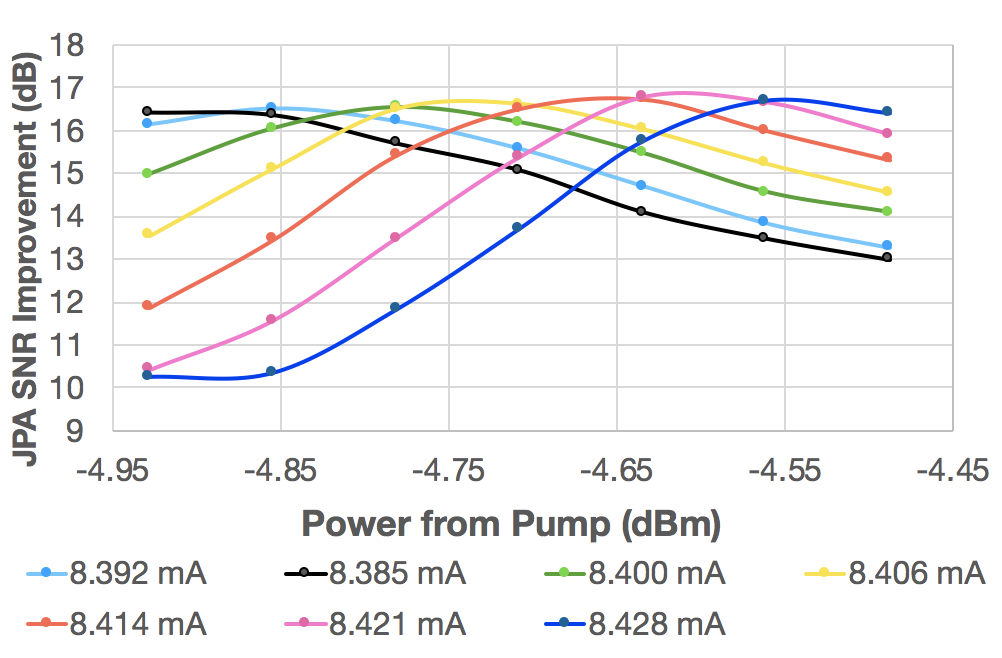}
\caption{JPA SNR improvement as a function of the absolute JPA pump tone at different flux bias values denoted in mA corresponding to different colors. Lines connect the individual points. After the JPA was biased for optimal gain through small steps of flux values, the SNR increase was optimized by scanning through various pump tones.}
\label{fig:JPA_SNRI}
\end{center}
\end{figure}

In the case of the JPA in Run 1B, since the JPA acts as a perfect reflector when the pump tone is disabled, the total SNRI could be measured very quickly with the pump on and off (Fig.~\ref{fig:JPA_SNRI}). Combined with a heated load HEMT measurement, this procedure provided a reliable and immediate measurement of the system noise temperature.

\subsubsection{On-off Resonance Measurements}

In Runs 1A and 1B, the physical temperature of the cavity was significantly different than the physical temperature of the millikelvin electronics. The former was close to $150$ mK ($130$ mK) for Run 1A (1B) and the latter was close to $300$ mK ($250$ mK) for Run 1A (1B). The relative thermal power on and off-resonance encoded sufficient information to determine the system noise in the same way as a heated load measurement.  Provided the attenuation in the millikelvin electronics space, $\alpha_b$, could be determined, and there were no reflections in the system, the expected noise power entering the first-stage amplifier was
\begin{align}
P \propto
(1-\alpha_b) T_m &+ \alpha_b [ (1-\Gamma) T_c \nonumber\\
&+\Gamma \big( (1-\alpha_b)T_m+\alpha_b T_a\big)],
\end{align}
where  $T_a$ is the physical temperature of attenuator A in Fig.~ \ref{fig:2017_fit}, $T_c$ is the physical temperature of the cavity, and $\Gamma(f)$ is the reflection coefficient of the cavity near resonance.  If the antenna is critically coupled, $\Gamma(f)$ is zero on resonance and unity far from resonance. Small reflections within the passive components in the millikelvin space could distort this shape, but an overall fit of a more sophisticated model to the power as a function of frequency enabled us to extract $T_{sys}$ using temperature sensor measurements of $T_m$, $T_a$ and $T_c$.
   \subsubsection{Run 1A Noise Temperature}\label{sec:1A_Noise_temp}

\begin{figure}
\begin{center}
\includegraphics[width=8 cm]{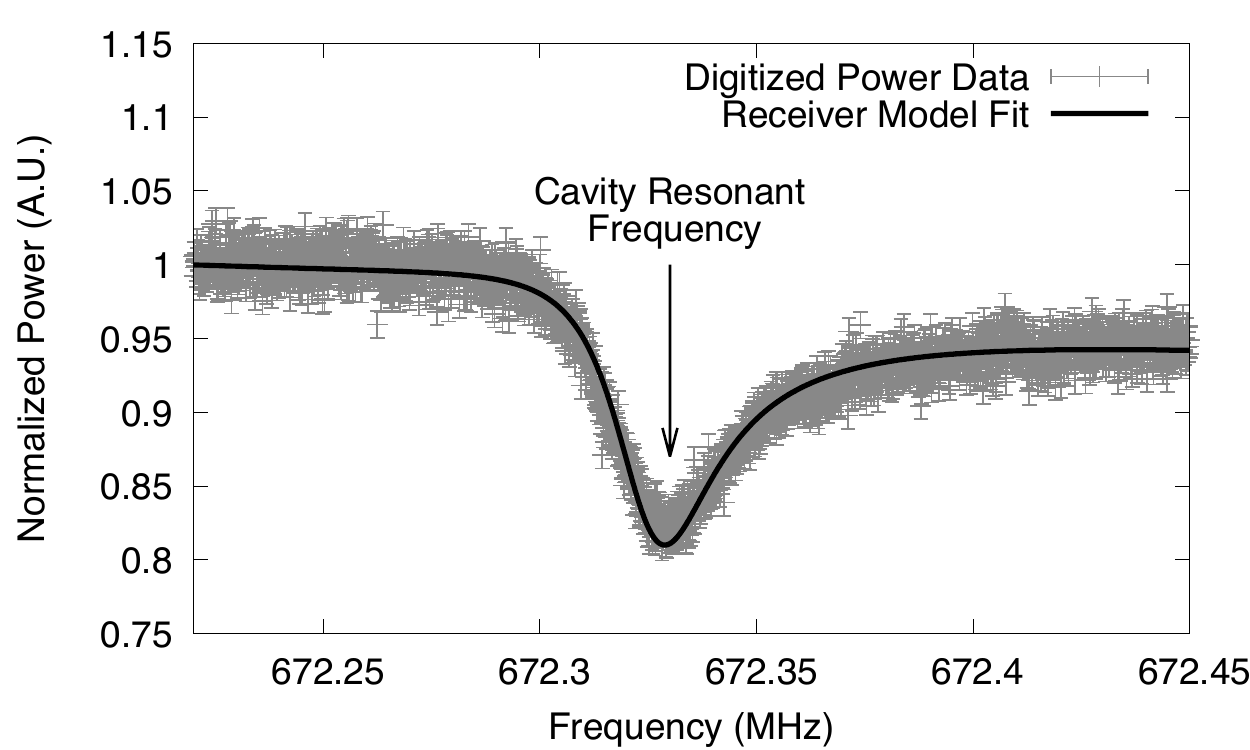}
\caption{On-off resonance measurement for Run 1A \cite{Du_2018}. The power is the sum of
the $300$ mK physical temperature of an attenuator and the
receiver noise temperature. On resonance, the power is
the sum of the $150$ mK physical temperature of the cavity and
the receiver noise temperature. The noise power on versus off
resonance acts as an effective hot-cold load, with the physical
temperatures measured with sensitive thermometers. The asymmetry of the shape is a result of interactions between components,
as described in the Supplemental Material in \cite{PhysRevLett.120.151301}.}
\label{fig:2017_fit}
\end{center}
\end{figure}

In practice, a combination of the above measurements yields a reliable system noise temperature. In Run 1A, the switch for the heated load malfunctioned, so all noise calibration measurements came from the ``on-off resonance" method. Typical parameters for this measurement were: the noise contribution off-resonance of $300$ mK, the noise contribution coming from the cavity and accounting attenuators of $100$ mK, and a dip of order $20\%$ in power seen at the cavity resonance in Fig.~\ref{fig:2017_fit}. This measurement yielded a system noise temperature of order $500$ mK for the run, and was measured at several frequencies.  There was significant variation in the system noise over frequency due to different gains of the MSA.  This variation was tracked by a SNRI measurement before each 100 second digitization using a network analyzer transmission measurement to track the gain, and the average power in the digitization itself to track the noise power.  More information on this is available in the supplemental material for Ref \cite{PhysRevLett.120.151301}.

\subsubsection{Run 1B Noise Temperature}
 In ADMX Run 1B, the hot load had an unintentional touch to the same temperature stage as the cavity and the millikelvin electronics and cavity were closer in temperature than Run 1A, so that the on-off resonance method could not be used effectively. We constructed a model with attenuation through millikelvin electronics $\alpha$, (which included circulators C$_{1}$, C$_{2}$, and C$_{3}$, along with any line losses between components), and noise temperature of the HEMT amplifier as free parameters. In addition, we used the hot load temperature $T_L$, and the millikelvin electronics temperature $T_m$ which we could measure and change independently. This model was  simultaneously fit to measurements where the hot load was heated and where the cryogenic electronics package's temperature was changed. The hot load and heated cryogenic electronics package's measurements were fit to Eq. \ref{eq:Eq.11}.  This yielded the system noise temperature with the JPA without the pump power. It is helpful to recognize that the total loss $\alpha$ can be decomposed into the loss between the cavity and JPA, $\alpha_{C-J}$, and the loss between the JPA and the HEMT, $\alpha_{J-H}$, such that $\alpha=\alpha_{J-H}\alpha_{J-C}$.  Because the components and line lengths are similar, $\alpha_{J-H} \simeq \alpha_{J-C}$.

We made an SNRI measurement frequently throughout the data run. The system noise with the JPA energized was
\begin{equation}
T_{sys,JPA}=T_{sys,HEMT}\frac{P_{JPA}}{P_{HEMT}}\frac{1}{\alpha_{J-H}G_{JPA}.}
\end{equation}
We derived this from Eq.~\ref{Eq:12}, noting that the JPA gain must be weighted by the attenuation between the JPA and the HEMT.  The power and gain of JPA ($P_{JPA}, G_{JPA}$) and HEMT ($P_{HEMT}, G_{HEMT}$) can be measured directly, but the uncertainty in how much of the loss is distributed upstream and downstream of the JPA leads to some uncertainty on this system noise as considered at the JPA input. However, to calculate the equivalent system noise from a signal generated in the cavity (such as an axion), the system noise as measured at the JPA input must take into account the attenuation between the cavity and the JPA which reduces the signal and replaces it with thermal noise from the attenuator, yielding
\begin{align}
T_{sys,Cavity}=~&T_{sys,HEMT}\frac{P_{JPA}}{P_{HEMT}}\frac{1}{(\alpha_{J-H}\alpha_{C-J}){G_{JPA}}}\nonumber\\
=~&T_{sys,HEMT}\frac{P_{JPA}}{P_{HEMT}}\frac{1}{\alpha {G_{JPA}}.}
\end{align}
Thus, the uncertainty in distribution of loss/attenuation does not significantly affect uncertainty in the system noise as compared to signal in the cavity.

Typical measurements for Run 1B were $\alpha$= $3.27\pm0.08$ dB, and a noise temperature at the HEMT input of $8.16\pm0.11$ K. These were independent of frequency below $770$ MHz. The ideal noise temperature of HEMT is notably lower ($2$ K) in the component datasheet, indicating the possibility that the HEMT noise was adversely affected by the magnetic field, an effect studied in \cite{doi:10.1063/1.366000}. SNRI measurements varied between $13$ dB and $16$ dB during the data run (Fig.~\ref{fig:JPA_SNRI}), depending on frequency, amplifier bias conditions, and the physical temperature of the amplifier and the cavity. This yielded noise temperatures of {$350$ to $500$ mK}.  

Above $770$ MHz, the transmission coefficient of the circulators decreased, increasing $\alpha$, and thus yielding slightly higher system noise temperature.   The increase of $\alpha$ was $3$ dB at $800$ MHz, consistent with both the component data sheet and separate cold measurements of the circulators.  This additional attenuation caused a proportional increase in system noise temperature.

\section{Data Analysis}
After acquiring the information regarding the experimental state through sensors and determining the system noise temperature, we analyzed the data by combining the individual $100$ second power spectra collected into a cumulative ``grand spectrum" which we used to search for axion-like signals. The details of the analysis procedure vary between runs but the general steps follow those outlined in Ref\cite{Brubaker:2017rna}. First, the background receiver shape in the individual scans is filtered out. In Run 1A, the background receiver shape was removed by applying a Savitsky-Golay filter (length $121$ and polynomial order $4$) to $95\%$ of the least-deviant power bins~\cite{PhysRevLett.120.151301}. In Run 1B, a six-order Pad\'{e} filter was used to remove the background ~\cite{braine2019extended}. The filtered power spectra were then scaled to the system noise temperature and individual bins of the spectra were weighted by their difference from the cavity’s resonant frequency via the Lorentzian line shape. This produced a spectrum of the excess power due to a potential axion signal. Each of these filtered and weighted scans were then combined into a single grand spectrum representing the excess power from the cavity across the entire frequency range covered in each run. The grand spectrum was then used to search for axion-like signals by a convolution with two different axion signal shapes: a boosted Maxwell-Boltzmann shape predicted from the standard halo model for axion dark-matter, as described in Ref\cite{PhysRevD.42.3572}, and a signal shape predicted from N-body simulations described in Ref\cite{0004-637X-845-2-121}. Data were taken usually in the order of $10$ MHz wide ``nibbles". After an initial sweep in frequency, any axion-like signals were flagged as possible axion candidates and were re-scanned (usually within a week of the orignal scan). Re-scanning of candidates consisted of tuning the cavity to the frequency of the candidate and integrating for a significantly longer time to improve the expected signal-to-noise for a possible axion signal. Any signals that persisted following the re-scan were subjected to individual analysis, as detailed in Ref\cite{PhysRevLett.120.151301},\cite{Du_2020} before moving to the next ``nibble".
\par A medium-resolution and high-resolution channel enabled the search of virialized and non-virialized axions, respectively. Virialized axions are defined as having been gravitationally thermalized, and are expected to follow a Maxwell-Boltzman lineshape (though other lineshapes are studied as well Ref\cite{0004-637X-845-2-121}), with a spectral width on the order of $1$ kHz~\cite{PhysRevD.94.082001}. Analyses of the medium-resolution channel was undertaken in Refs\cite{PhysRevLett.120.151301},\cite{braine2019extended}. A more detailed outline of the analysis procedure which uses the medium-resolution channel and a second lineshape derived from the N-body simulations described is in preparation. Non-virialized axions, on the other hand, have not reached a steady state of motion. Axions would be non-virialized if they have only recently entered the galactic halo, or have been pulled out of tidally disrupted subhalos~\cite{PhysRevD.94.082001}.

\section{Experimental sensitivity}
\begin{figure*}
    \centering
    \includegraphics[width=13cm]{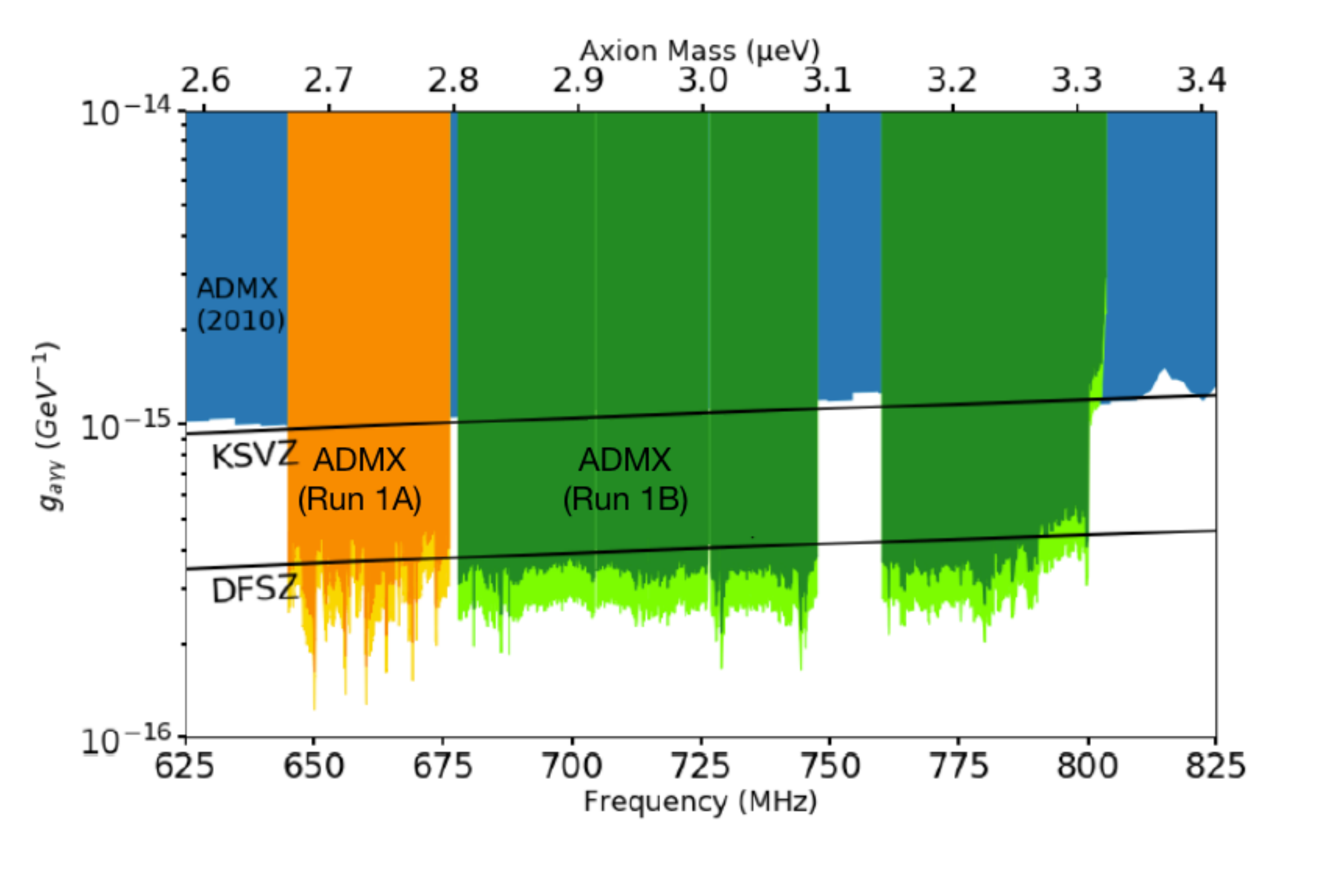}
    \caption{Recent limits set by Runs 1A and Run 1B. 90\% confidence exclusion on axion-photon coupling as a function of axion mass for the Maxwell-Boltzmann (MB)
dark-matter model (dark green) and N-body model (light green) from Ref\cite{Du_2020}. Blue and Orange denote limits reported in ~\cite{PhysRevD.64.092003} and ~\cite{PhysRevLett.120.151301} respectively.}
    \label{fig:limit_plot}
\end{figure*}

With the above discussed hardware upgrades and analysis techniques including the system noise temperature determination, ADMX was able to achieve DFSZ sensitivity in Run 1A in 2017. In this run, axion masses covering $2.66$ to $2.81$ $\mu$eV  corresponding to the frequency range of $645$ to $680$ MHz were probed using the MSA technology. The results are highlighted in Ref\cite{Du_2018}. Similarly, in Run 1B in 2018, we were able to maintain the DFSZ sensitivity while probing axion masses from $2.81$ to $3.31$ $\mu$eV  corresponding to $680$ to $790$ MHz using the JPA technology. This was an improvement of about a factor of three in the frequency coverage from Run 1A in 2017. A detailed summary of these result is plotted in Fig. ~\ref{fig:limit_plot} ~\cite{Du_2020}. These results were a factor of $7$ improvement in the sensitivity from results in 2010 ~\cite{Asztalos:2009yp}. To date, ADMX is the only axion dark matter experiment that has achieved this sensitivity.

\section{Future and Conclusion}

While ADMX has made progress in ultra low noise detector development technology for axion dark matter searches, much of the QCD axion parameter space still remains unexplored. 

As it stands, ADMX uses a single cavity to probe for axions in the $2.4$ -- $6.2$ $\mu$eV ($580$ - $1.5$ GHz) range.  Subsequent phases will search for $6.2$ -- $40$ $\mu$eV ($1.5$ - $10$ GHz) axions,  requiring microwave cavities of smaller volume. To compensate for the reduced axion power deposited in a smaller cavity from axions, multiple cavities will be used that tune together in frequency.  This strategy trades a small volume, yielding reduced sensitivity to axions for the added complexity of frequency-locking multiple cavities at each axion detection frequency step. 

The first ever $4$-cavity array proof of concept for ADMX was tested in the work of Ref\cite{thesisKinion}. A 4-cavity array prototype was implemented at UF to provide a testbed for a multi-cavity array for the ADMX experiment. Designed for a frequency range of $4.4$ to $6.3$ GHz, the prototype array uses a cavity cross section that is roughly $1:3$ scale to that of the planned full size experiment, which will achieve resonances in the range of $1$ to $2$ GHz. The principal design challenge of a multi-cavity system is to maintain a frequency lock between the cavities in order to use the array as a single higher frequency cavity that exploits the maximum volume possible within the ADMX magnet bore. A preprint discussing the details of this design and challenges is in progress. 

In the near future, there are two potential enhancements being developed in parallel to increase the scan rate. The first is exploring increasing cavity quality factor by employing superconducting cavities that can maintain their low RF-losses in a high magnetic field. The second is to explore using squeezing techniques to lower noise beyond the quantum limit, which from $2$ to $4$ GHz, rises from $100$ mk to $200$ mK and begins to dominate the thermal noise of the ADMX system. With these enhancements, ADMX hopes to cover several GHz in frequency over the next few years. Needless to say, systems operating in lower than the quantum-noise-limit is a must before the search can be extended to higher axion mass in a reasonable amount of time. Therefore, superconducting Josephson Junction (JJ) based single photon sensors which count photons thereby eliminating the quantum-noise-limit seem to be a promising avenue to pursue. Furthermore, significant strides in technological advancement is necessary before the complete QCD axion parameter space can be examined.  
\par

\begin{acknowledgements}
This research was supported by the U.S. Department of Energy through Grants Nos. DE-SC0009723, DE-SC0010296, DE-SC0010280, DE-SC0010280, DE-FG02-97ER41029, DE-FG02-96ER40956, DE-AC52-07NA27344, DE-C03-76SF00098 DE-SC0009800
and DE-SC00116655. This paper was written by Fermi Research Alliance, LLC under Contract No. DE-AC02-07CH11359 with the U.S. Department of Energy, Office of Science, Office of High Energy Physics. Support from the Heising-Simons Foundation along with Lawrence Livermore and Pacific Northwest National Laboratories LDRD was also received. Report Numbers: FERMILAB-PUB-20-331-AD-E-QIS, LLNL-JRNL-814312, PNNL-SA-154970.
\end{acknowledgements}
\bibliographystyle{apsrev4-1}
\bibliography{references}
\end{document}